\documentclass[twocolumn,showpacs,prc,preprintnumbers,superscriptaddress,multicol]{revtex4}
\usepackage{graphicx, fancybox}
\usepackage{amsmath,amssymb}
\usepackage{subfigure,dcolumn}
\usepackage{enumerate}
\usepackage{epstopdf}
\usepackage{upgreek}
\usepackage{textcomp}
 \usepackage{multirow}
\usepackage[colorlinks,
            linkcolor=blue,
            anchorcolor=blue,
            citecolor=blue
            ]{hyperref}
\newcommand{\KV}{{\mbox{$\kappa \sigma^{2}$}}}
\newcommand{\SD}{{\mbox{$S \sigma$}}}
\newcommand{\VM}{{\mbox{$\sigma^{2}/M$}}}

\newcommand{\sNN}{{{$\sqrt{s_{_{\mathrm{NN}}}}$}}}

\newcommand{ \be }{\begin{equation}}
\newcommand{ \ee }{\end{equation}}

\newcommand\la{\langle}
\newcommand\ra{\rangle}

\begin{document}
\title{Search for the QCD Critical Point with Fluctuations of Conserved Quantities in Relativistic Heavy-Ion Collisions at RHIC : An Overview }

\author{Xiaofeng Luo}
\email{xfluo@mail.ccnu.edu.cn}
\affiliation{Institute of Particle Physics and Key Laboratory of Quark \& Lepton Physics (MOE), \\Central China Normal University, Wuhan, 430079, China.}
\affiliation{Department of Physics and Astronomy, University of California, Los Angeles, California 90095, USA}
\author{Nu Xu}
\email{nxu@lbl.gov}
\affiliation{Institute of Particle Physics and Key Laboratory of Quark \& Lepton Physics (MOE), \\Central China Normal University, Wuhan, 430079, China.}
\affiliation{Nuclear Science Division, Lawrence Berkeley National Laboratory, Berkeley, CA 94720, USA.}
\begin{abstract}
Fluctuations of conserved quantities, such as baryon, electric charge and strangeness number, are sensitive observables in relativistic heavy-ion collisions to probe the QCD phase transition and search for the QCD critical point. In this paper, we review the experimental measurements of the cumulants (up to fourth order) of event-by-event net-proton (proxy for net-baryon), net-charge and net-kaon (proxy for net-strangeness) multiplicity distributions in Au+Au collisions at $\sqrt{s_{NN}}=7.7, 11.5, 14.5, 19.6, 27, 39, 62.4, 200$ GeV from the first phase of beam energy scan program at the Relativistic Heavy-Ion Collider (RHIC). We also summarize the data analysis methods of suppressing the volume fluctuations, auto-correlations and the unified description of efficiency correction and error estimation.  Based on theoretical and model calculations, we will discuss the characteristic signatures of critical point as well as backgrounds for the fluctuation observables in heavy-ion collisions. The physics implications and the future second phase of the beam energy scan (2019-2020) at RHIC will be also discussed. 
\end{abstract}
\maketitle

%\newpage
\begin{itemize}
\item[\ref{sec:intro})] Introduction 
\item[\ref{sec:CP})] The QCD Critical Point
\item[\ref{sec:Flu_Corr})] Fluctuations and Correlations
\item[\ref{sec:HRG})] Hadron Resonance Gas Model
\item[\ref{sec:Lattice})] Lattice QCD
\item[\ref{sec:observable})] Experimental Observables
\item[\ref{sec:signature})] Fluctuations Signature near the QCD Critical Point
\item[\ref{sec:sigmafield})] $\sigma$ Field Model
\item[\ref{sec:NJL})] NJL Model
\item[\ref{sec:baselines})] Baselines and Background Effects for Net-Particle Cumulants
\item[\ref{sec:Poisson})] Expectations from Poisson, Binomial and Negative Binomial Statistics 
\item[\ref{sec:JAM})] Effects of Baryon Number Conservation and Nuclear Potential on Net-Proton (Baryon) Cumulants
\item[\ref{sec:UrQMD_AMPT})] Fluctuations of Net-Proton (Baryon) from UrQMD and AMPT
\item[\ref{sec:corr_fun})] Cumulants and Correlation Functions
\item[\ref{sec:DataAnalysis})] Data Analysis Methods
\item[\ref{sec:Centrality})] Collision Geometry and Centrality Definition
\item[\ref{sec:CBWC})] Centrality Bin Width Correction
\item[\ref{sec:volume})] Volume Fluctuations Effects
\item[\ref{sec:autocorrelation})] Auto Correlation Effects
\item[\ref{sec:EfficiencyCorrection})] Efficiency Corrections for Cumulants
\item[\ref{sec:ErrorEstimation})] Error Estimations for Efficiency Corrected Cumulants
\item[\ref{sec:results})] Experimental Results
\item[\ref{sec:BESII})]  Beam Energy Scan Phase-II and STAR Detector Upgrades
\item[\ref{sec:summary})] Summary
\end{itemize}
\begin{figure*}[htbp]
\begin{center}
\hspace{-1cm}
\includegraphics*[scale=0.55]{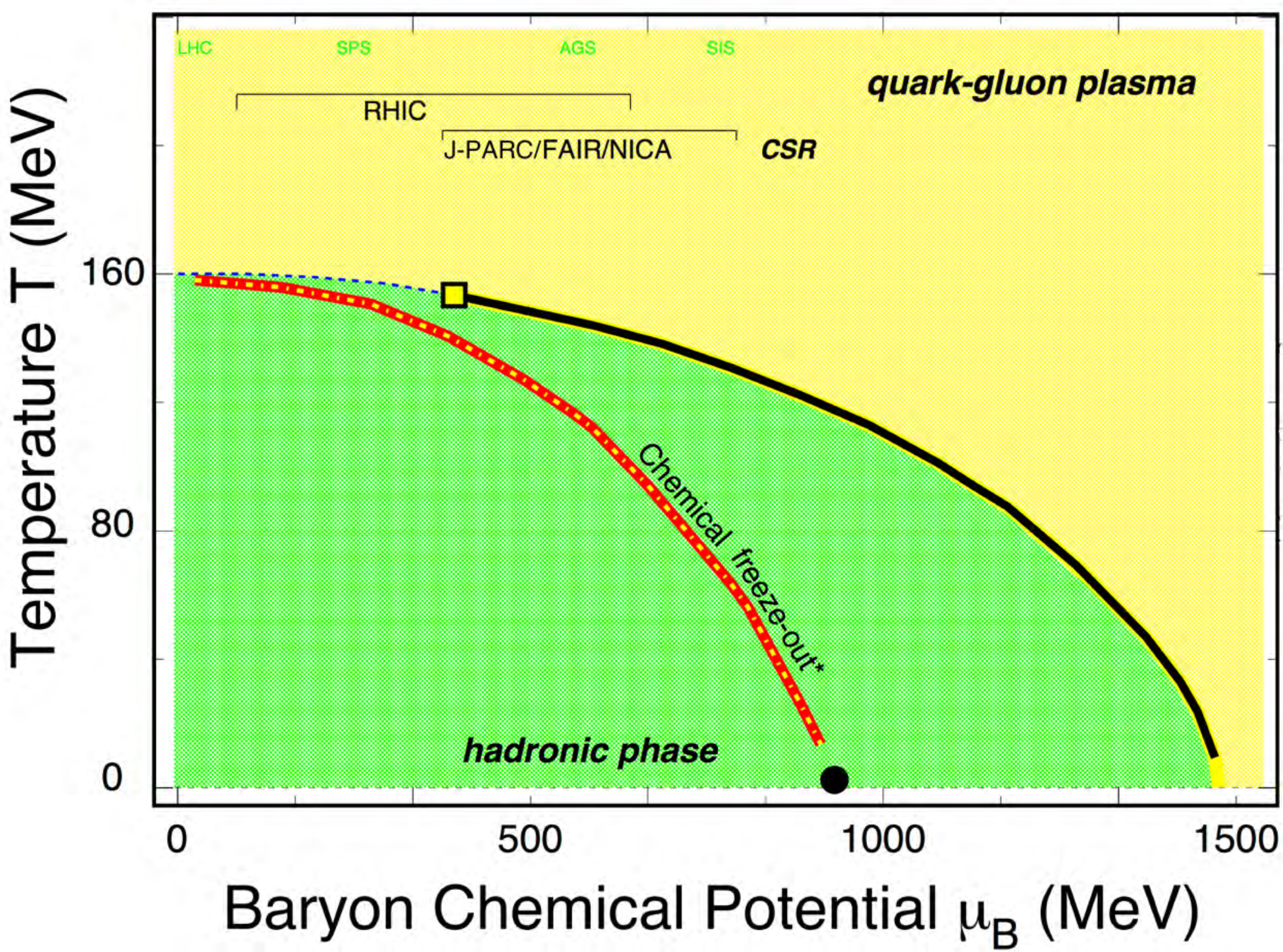}\\
%\vspace{-0.5cm}
\caption{ (Color Online) The conjectured QCD phase diagram~\cite{science} temperature $T$ as a function of baryon chemical potential ($\mu_B$). The red-line is the empirical chemical freeze-out line determined by the experimental data of heavy-ion collisions. The solid circle is located at $T=0$ and $\mu_{B}=938$ MeV, the rest mass of nucleon. 
The solid black line is the speculated first-order phase boundary and the end point (solid square) of this boundary is the QCD critical point. At $\mu_{B} \sim 0$, the transition from hadronic gas to quark gluon plasma (QGP) becomes a smooth crossover, which is represented by the dashed line.}
\label{fig:phase}
\end{center}
\end{figure*}
\section{Introduction}\label{sec:intro}
A major uncertainty in our understanding of strongly interacting nuclear matter is the so called Quantum Chromodynamics (QCD) phase structure and the possible existence of a critical point in the QCD phase diagram, located at high temperature and non-zero baryon chemical potential. It is one of the main goals of the Beam Energy Scan (BES) program at the Relativistic Heavy-Ion Collider (RHIC)~\cite{bes,BESII_WhitePaper}, which is located at the Brookhaven National Laboratory (BNL), US.  This also serves as a main motivation for the research programs
at the future accelerator facilities FAIR in Darmstadt and NICA in Dubna. As shown in Fig.\ref{fig:phase}, the conjectured QCD phase diagram, it can be displayed in the two dimensional phase diagram (temperature, $T$ vs. baryon chemical potential, $\mu_{B}$). Finite temperature Lattice QCD calculations has shown that at zero $\mu_B$ ($\mu_B=0$) region, it is a crossover transition between hadronic phase and quark-gluon plasma (QGP) phase~\cite{crossover}. At large $\mu_B$ region, the QCD based models predicted that the phase transition is of the first order~\cite{de2002qcd,endrHodi2011qcd} and there should exist a so called QCD Critical Point (CP) as the endpoint of the first order phase boundary~\cite{QCP_Prediction,fodor2004critical}. Due to sign problem at finite $\mu_B$ region,  it is difficult to precisely determine the location of the CP or even its existence~\cite{gavai2015qcd}.  Experimental confirmation of the existence of the CP will be an excellent verification of QCD theory in the non-perturbative region and a milestone of exploring the QCD phase structure. Please note that the first-order phase boundary, the critical point and the smooth crossover are closely related thermodynamically. For example, if the smooth crossover and the first-order phase boundary exist, there must be a critical point at the end of the first-order line. To some extent, the burden is on the experimental side who should determine the location of the QCD critical point or the first-order phase boundary. To access a broad region of the QCD phase diagram, experimentalists vary the temperature ($T$) and baryon chemical potential ($\mu_B$) of the nuclear matter created in heavy-ion collisions~\cite{bes} by tuning the colliding energies of two nuclei. It is expected that fluctuations of conserved charges yield information on the phase structure of QCD matter~\cite{stephanov1998signatures,stephanov1999event,jeon1999fluctuations,asakawa2000fluctuation,koch2005baryon,ejiri2006hadronic,ratioCumulant}, provided the freeze-out is sufficiently close to the phase boundary. These conserved quantities have been long time predicted to be sensitive to the correlation length~\cite{qcp_signal,ratioCumulant,Neg_Kurtosis,Hatta} and directly connected to the susceptibilities computed in the first principle Lattice QCD calculations~\cite{science,ding2015thermodynamics,Lattice,MCheng2009,bazavov2012fluctuations,bazavov2012freeze,BFriman_EPJC,2014_Bengt_flu,chiral_HRG,kenji}. Consequently, the analysis of event-by-event fluctuations of the net baryon number ($B$), electric charge ($Q$), and strangeness ($S$), in particular their higher order cumulants, play a central role in the efforts to reveal the thermodynamics of the matter created in heavy-ion collisions at RHIC and LHC. Thus, it can serve as a powerful observables to study the phase transition and search for the CP in heavy-ion collisions~\cite{Misha2009_PRL,Asakawa},. 

During the first phase of the RHIC BES (2010 to 2014), the STAR experiment has measured the cumulants (up to  the fourth order) of net-proton (proton minus anti-proton number, proxy of net-baryon~\cite{Hatta}) , net-charge and net-kaon multiplicity distributions in Au+Au collisions at {\sNN}= 7.7, 11.5, 14.5, 19.6, 27, 39, 62.4 and 200 GeV.  In those energies, the data of 14.5 GeV is taken in the year 2014, 19.6,  27 GeV are taken in the year 2011, and the other energies are collected in the years 2010.  In this paper, we will review the experimental results on fluctuations of conserved quantities from the BES data measured by the STAR and PHENIX experiments. The corresponding physics implications will be also discussed.
\begin{figure}[htbp]
\begin{center}
%\vspace{-0.8cm}
\includegraphics[scale=0.85]{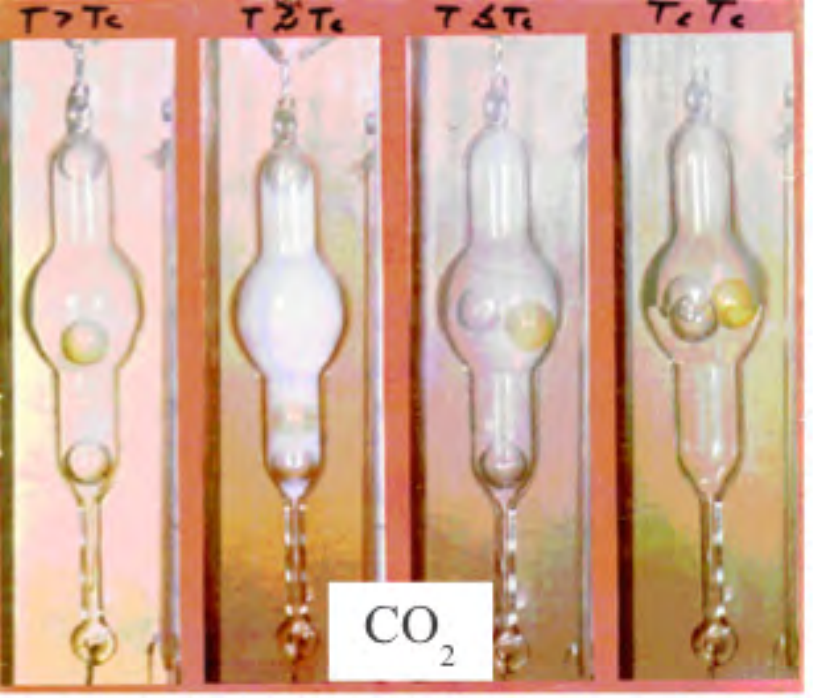}
%\hspace{0.2in}
\caption[]{(Color online) The critical opalescence near the critical point of $\mathrm{CO_{2}}$~\cite{andrews1869bakerian}. } \label{fig:CO2}
\end{center}
\end{figure}
\begin{figure}[htbp]
\begin{center}
\hspace{-0.3in}
%\vspace{-0.8cm}
\includegraphics[scale=0.88]{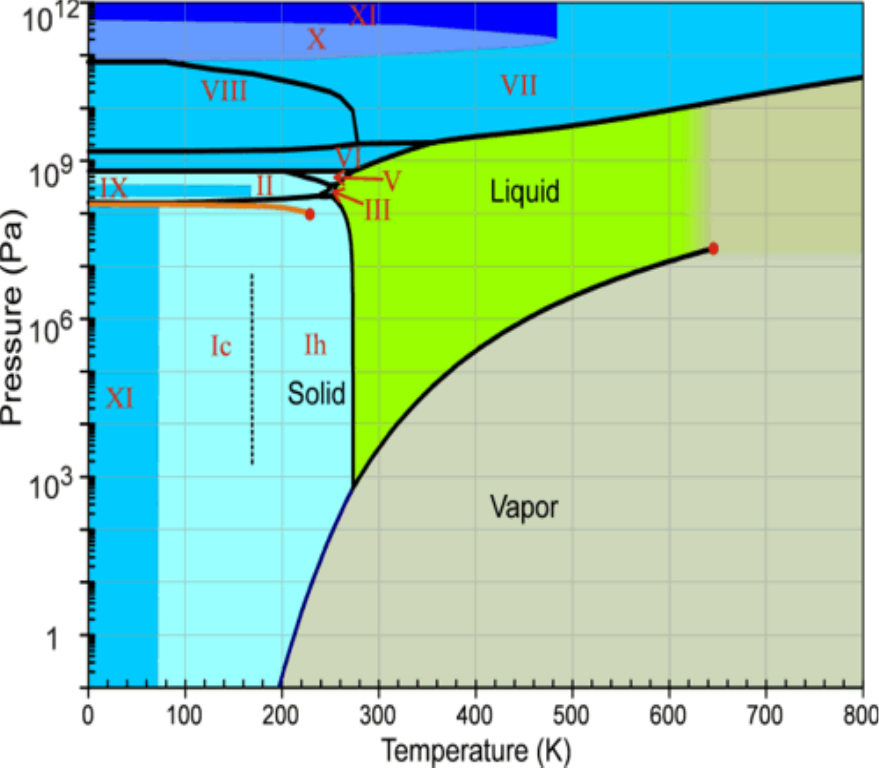}
\caption[]{(Color online) The phase diagram of water~\cite{braun2009colloquium}.} \label{fig:water}
\end{center}
\end{figure}

\begin{table*}[t]
\centering
\caption{Locations of the QCD critical point from Lattice QCD and DSE, respectively.}
\vspace{0.2cm}
\label{table:LCP}
\begin{tabular}{|c|c|c|c|c|c|c|}
\hline
 & \multicolumn{3}{c|}{Lattice} & \multicolumn{3}{c|}{DSE} \\ \hline
\multirow{2}{*}{\begin{tabular}[c]{@{}c@{}}($\mu_{B}^{E}, T^{E}$)\\ MeV\end{tabular}} & I~\cite{qcp_zoltan} & II~\cite{qcp_Gupta2,Gupta_Lattice2016} & III~\cite{Karsch_QM2015,Karsch_CPOD2016,Karsch_SQM2016,Karsch_INT_2016} & I~\cite{xin2014quark} & II~\cite{shi2014locate} & III~\cite{fischer2014phase} \\ \cline{2-7} 
 & (360,162) & (285,155) & $\mu_{B}^{E}/T^{E}$\textgreater2 & (372,129) & (405,127) & (504,115) \\ \hline
\end{tabular}
\end{table*}
\section{The QCD Critical Point}  \label{sec:CP}
A critical point is the end point of the first order phase transition boundary in the phase diagram, at which, the phase transition is of the second order and one cannot distinguish difference between the two phases. For e.g., in the liquid-gas phase transition of water, one cannot distinguish vapor and liquid of the water when the temperature is above the critical temperature $T_c$ (373.946~\textcelsius). In equilibrated matter in the vicinity of a critical point, various thermodynamic quantities exhibit large critical fluctuations, which in laboratory systems give rise to e.g. critical opalescence. The critical phenomena (critical opalescence) is discovered by Baron Charles Cagniard de la Tour in 1822 in the study of the liquid-gas phase transition for the mixtures of alcohol and water~\cite{berche2009critical}. The term "critical point" is firstly named by Thomas Andrews in 1869~\cite{andrews1869bakerian} when he studied the liquid-gas transition in carbon dioxide ($\mathrm{CO_2}$), the critical temperature is about 31~\textcelsius. When the thermodynamic condition of system is approaching the critical point, the correlation length of system will diverge.  The divergency of the correlation length ($\xi$) is one of the most important characteristic feature of the critical point and it is also related to the divergency of the specific heat ($\mathrm{C_{v}}$), susceptibility ($\chi$), compressibility ($\kappa$) and critical opalescence. In Fig.~\ref{fig:CO2}, it demonstrates the well-known critical opalescence, the visible cloudy phenomena near the critical point of liquid-gas phase transition. When the lights are passing through the $\mathrm{CO_{2}}$ near the critical point, the light will undergo large scattering due to its wavelength is comparable to the length scale (correlation length) of the density fluctuations in the phase transition of the liquid-gas system. 

Those critical behaviors can be described by power law divergence with a set of critical exponents. The critical exponents of the critical point for various systems with same symmetry and dimension belong to the same universality class. Due to self-similarity and scaling properties of the critical point, those critical exponents can be precisely calculated by the renormalization group theory~\cite{wilson1974renormalization}. Another important feature of the critical point is the so called finite size effect, which is originated from that the correlation length is comparable with the size of system and the system size limits the growing of the correlation length. This leads to an observable effects when one varies the system size.

The phase diagram of water is shown in Fig.~\ref{fig:water}~\cite{braun2009colloquium}. It can be found that the phase structure of water are very rich, which is the emergent properties of quantum electrodynamics (QED). Due to the easily realized phase transition conditions, the water phase diagram are precisely known. On the other hand, the phase structure of the hot and dense nuclear matter, which is governed by the strong force described by the QCD theory, is rarely known to us. Thus, it is very important to explore the QCD phase structure and search for the QCD critical point theoretically and experimentally.
From theoretical side, it is still very difficult to precisely determine the location of the critical point due to its non-perturbative feature. The QCD based models, such
as NJL, PNJL, PQM, have given many results of the location of the QCD critical point, which are summarized in the reference~\cite{location}. The locations of the QCD critical
point obtained from the first principle Lattice QCD and Dyson-schwinger equation (DSE) calculations are summarized in the table~\ref{table:LCP}. One can see that the baryon chemical potential ($\mu_{B}^{E}$) of the QCD critical point are ranging from 266 to 504 MeV, the critical temperature is from 115 to 162 MeV. There still has big difference between the results from different methods and groups. Experimentally, we aim to search for the critical point with the strongly interacting QCD matter created in the relativistic heavy-ion collisions. It is very challenging due to the following reasons: 1. The hot dense medium created in the heavy-ion collisions are not static but expanding rapidly.
Thus, the correlation length of the system is not only limited by the size of the system, but also by the finite expansion time and it is predicted to be 2-3 $fm$ by assuming the existence of a critical point~\cite{correlationlength}. One has to consider finite time and finite size effects in order to determine the exact location of the critical point. 2. What's the sensitive observables and what's the smoking gun signature of the QCD critical point in heavy-ion collisions. 3. One has to understand the non-critical contributions to the experimental observables and the signal to background ratio
should not be too small. 4. One needs that the freeze-out thermodynamic conditions of the QCD matter created in heavy-ion collisions should be close enough to the phase  boundary that the phase transition signals weren't washed out after the expansion.  

Due to the difficulties and challenges discussed above, we should set up good strategies to search for the QCD critical point. Firstly, we need to have good quality experimental data of heavy-ion collisions at a wide range of energies. This allows us to scan a broad region of the QCD phase diagram. Then, we use sensitive observables to find the smoking gun signatures and confirm the existence of the QCD critical point before determining its location. In order to extract critical signature and understand the background contributions, careful modelling of the critical phenomena and dynamical evolution of the heavy-ion collisions are needed. It requires close collaboration between theorists and experimentalists. If the QCD critical point is given and hidden in nature, we will finally discover it and put a permanent landmark in the phase diagram of the strongly interacting nuclear matter.

\section{Fluctuations and Correlations} \label{sec:Flu_Corr}
Fluctuations and correlations have long been considered to be sensitive observables in heavy-ion collisions
to explore the phase structure of the strongly interacting QCD matter~\cite{jeon2004event,koch2010hadronic,asakawa2000fluctuation}. They have a well defined physical interpretation for a system in thermal equilibrium and can provide essential information about the effective degrees of freedom. The well known phenomenon of critical opalescence is a result of fluctuations at all length scales due to a second order phase transition. The most efficient way to study the fluctuations of a system created in a heavy-ion collision is to measure an observable on the event-by-event basis and the fluctuations are studied over the ensemble of the events. In strong interaction, the net number of charges in a closed system is conserved. The magnitude of these fluctuations in a grand canonical ensemble at finite temperature are distinctly different in the hadronic and quark gluon plasma phases. Event-by-event fluctuation and correlation of the conserved charges is one of the observables to study the properties of the QCD medium created in relativistic heavy-ion collisions. Although these observables are hadronic ones, it is believed that they can reflect the thermal property in the early stage. A system in thermal equilibrium (for a grand-canonical ensemble) can be characterized by its dimensionless pressure, which is the logarithm of the QCD partition function~\cite{ding2015thermodynamics}:
\begin{equation}
\frac{P}{{{T^4}}} = \frac{1}{{V{T^3}}}\ln [Z(V,T,{\mu _B},{\mu _Q},{\mu _S})]
\end{equation}
where $V$ and $T$ are the system volume and temperature. The  $\mu_B$, $\mu_{Q}$ and ${\mu _S}$ are baryon, charge and
strangeness chemical potential, respectively. The equation of state is very different for thermodynamical system with different degree of 
freedom and interactions. The susceptibility of the conserved charges ($B,Q,S$) are defined as the derivative of the dimensionless pressure with respected to the reduced chemical potential.
\begin{equation} \label{equ:sus}
\chi _{ijk}^{BQS} = \frac{{{\partial ^{(i + j + k)}}[P/{T^4}]}}{{\partial \hat \mu _B^i\partial \hat \mu _Q^j\partial \hat \mu _S^k}} 
\end{equation}
where $\hat{\mu_{q}}=\mu_{q}/T, q=B,Q,S$. The cumulants of these conserved quantities ($B,Q,S$) distributions are connected to the corresponding susceptibilities by 
\begin{equation}
\begin{split}
C_{ijk}^{BQS} &= \frac{{{\partial ^{(i + j + k)}}\ln [Z(V,T,{\mu _B},{\mu _Q},{\mu _S})]}}{{\partial \hat \mu _B^i\partial \hat \mu _Q^j\partial \hat \mu _S^k}} \\
&= V{T^3}\chi _{ijk}^{BQS}(T,{\mu _B},{\mu _Q},{\mu _S})
\end{split}
\end{equation}
where the $C_{ijk}^{BQS}$ denotes both diagonal and off-diagonal cumulants of conserved quantities ($B,Q,S$) ($i,j,k=1,2,3,4...n$). 
Experimentally, we construct the ratios of cumulants as the experimental observables, which cancel the volume dependent and can
be directly compared with the ratios of susceptibilities from theoretical calculations. To obtain the ratio of cumulants, we firstly introduce various order cumulants up to sixth order and their relations to the central moments as:
\begin{widetext}
\begin{eqnarray}
{M_q} &=&  < {N_q} >  = V{T^3}\chi _1^q,\begin{array}{*{20}{c}}{}&{}&{}&{}&{}\end{array}\sigma _2^q = C_2^q =  < {(\delta {N_q})^2} >  = V{T^3}\chi _2^q\\
C_3^q &=&  < {(\delta {N_q})^3} >  = V{T^3}\chi _3^q, \begin{array}{*{20}{c}}{}&{}\end{array}C_4^q =< {(\delta {N_q})^4} >  - 3 < {(\delta {N_q})^2}{ > ^2} = V{T^3}\chi _4^q\\
C_5^q &=&  < {(\delta {N_q})^5} >  - 10 < {(\delta {N_q})^3} >  < {(\delta {N_q})^2} >  = V{T^3}\chi _5^q\\
C_6^q &=&  < {(\delta {N_q})^6} >  - 15 < {(\delta {N_q})^4} >  < {(\delta {N_q})^2} >  - 10 < {(\delta {N_q})^3}{ > ^2} + 30 < {(\delta {N_q})^2}{ > ^3} = V{T^3}\chi _6^q
\end{eqnarray}
\end{widetext}
where $M_{q}$, $\sigma _2^q $ are the mean and variance, respectively.  The $C_n^q (n=2,3,4,...)$ are the $n^{th}$ order cumulants with $q=B,Q,S$ and $\delta N_q=N_q-<N_q>$. We didn't consider the correlations between different conserved charges. 

On the other hand, we introduce two well known statistic quantities, the so called $skewness (S)$ and $kurtosis (\kappa)$. In statistics, those two quantities can be used to describe the shape of distributions and they are defined as :
\begin{eqnarray}
{S_q} &=& \frac{{ < {{(\delta {N_q})}^3} > }}{{ < {{(\delta {N_q})}^2}{ > ^{3/2}}}} = \frac{{C_3^q}}{{{{(\sigma _2^q)}^{3/2}}}}\\
{\kappa _q} &=& \frac{{ < {{(\delta {N_q})}^4} > }}{{ < {{(\delta {N_q})}^2}{ > ^2}}} - 3 = \frac{{C_4^q}}{{{{(\sigma _2^q)}^2}}}
\end{eqnarray}
For gaussian distribution, both of the two quantities are equal to zero. Thus, they are widely used to quantify the non-gaussianity.
With above definition of the mean, variance, skewness, kurtosis and various order cumulants, we can have the
following relations: 
\begin{eqnarray}
\frac{{\sigma _q^2}}{{{M_q}}} &=& \frac{{C_2^q}}{{{M_q}}} = \frac{{\chi _2^q}}{{\chi _1^q}}, \begin{array}{*{20}{c}}{}&{}&{}\end{array} S_q^{}\sigma _q^{} = \frac{{C_3^q}}{{C_2^q}} = \frac{{\chi _3^q}}{{\chi _2^q}}\\
\kappa _q^{}\sigma _q^2 &=& \frac{{C_4^q}}{{C_2^q}} = \frac{{\chi _4^q}}{{\chi _2^q}}, \begin{array}{*{20}{c}}{}&{}&{}\end{array} \frac{{\kappa _q^{}\sigma _q^{}}}{{S_q^{}}} = \frac{{C_4^q}}{{C_3^q}} = \frac{{\chi _4^q}}{{\chi _3^q}}
\end{eqnarray}
Those equations connect the experimental measurements ($l.h.s.$) and theoretically calculations ($r.h.s.$).
In the following, we will discuss the results calculated from the Hadron Resonance Gas Model and Lattice QCD. 
\begin{figure*}[htbp] 
\begin{center}
\hspace{-0.1in}
\includegraphics[scale=0.355]{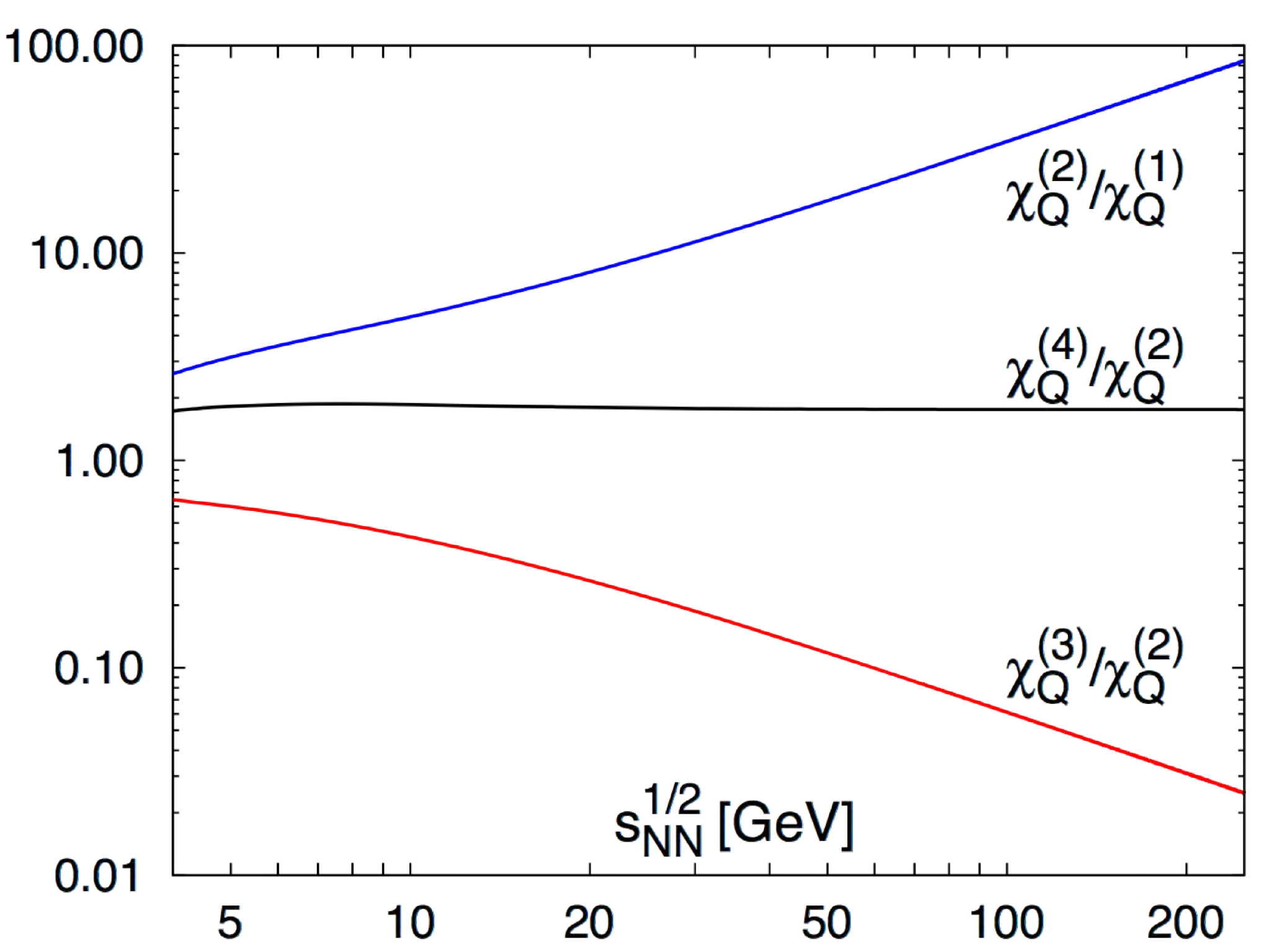}
\hspace{0.8cm}
\includegraphics[scale=0.36]{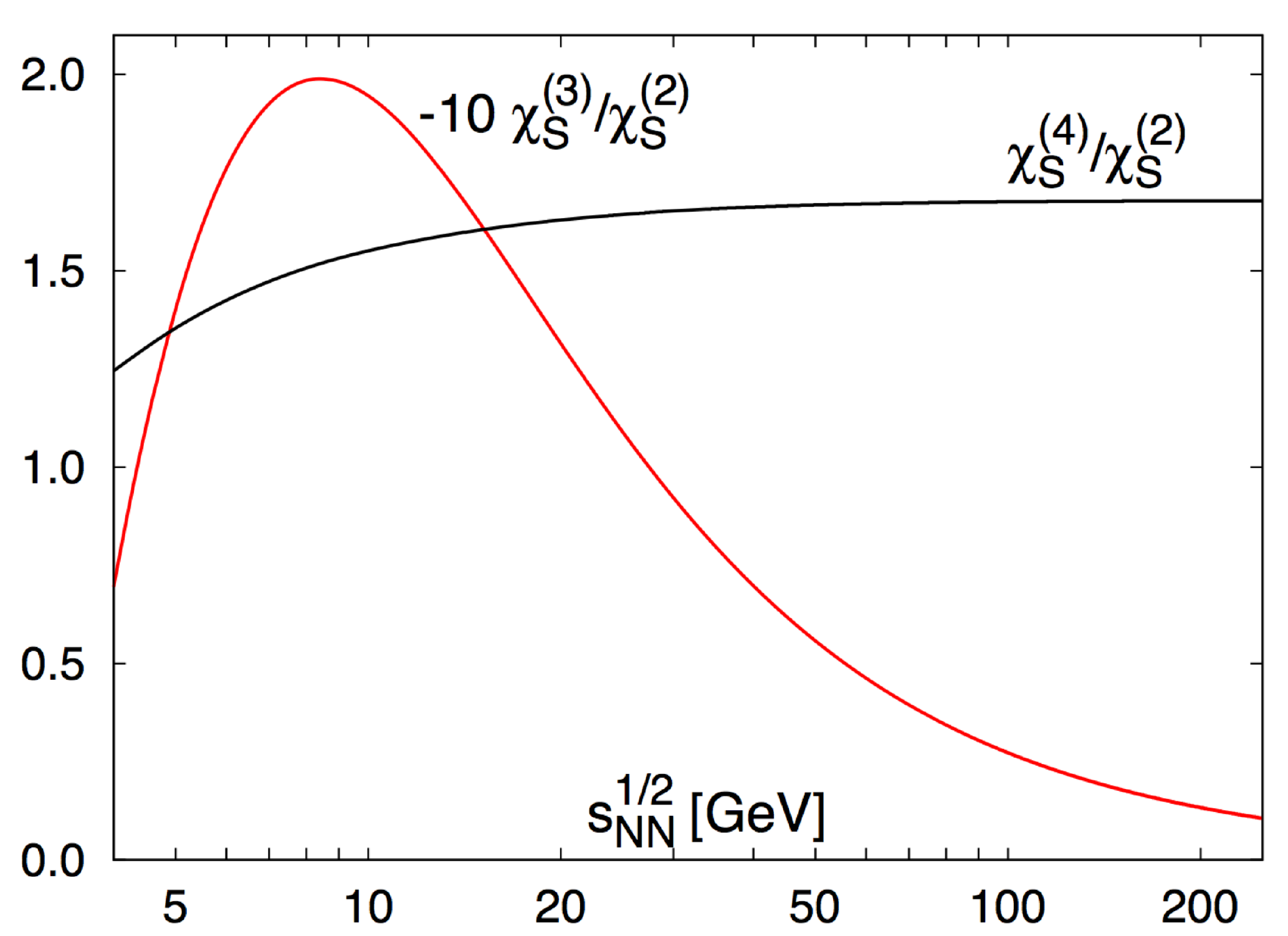}
\vspace{0.3cm}
\caption[]{(Color online) The ratio of susceptibilities of charge (left) and strangeness (right) fluctuations as a function of colliding energy along the parameterized freeze-out curve in heavy-ion collisions.  The results are calculated from the HRG model~\cite{HRG_Karsch}.} \label{fig:QS}
\end{center}
\end{figure*}
\subsection{Hadron Resonance Gas Model } \label{sec:HRG}
In the Hadron Resonance Gas (HRG) model, non-interacting hadrons and their resonance are the basic degree of freedom. The interactions are encoded in the thermal creation of hadronic resonances based on their Boltzmann factor. The HRG can successfully describe the observed particle abundances in heavy ion collisions. For simplify and discussion purpose, we use the Boltzmann approximation and the pressure can be expressed~\cite{HRG_Karsch,HRG_fjh,HRG_FJH2}:
\begin{eqnarray}
\frac{P}{T^4}&=&\frac{1}{VT^{3}}\mathrm{ln}[Z(V,T,\mu_{B},\mu_{Q},\mu_{S})]\nonumber\\
&=&\frac{1}{\pi^{2}}\sum_{i\in{X}}g_{i}(\frac{m_{i}}{T})^2K_{2}(\frac{m_{i}}{T})\nonumber\\
&\times &{\mathrm{cosh}(B_{i}\hat{\mu}_{B}+Q_{i}\hat{\mu}_{Q}+S_{i}\hat{\mu}_{S})}.
\end{eqnarray}
where $g_{i}$ is the degeneracy factor for hadrons of mass $m_{i}$, and $\hat{\mu}_{q}\equiv\frac{\mu_q}{T}$, with $q=B$, $S$, $Q$ denote the net-baryon number, net-strangeness and the net-charge, and $\mu_{B}$, $\mu_{S}$, $\mu_{Q}$ are the corresponding chemical potentials respectively. The $K_{2}(x)$ is the modified Bessel function and the summation is taking over all stable hadrons and resonance and thus the contribution of anti-particles are automatically included. The results from HRG model are usually served as a baseline for finding the signature of phase transition and QCD critical point in heavy-ion collisions. For net-baryon number fluctuations, the ratios of cumulants from HRG model are simple. With Boltzmann approximation, the baryon number susceptibility can be expressed as:
%\begin{widetext}
\begin{eqnarray}
\chi _{2n}^B&=& \frac{{{\partial ^{2n}}[P/{T^4}]}}{{\partial {{\hat \mu }^{2n}}_B}} = {\sum\limits_{i \in B} {{g_i}(\frac{{{m_i}}}{T})} ^2}{K_2}(\frac{{{m_i}}}{T}) \nonumber \\
&\times &\cosh [{{\hat \mu }_B} + {Q_i}{{\hat \mu }_Q} + {S_i}{{\hat \mu }_S}] \\
\chi _{2n - 1}^B &=& \frac{{{\partial ^{2n - 1}}[P/{T^4}]}}{{\partial {{\hat \mu }^{2n - 1}}_B}} = {\sum\limits_{i \in B} {{g_i}(\frac{{{m_i}}}{T})} ^2}{K_2}(\frac{{{m_i}}}{T}) \nonumber \\
&\times&\sinh [{{\hat \mu }_B} + {Q_i}{{\hat \mu }_Q} + {S_i}{{\hat \mu }_S}]
\end{eqnarray}
%\end{widetext}
Thus, the ratio of baryon number susceptibilities can be easily obtained:
\begin{eqnarray}
\frac{{C_{\mathrm{even}}^B}}{{C_\mathrm{even}^B}} &=& \frac{{\chi _\mathrm{even}^B}}{{\chi _\mathrm{even}^B}} = 1,\frac{{C_{\mathrm{odd}}^B}}{{C_\mathrm{odd}^B}} = \frac{{\chi _\mathrm{odd}^B}}{{\chi _\mathrm{odd}^B}} = 1 \label{equ:15} \\
\frac{{C_\mathrm{odd}^B}}{{C_\mathrm{even}^B}} &=& \frac{{\chi _\mathrm{odd}^B}}{{\chi _\mathrm{even}^B}} = {\left. {\tanh({\mu _B}/T)} \right|_{{\mu _Q} = {\mu _S} = 0}} \label{equ:16}
\end{eqnarray}
Based on the Eq.(\ref{equ:15}) and (\ref{equ:16}), we obtain:
\begin{eqnarray}
\frac{{M_B}}{{{\sigma _B^2}}} &=& S_q^{}\sigma _q^{} = {\left. {\tanh({\mu _B}/T)} \right|_{{\mu _Q} = {\mu _S} = 0}}\\
\kappa _B^{}\sigma _B^2 &=& \frac{{S_B \sigma^{3}_{B}}}{{M_B}} = 1
\end{eqnarray}
where $n=1,2,3...$, $\mu_{B}$ and $T$ are the baryon chemical potential and temperature of the thermal system. 
This simple result arises from the fact that only baryons with baryon number $B = 1$ contribute to the various
cumulants in the HRG model. However, due to the contribution of the multi-charge states $Q = 2$ or $S = 2,3$
for net-charge and net-strangeness fluctuations, respectively, thus the results of net-charge and net-strangeness
fluctuations are more complicated than net-baryon number fluctuations from the HRG model. Fig.\ref{fig:QS} shows the 
ratio of susceptibilities of charge (left) and strangeness (right) from the HRG model calculations along the chemical freeze-out curve 
in heavy-ion collisions. It can be found that the susceptibilities ratios $\chi_2/\chi_1$ or  $\chi_3/\chi_2$ of charge and strangeness show strong energy dependence whereas the  $\chi_4/\chi_2$ has small variation with energies. Due to the contributions from the multi-charge states $Q = 2$ or $S = 2,3$, the charge and strangeness $\chi_4/\chi_2$ deviate from unity.

\subsection{Lattice QCD} \label{sec:Lattice}
Lattice QCD is a well-established non-perturbative approach to solve the QCD theory of quarks and gluons exactly from first principles and without any assumptions~\cite{wilson1974confinement}. It can be used to study the thermodynamic properties of a strongly interacting system in thermal equilibrium. Most importantly, lattice QCD provides a framework for investigation of non-perturbative phenomena such as confinement and quark-gluon plasma formation, which are intractable by means of analytic field theories.
\begin{figure}[htbp]
\begin{center}
\hspace{-0.1in}
%\vspace{-0.8cm}
\includegraphics[scale=0.8]{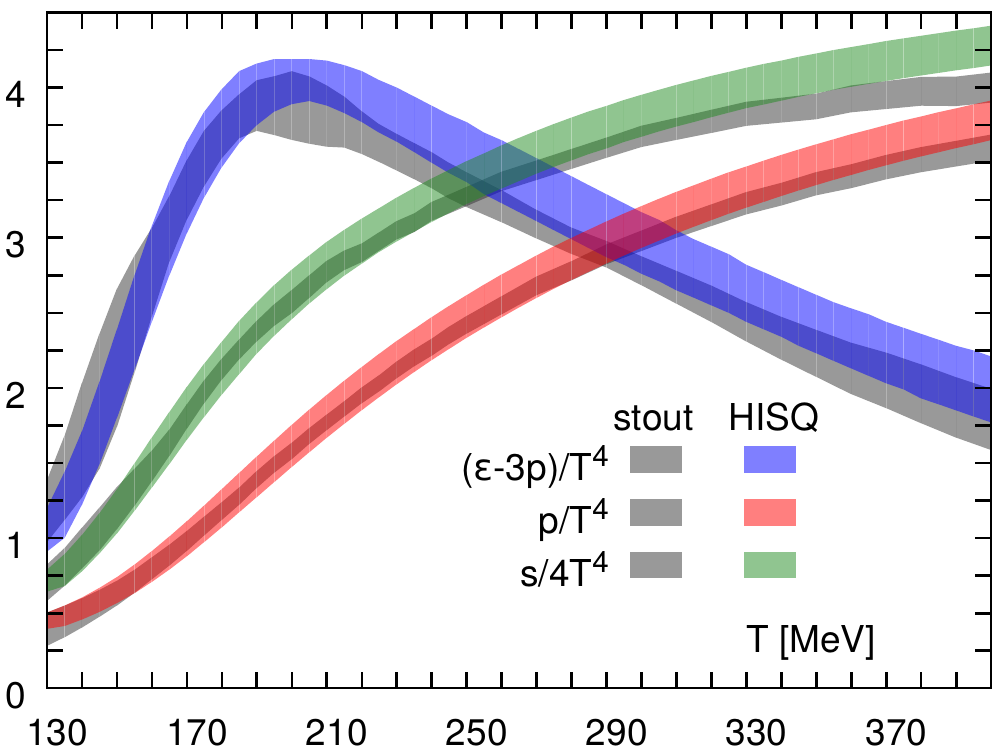}
\caption[]{(Color online) The comparison of the HISQ/tree (Hot QCD Collaboration) and stout (Wuppertal-Budapest Collaboration) results for the trace anomaly, the pressure, and the entropy density in the Lattice QCD calculation at vanishing baryon chemical potential~\cite{bazavov2014equation}.} \label{fig:wb_eos}
\end{center}
\end{figure}
\begin{figure*}[htbp]
\begin{center}
\hspace{-0.1in}
%\vspace{-0.8cm}
\includegraphics[scale=0.7]{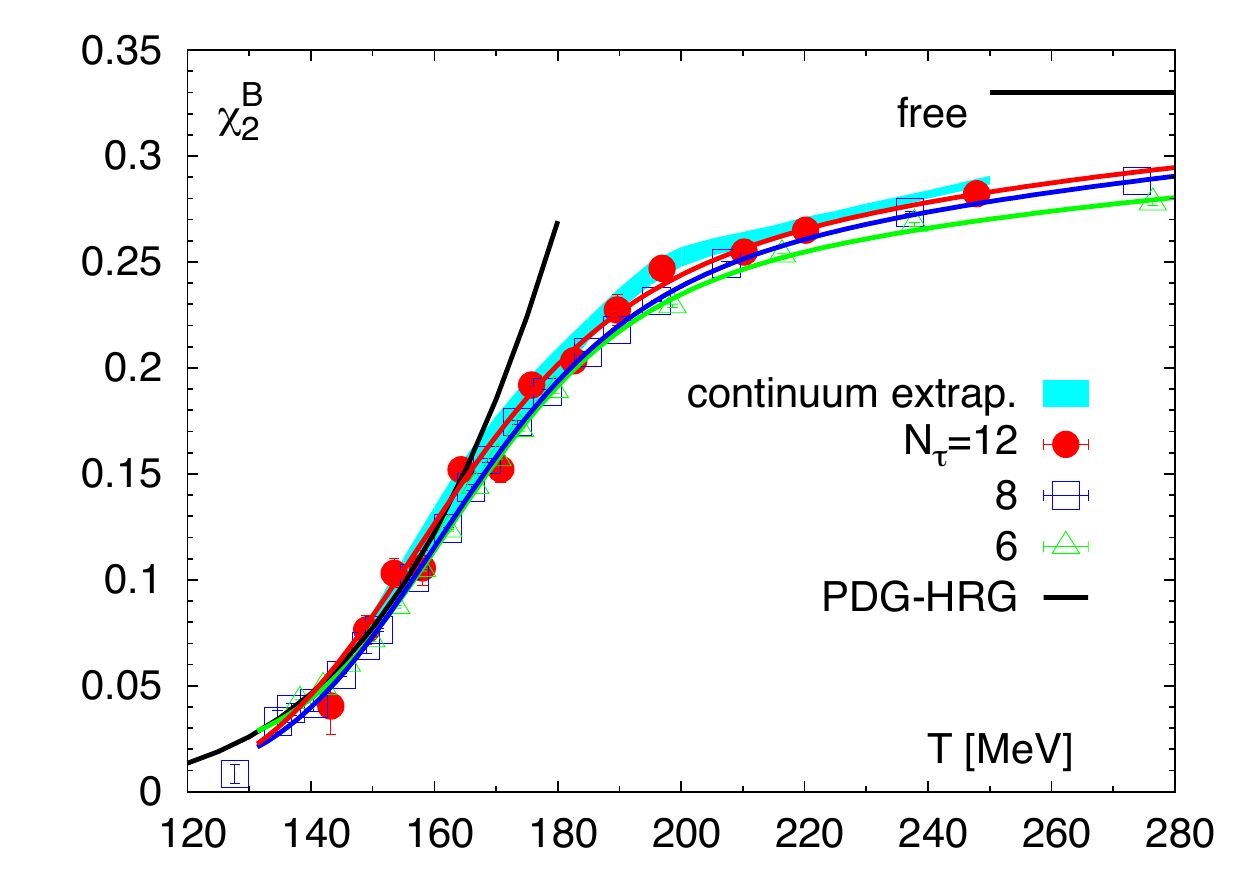}
\includegraphics[scale=0.7]{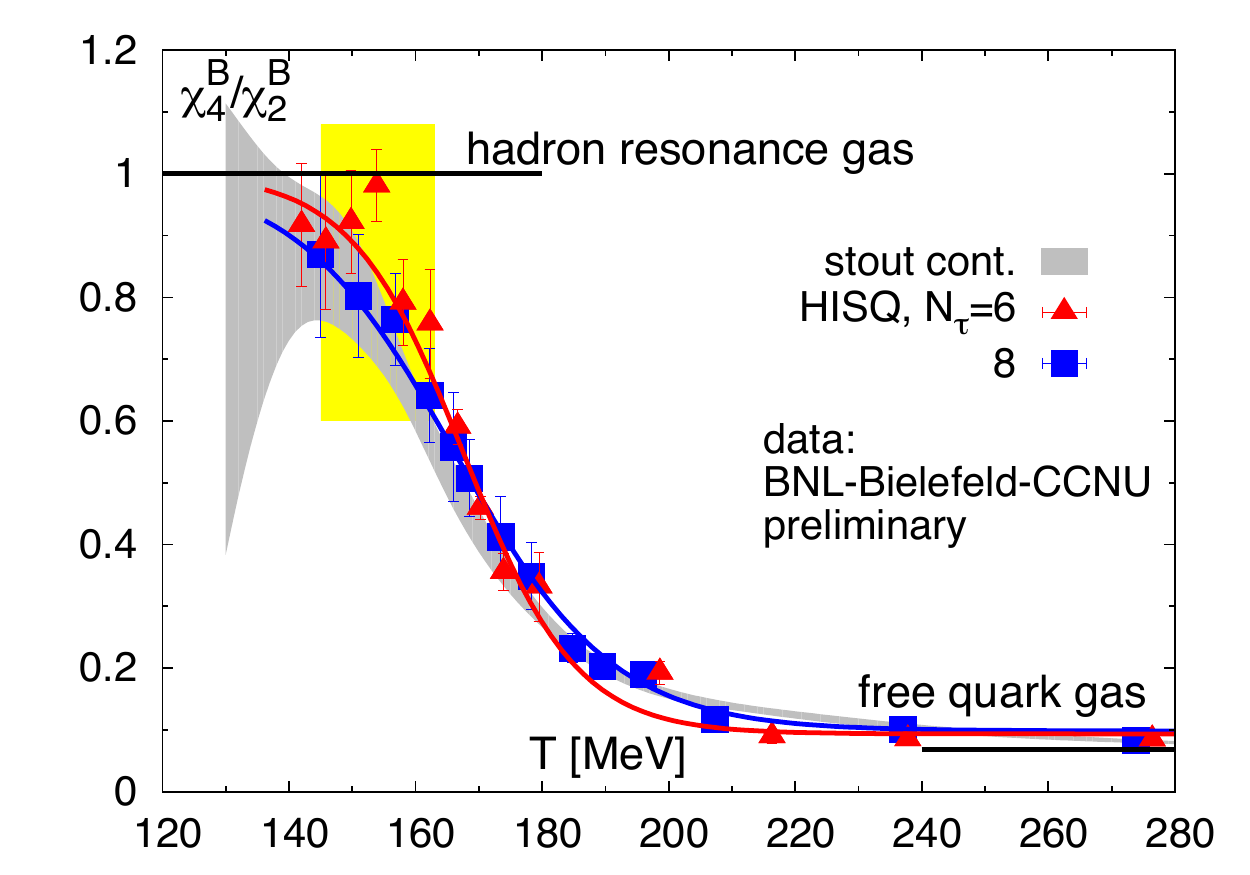}
\caption[]{(Color online)  Lattice QCD results from BNL-Bielefeld-CCNU collaboration~\cite{C6_LatticeQCD,ding2015thermodynamics,Karsch_INT_2016}: the second order baryon number susceptibility ($\chi^{B}_{2}$) (left) and the fourth to second order baryon number susceptibilities ratio ($\chi^{B}_{4}/\chi^{B}_{2}$) (right) as a function of temperature calculated from Lattice QCD at vanishing chemical potential ($\mu_{q}=0, q=B,Q,S$) } \label{fig:C4C2}
\end{center}
\end{figure*}
\begin{figure*}[htbp]
\begin{center}
%\hspace{-0.1in}
%\vspace{-0.8cm}
\includegraphics[scale=0.41]{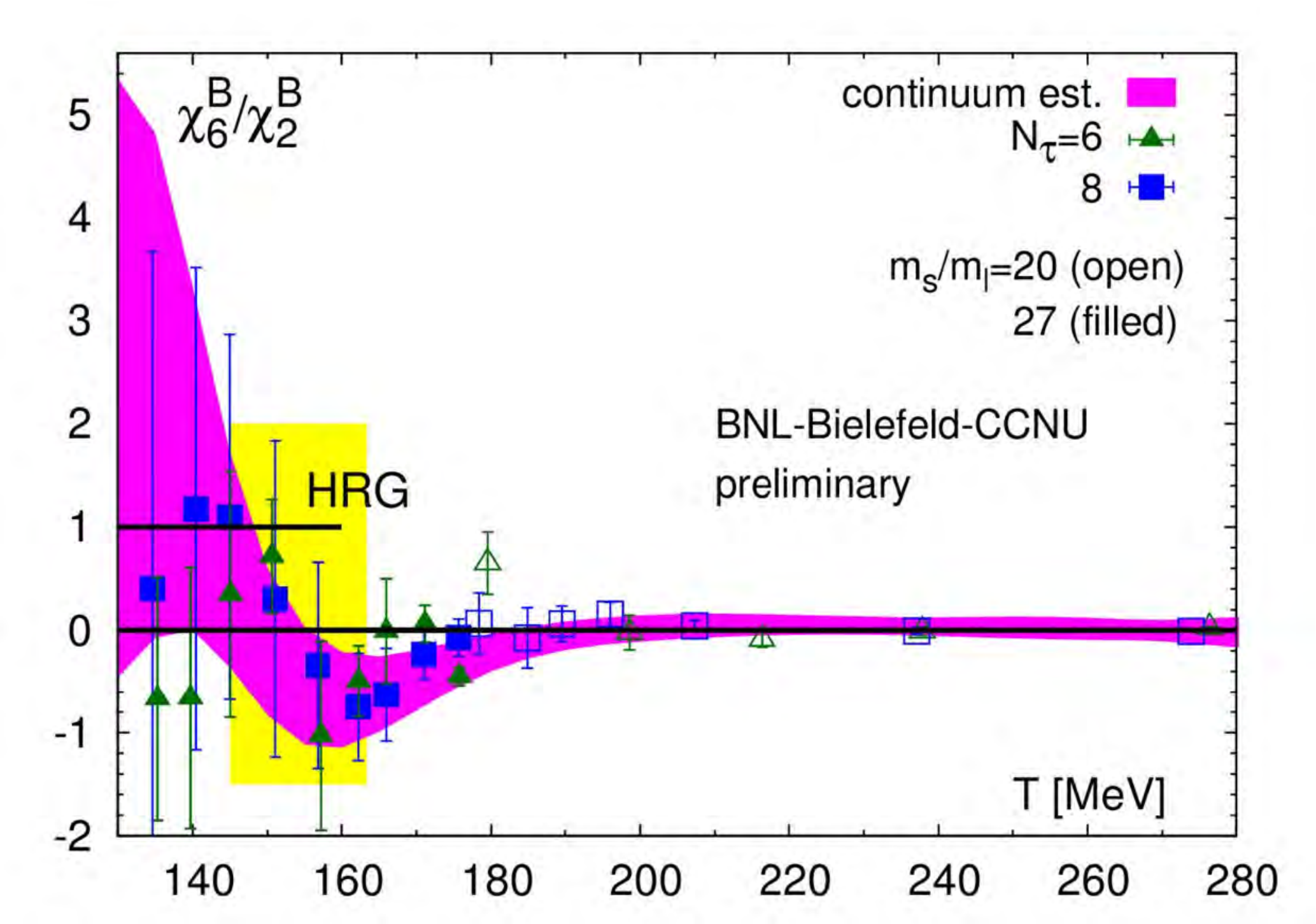}
\includegraphics[scale=0.41]{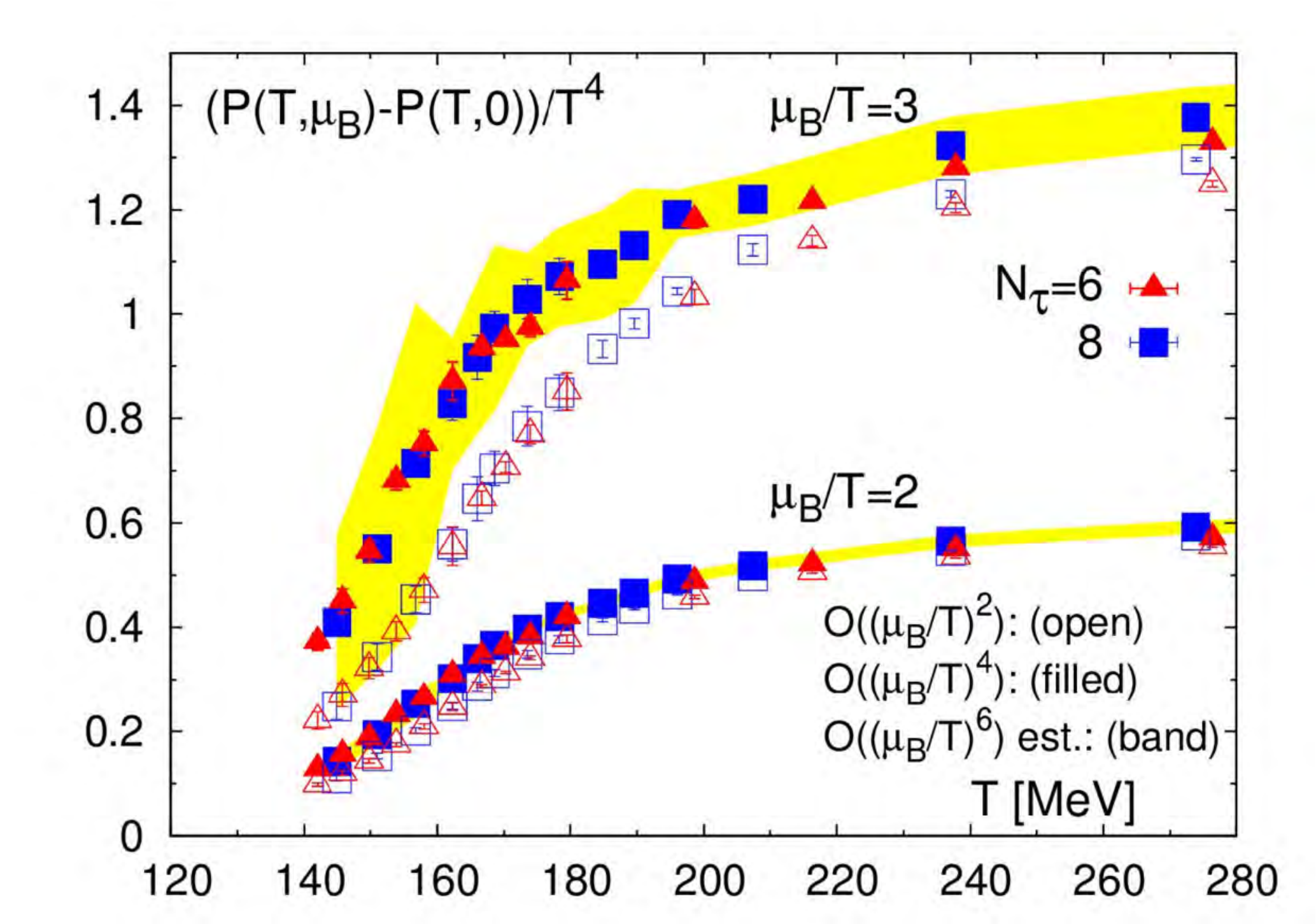}
\caption[]{(Color online)  Lattice QCD results from BNL-Bielefeld-CCNU collaboration~\cite{C6_LatticeQCD,ding2015thermodynamics,Karsch_INT_2016}: the sixth to second order baryon number susceptibilities ratio ($\chi^{B}_{6}/\chi^{B}_{2}$) at vanishing chemical potential (left) and the pressure as a function of temperature for different baryon chemical potential and precision level (right).} \label{fig:C6C2}
\end{center}
\end{figure*}

Fig.~\ref{fig:wb_eos} shows the results of QCD equation of state (the trace anomaly, the pressure and the entropy density) from two independent groups: HotQCD and Wuppertal-Budapest Collaboration, which used the different actions. The results from the two groups got good agreement with each other. On the other hand,  the pressure ($P/T^4$) at finite $\mu_B$ region can be calculated by using the Taylor expansion techniques. By putting the $\mu_Q$=$\mu_S$=0, we can expand the pressure ($P/T^4$) into finite $\mu_{B}$ as~\cite{C6_LatticeQCD,ding2015thermodynamics,Karsch_INT_2016}:
\begin{equation}
\begin{split}
&\frac{{P(T,{\mu _B}) - P(T,0)}}{{{T^4}}} = \frac{1}{2}\chi _2^B(T){\left( {\frac{{{\mu _B}}}{T}} \right)^2} \times \\
&\left[ {1 + \frac{1}{{12}}\frac{{\chi _4^B(T)}}{{\chi _2^B(T)}}{{\left( {\frac{{{\mu _B}}}{T}} \right)}^2} + \frac{1}{{360}}\frac{{\chi _6^B(T)}}{{\chi _2^B(T)}}{{\left (\frac{{{\mu _B}}}{T}\right)}^4}} \right] + {\cal O}(\mu _B^8)
\end{split}
\end{equation}
Due to the symmetry of QCD, the odd terms are vanishing and only even terms are left. It shows various order corrections to the pressure. The coefficients of leading order (LO), next leading order (NLO) and next next leading order (NNLO) are related to the baryon number susceptibilities $\chi_{2}^{B}$, $\chi_{4}^{B}/\chi_{2}^{B}$ and $\chi_{6}^{B}/\chi_{2}^{B}$, respectively. Those susceptibilities are defined in Eq. (\ref{equ:sus}) and can be evaluated at $\mu_{B}=\mu_{Q}=\mu_{S}=0$. Fig.\ref{fig:C4C2} and \ref{fig:C6C2} left show the preliminary BNL-Bielefeld-CCNU results of baryon number susceptibility ($\chi_{2}^{B}$) and the susceptibilities ratios ( $\chi_{4}^{B}/\chi_{2}^{B}$ and $\chi_{6}^{B}/\chi_{2}^{B}$) as a function of temperature computed from Lattice QCD at vanishing chemical potentials. It can be found that at low temperature, the results from Lattice QCD are consistent with the results from HRG whereas it shows large discrepancies between those two at high temperature. The ratio $\chi_{6}^{B}/\chi_{2}^{B}$ shows negative values near and above transition temperature and positive values around unity at low temperatures, but there are still large uncertainties and more statistics are needed. Fig.~\ref{fig:C6C2} right shows the Lattice calculation of the pressure at finite baryon density and the effects of the correction with different orders.
The correction of the NNLO term on the pressure is found to be very small ($<5\%$) when $\mu_{B}/T<2$. It means the Taylor expansion up to NNLO order is under control with $\mu_{B}/T<2$. The susceptibility of conserved charges ($B, Q, S$) can be also computed at finite baryon density region by using the Taylor series in terms of the baryon chemical potential ($\mu_B$) at $\mu_Q=\mu_{S}=0$~\cite{C6_LatticeQCD,ding2015thermodynamics,Karsch_INT_2016}: 
\begin{eqnarray}
\chi _n^B(T,{\mu _B}) &=& \sum\limits_{k = 0}^\infty  {\frac{1}{{k!}}} \chi _{n + k}^B(T){\left( {\frac{{{\mu _B}}}{T}} \right)^k}\\
\chi _n^Q(T,{\mu _B}) &=& \sum\limits_{k = 0}^\infty  {\frac{1}{{k!}}} \chi _{k,n}^{BQ}(T){\left( {\frac{{{\mu _B}}}{T}} \right)^k}\\
\chi _n^S(T,{\mu _B}) &=& \sum\limits_{k = 0}^\infty  {\frac{1}{{k!}}} \chi _{k,n}^{BS}(T){\left( {\frac{{{\mu _B}}}{T}} \right)^k}
\end{eqnarray}
In the following, we focus on discussing the next leading order (NLO) Taylor expansion of the baryon number susceptibilities. Due to the QCD sysmetry for matter and anti-matter, the NLO Taylor expansion for the odd and even order baryon number susceptibilities can be expressed as:
\begin{eqnarray}
\hspace{-3cm}\chi _{2n - 1}^B(T,{\mu _B}) &=& \chi _{2n}^B(T)\left( {\frac{{{\mu _B}}}{T}} \right) + \frac{1}{6}\chi _{2n + 2}^B(T){\left( {\frac{{{\mu _B}}}{T}}\right)^3}\\
\chi _{2n}^B(T,{\mu _B}) &=& \chi _{2n}^B(T) + \frac{1}{2}\chi _{2n + 2}^B(T){\left( {\frac{{{\mu _B}}}{T}} \right)^2}
\end{eqnarray}
where the $\chi _{2n}^B(T)$ and $ \chi _{2n+2}^B(T)$ are the baryon number susceptibilities evaluated at $\mu_B$=$\mu_Q=\mu_{S}=0$ with $n=1,2,3,4,...,N$.
If we define a dimensionless quantity $L_n$:
\begin{equation}
{L_n} = \frac{1}{6}\frac{{\chi _{2n + 2}^B(T)}}{{\chi _{2n}^B(T)}}{\left( {\frac{{{\mu _B}}}{T}} \right)^2}
\end{equation}
With this definition, the Taylor expansion of the odd and even order susceptibility at the next leading order can be re-written as:
\begin{eqnarray}
\chi _{2n - 1}^B(T,{\mu _B}) &=& \chi _{2n}^B\frac{{{\mu _B}}}{T}(1 + {L_n})\\
\chi _{2n}^B(T,{\mu _B}) &=& \chi _{2n}^B(1 + 3{L_n})
\end{eqnarray}
Then, we can express the baryon number susceptibilities ratios as:
\begin{eqnarray}
\frac{{\chi _{2n}^B(T,{\mu _B})}}{{\chi _{2n - 1}^B(T,{\mu _B})}} &=& \frac{T}{{{\mu _B}}}\left( {1 + \frac{2}{{1 + 1/{L_n}}}} \right) \label{equ:28}\\
\frac{{\chi _{2n + 1}^B(T,{\mu _B})}}{{\chi _{2n}^B(T,{\mu _B})}} &=& 6\frac{T}{{{\mu _B}}}{L_n}\left( {1 + \frac{{{L_{n + 1}} - 3{L_n}}}{{1 + 3{L_n}}}} \right)\\
\hspace{-2cm}\frac{{\chi _{2n + 2}^B(T,{\mu _B})}}{{\chi _{2n}^B(T,{\mu _B})}} &=& 6{\left( {\frac{T}{{{\mu _B}}}} \right)^2}{L_n}\left[ {1 + \frac{{3({L_{n + 1}} - {L_n})}}{{1 + 3{L_n}}}} \right]\\
\frac{{\chi _{2n + 1}^B(T,{\mu _B})}}{{\chi _{2n - 1}^B(T,{\mu _B})}} &=& 6{\left( {\frac{T}{{{\mu _B}}}} \right)^2}{L_n}\left({1 + \frac{{{L_{n + 1}} - {L_n}}}{{1 + {L_n}}}} \right) \label{equ:31}
\end{eqnarray}
If we consider $L_{n}<<1$, the $r.h.s.$ of the Eq.(\ref{equ:28}) to (\ref{equ:31}) can be simplified as :
\begin{eqnarray}
\frac{{\chi _{2n}^B(T,{\mu _B})}}{{\chi _{2n - 1}^B(T,{\mu _B})}}& =& \frac{T}{{{\mu _B}}} \label{equ:32}\\
\frac{{\chi _{2n + 1}^B(T,{\mu _B})}}{{\chi _{2n}^B(T,{\mu _B})}} &=& 6\frac{T}{{{\mu _B}}}{L_n}\left( {1 + {L_{n + 1}} - 3{L_n}} \right)\\
\frac{{\chi _{2n + 2}^B(T,{\mu _B})}}{{\chi _{2n}^B(T,{\mu _B})}} &=& 6{\left( {\frac{T}{{{\mu _B}}}} \right)^2}{L_n}\left[ {1 + 3({L_{n + 1}} - {L_n})} \right] \label{equ:34}\\
\frac{{\chi _{2n + 1}^B(T,{\mu _B})}}{{\chi _{2n - 1}^B(T,{\mu _B})}} &=& 6{\left( {\frac{T}{{{\mu _B}}}} \right)^2}{L_n}(1 + {L_{n + 1}} - {L_n}) \label{equ:35}
\end{eqnarray}
Based on the Eq.(\ref{equ:32}), (\ref{equ:34}) and (\ref{equ:35}), we have:
%\begin{widetext}
\begin{equation} \label{equ:36}
\begin{split}
&\frac{{\chi _{2n + 2}^B(T,{\mu _B})}}{{\chi _{2n}^B(T,{\mu _B})}} - \frac{{\chi _{2n + 1}^B(T,{\mu _B})}}{{\chi _{2n - 1}^B(T,{\mu _B})}}= \\
& \frac{1}{3}\left[ {\frac{{\chi _{2n + 4}^B(T)}}{{\chi _{2n}^B(T)}} - {{\left( {\frac{{\chi _{2n + 2}^B(T)}}{{\chi _{2n}^B(T)}}} \right)}^2}} \right]{\left( {\frac{{{M_B}}}{{\sigma^B_2}}} \right)^2}
\end{split}
\end{equation}
%\end{widetext}
where we use a leading order approximation $\frac{\sigma^B_{2}}{M_{B}}(T,\mu_B)=\frac{T}{\mu_B}$. We define a temperature dependent quantity $r_{n}(T)$:
\begin{equation} \label{equ:37}
{r_n}(T) = \frac{1}{3}\left[ {\frac{{\chi _{2n + 4}^B(T)}}{{\chi _{2n}^B(T)}} - {{\left( {\frac{{\chi _{2n + 2}^B(T)}}{{\chi _{2n}^B(T)}}} \right)}^2}} \right] 
\end{equation}
Then, deriving from Eq. (\ref{equ:36}) and (\ref{equ:37}), we get:
\begin{equation}
{r_n}(T) = \left( {\frac{{\chi _{2n + 2}^B(T,{\mu _B})}}{{\chi _{2n}^B(T,{\mu _B})}} - \frac{{\chi _{2n + 1}^B(T,{\mu _B})}}{{\chi _{2n - 1}^B(T,{\mu _B})}}} \right){\left( {\frac{{\sigma _B^2}}{{{M_B}}}} \right)^2}
\end{equation}
For the lowest order with $n = 1$, we obtain:
\begin{eqnarray}
\frac{{\sigma _B^2}}{{{M_B}}}(T,{\mu _B}) &=& \frac{T}{{{\mu _B}}}\\
{S_B}{\sigma _B}(T,{\mu _B}) &=& 6\frac{T}{{{\mu _B}}}{L_1}\left( {1 + {L_2} - 3{L_1}} \right)\\
{\kappa _B}\sigma _B^2(T,{\mu _B}) &=& 6{\left( {\frac{T}{{{\mu _B}}}} \right)^2}{L_1}\left[ {1 + 3({L_2} - {L_1})} \right]\\
\frac{{{S_B}\sigma _B^3}}{{{M_B}}}(T,{\mu _B}) &=& 6{\left( {\frac{T}{{{\mu _B}}}} \right)^2}{L_1}\left( {1 + {L_2} - {L_1}} \right)
\end{eqnarray}
where ${L_1} = \frac{1}{6}\frac{{\chi _4^B(T)}}{{\chi _2^B(T)}}{\left( {\frac{{{\mu _B}}}{T}} \right)^2}$ and ${L_2} = \frac{1}{6}\frac{{\chi _6^B(T)}}{{\chi _4^B(T)}}{\left( {\frac{{{\mu _B}}}{T}} \right)^2}$. We may find that $\kappa _B\sigma _B^2$ and $S_B\sigma _B^3/M_B$ are closely related and their difference is:
\begin{equation}
\begin{split}
{\kappa _B}\sigma _B^2 - \frac{{{S_B}\sigma _B^3}}{{{M_B}}} &= 12{\left( {\frac{T}{{{\mu _B}}}} \right)^2}{L_1}\left( {{L_2} - {L_1}} \right)\\
 &= \frac{1}{3}\left[ {\frac{{\chi _6^B}}{{\chi _2^B}} - {{\left( {\frac{{\chi _4^B}}{{\chi _2^B}}} \right)}^2}} \right]{\left( {\frac{{{\mu _B}}}{T}} \right)^2} \\
 &= \frac{1}{3}\left[ {\frac{{\chi _6^B}}{{\chi _2^B}} - {{\left( {\frac{{\chi _4^B}}{{\chi _2^B}}} \right)}^2}} \right]{\left( {\frac{{{M_B}}}{{\sigma _B^2}}} \right)^2}\\
& = {r_1}(T){\left( {\frac{{{M_B}}}{{\sigma _B^2}}} \right)^2}
\end{split}
\end{equation}
where 
\begin{equation}
{r_1}(T) = \frac{1}{3}\left[ {\frac{{\chi _6^B}}{{\chi _2^B}} - {{\left( {\frac{{\chi _4^B}}{{\chi _2^B}}} \right)}^2}} \right]
\end{equation}
In above Taylor expansions of the baryon number susceptibilities in Lattice QCD,  we always assume $\mu_{Q}=\mu_{S}=0$ and expand up to next to leading order. For more realistic, one need to consider the case $\mu_Q \neq \mu_S \neq 0$. In order to compare with experimental data, we need additional constrains. For eg., strangeness neutrality ($N_S$=0) and baryon to charge number ratios equals to 0.4 ($N_{Q}/N_B$=0.4) in Au+Au and Pb+Pb collisions. Furthermore, the self-consistent determination of the freeze-out in QCD thermodynamics for heavy-ion collisions are needed and makes the comparison between Lattice QCD and experimental data with more complication~\cite{Karsch_SQM2016,Karsch_INT_2016}.

\section{Experimental Observables} \label{sec:observable}
Event-by-event particle multiplicity fluctuations can be characterized by the cumulants of the event-by-event multiplicity distributions. It can be calculated as
\begin{align}
	C_1 &= \left<N\right>,C_2 = \left<(\delta N)^2\right>, \\
	C_3& = \left<(\delta N)^3\right> , C_4 = \left<(\delta N)^4\right> - 3\left<(\delta N)^2\right>^2 
\end{align}
where $N$ is particle or net-particle number measured on the event-by-event bias and the $\left<N\right>$  is average over entire event ensemble, $\delta N = N-\left<N\right>$.  With the definition of cumulants, we can also define mean ($M$), variance ($\sigma^{2}$), skewness ($S$) and kurtosis ($\kappa$) as:
\begin{equation}
M = C_{1}, \sigma^2 = C_{2}, S=\frac{C_{3}}{(C_{2})^\frac{3}{2}}, \kappa = \frac{C_{4}}{(C_{2})^2}
\end{equation}
In addition, the moments product $\kappa\sigma^2$ and $S\sigma$ can be expressed in terms of the ratios of cumulants:
\begin{equation}
\label{eq6} \kappa\sigma^2 = \frac{C_{4}}{C_{2}}, S\sigma = \frac{C_{3}}{C_{2}}, \sigma^{2}/M=\frac{C_{2}}{C_{1}}
\end{equation}
The ratios of cumulants are independent on system volume. 
The statistical errors of those cumulants and cumulants ratios are estimated by the Delta theorem~\cite{Delta_theory,Unified_Errors}. In general, the statistical errors strongly depend on the shape of the distributions, especially the width. For gaussian distributions,  the statistical errors of cumulants ($C_{n}$) can be approximated as $error(C_{n}) \propto \sigma^{n}/(\sqrt{N} \epsilon^{n})$, where $\sigma$ is the measured width of the distribution, $N$ represents the number of events and $\epsilon$ is the particle detection efficiency. Theoretically, conserved charge fluctuations are sensitive to the correlation length ($\xi$) of system, which is about 2-3 $fm$ near the QCD critical point in heavy-ion collisions. The fourth order cumulant are proportional to seventh power of the correlation length as $C_4 \propto \xi^7$. Experimentally, the various order cumulants of net-proton (proton number minus anti-proton number), net-charge and net-kaon multiplicity distributions are measured with the data of the beam energy scan at RHIC to search for the signature of the QCD critical point. Since the STAR detector at RHIC can not measure the neutron at mid-rapidity,  the net-proton fluctuations are used to approximate the fluctuations of net-baryons. And the net-kaon multiplicity fluctuations also is used to approximate to the fluctuations of the net-strangeness. To search for the CP in heavy-ion collisions, the event-by-event multiplicity fluctuations are not necessary conserved quantities, for eg., the proton number fluctuations itself can reflect the singularity of the critical point and can be directly used to search for the CP. However, there are two advantages in using the conserved quantities, one is that it can be directly connected to the susceptibilities of system, which can be computed in first principle Lattice QCD, the other one is that due to the dynamical expansion of the QCD medium created in heavy-ion collisions,  the signal of conserved quantities will not be easily washed out by the diffusion process and can be preserved in the final state~\cite{Asakawa}. 

In the following, we will discuss the fluctuation signatures of the QCD critical point from various theoretical calculations, such as $\sigma$ field and NJL model. Finally, we will also show the effects of nuclear potential and baryon number conservations on the cumulants of net-proton (baryon) distributions.

\begin{figure*}[htpb] 
\centering
\begin{minipage}[c]{0.5\textwidth}
\centering
\includegraphics[width=1\textwidth]{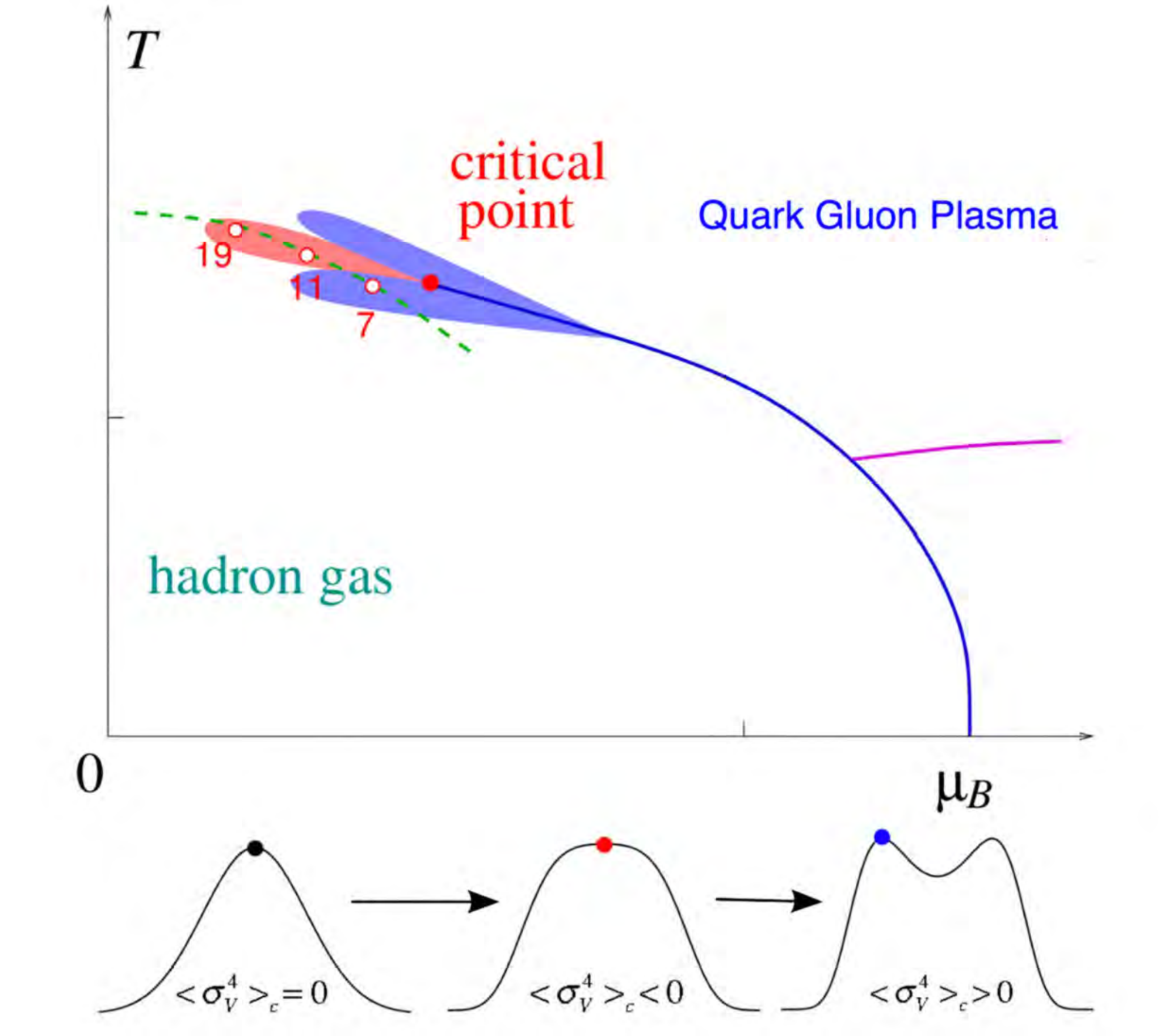} 
\end{minipage} 
\hspace{-1cm}
\begin{minipage}[c]{0.5\textwidth}
\centering
\includegraphics[width=0.8\textwidth]{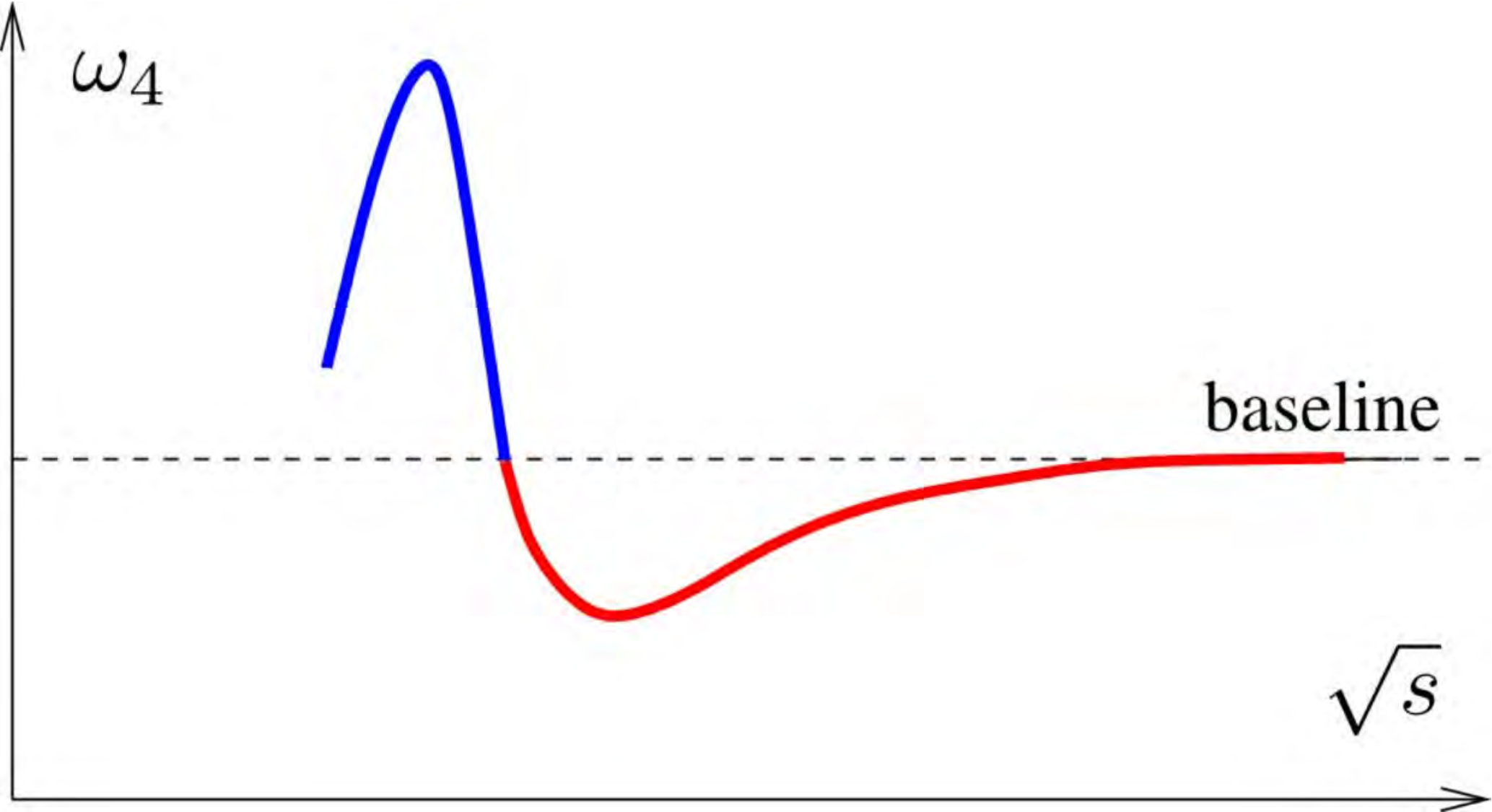}
\end{minipage}
\caption{(Color online) (Top left) The sketch of the QCD phase diagram with sign of the fourth order cumullants of the $\sigma$ field due to critical contributions~\cite{Neg_Kurtosis}. The red region represents negative values and the blue region are positive values. The green dashed line is the chemical freeze-out lines in heavy-ion collisions. (Bottom left) The probability distributions of the $\sigma$ field and the corresponding sign of the fourth order cumulants of the $\sigma$ field distributions. (Right) The expected non-monotonic energy dependence for the normalized fourth order cumulant of multiplicity distributions ($\omega_{4}=\la(\delta N)^4\ra_c/\la N\ra$) when the chemical freeze-out line passes the critical region indicated in the left-top plot.} 
\label{fig:sigma_field}
\end{figure*}
\subsection{Fluctuation Signature near QCD Critical Point} \label{sec:signature}
The characteristic feature of critical point is the divergence of the correlation length, which is limited by the system size and finite time effects due to the critical slowing down. When the critical point is passed by the thermodynamic condition of the matter created in heavy-ion collisions, the expected signature is the non-monotonic variation of the observables with the colliding energy. Many theoretical and model calculations including critical fluctuations have been done for the fluctuations of conserved charges ($B,Q,S$) along the chemical freeze-out lines in heavy-ion collisions. Those can provide predictions on the energy dependence of the fluctuation observables when passing by the critical point.

\subsubsection{$\sigma$ Field Model}  \label{sec:sigmafield}
One of the most important calculations is done with the $\sigma$ field model~\cite{Neg_Kurtosis}. This calculations first time qualitatively discussed the universal critical behavior of the the fourth order ($kurtosis$) of multiplicity fluctuations near the QCD critical point, which are realized by the coupling of particles with the order parameter $\sigma$ field. 
The fluctuations of order parameter field $\sigma(x)$ near a critical point can be described by the probability distributions as:
\begin{equation}
P[\sigma ] \sim \exp \{  - \Omega [\sigma(x)]/T\} 
\end{equation}
where $\Omega$ is the effective action functional for the field $\sigma$ and can be expanded in the powers of $\sigma$:
\begin{equation}
  \label{eq:Omega-sigma}
  \Omega%[\sigma] 
=\! \int\!d^3\bm x\left[
\frac{(\bm\nabla\sigma)^2}{2} +
\frac{m_\sigma^2}2 \sigma^2 
+ \frac{\lambda_3}{3}\sigma^3
+ \frac{\lambda_4}{4}\sigma^4 + \ldots
\right]
\,.
\end{equation}
where $m_{\sigma}=1/\xi$ and the critical point is characterized by $\xi \to \infty$. For the moments of the zero momentum mode ${\sigma _V} = \int {\sigma (x){d^3}} x$. Then, we have
\begin{eqnarray}
&& \la \sigma _V^2 \ra= VT{\xi ^2}\\
 &&\la \sigma _V^3 \ra  = 2{\lambda _3}VT{\xi ^6}\\
&& \la \sigma _V^4 \ra_c= 6V{T^3}[2{({\lambda _3}\xi )^2} - {\lambda _4}]{\xi ^8}
\end{eqnarray}
where $ \la \sigma _V^4 \ra_c$ is the fourth order cumulants of the $\sigma$ field. It is found that the higher order fluctuations are with higher power of the correlation length and diverge faster. 
If we introduce the coupling of the particles with the $\sigma$ field, the fourth order cumulants of the particle multiplicity distributions can be obtained as:
\begin{equation}
  \label{eq:N4}
  \la(\delta N)^4\ra_c = \la N\ra + \la\sigma_V^4\ra_c
\left(\frac{g\,d}{T} \int_{\bm p}\frac{ n_{\bm p}
%(1\pm n_{\p})
}{\gamma_{\bm p}}\right)^4 + \ldots,
\end{equation}
where $n_p$ is the equilibrium distributions for a particle of a given mass, $\gamma_p=(dE_p/dm)^{-1}$ is the relativistic gamma factor of a particle with momentum $p$ and mass $m$, $g$ is the coupling constant and $d$ is the degeneracy factor.
The mean value  $\la N \ra$ in the $r.h.s.$ of the Eq.(\ref{eq:N4}) 
is the pure statistical contribution (Poisson). 

\begin{figure*}[thpb]
\centering
\hspace{-0.5cm}
\includegraphics[width=0.8\textwidth]{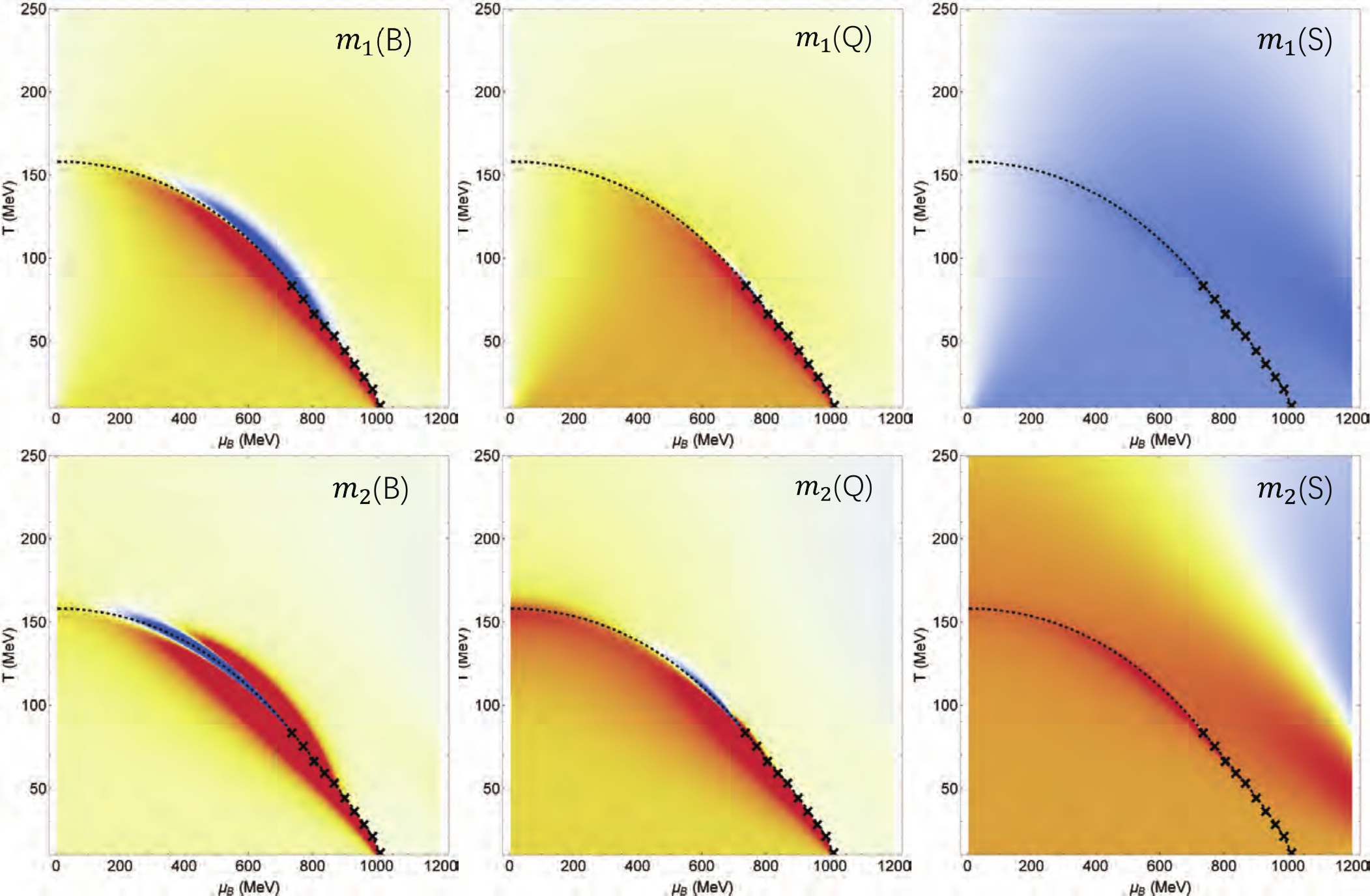}
\caption{(Color online) The sign of the $m_1$ (top) and $m_2$ (bottom) of the baryon ($B$), charge ($Q$) and strangeness ($S$) number. The red region represents positive value while blue region represents negative value. The dashed line is the crossover line while the crosses denotes the first order phase transition boundary~\cite{Wenkai_NJL}.}
\label{fig:m2m1}
\end{figure*}

\begin{figure}[hptb]
\centering
\includegraphics[width=0.45\textwidth]{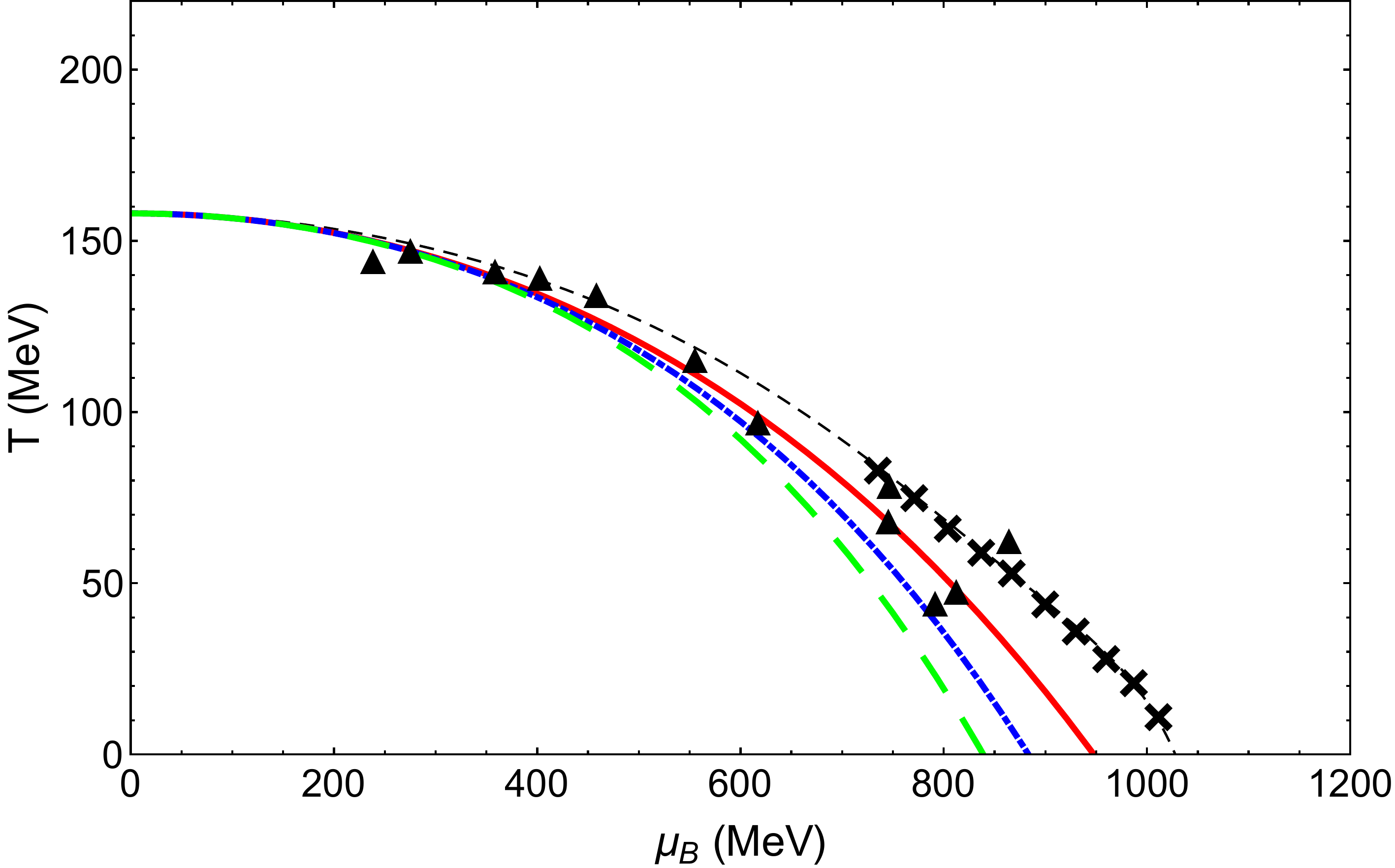}
  \caption{The red soild, blue dot-dashed and green dash lines represent three hypothetical freeze-out curves~\cite{Wenkai_NJL}. The black dashed line is the crossover line and the crosses denote the curve of the first order phase transition boundary. The triangles are experimental chemical freeze-out data.}\label{fig:freezeout}
\end{figure}

\begin{figure*}[hptb]
\centering
\includegraphics[width=0.6\textwidth]{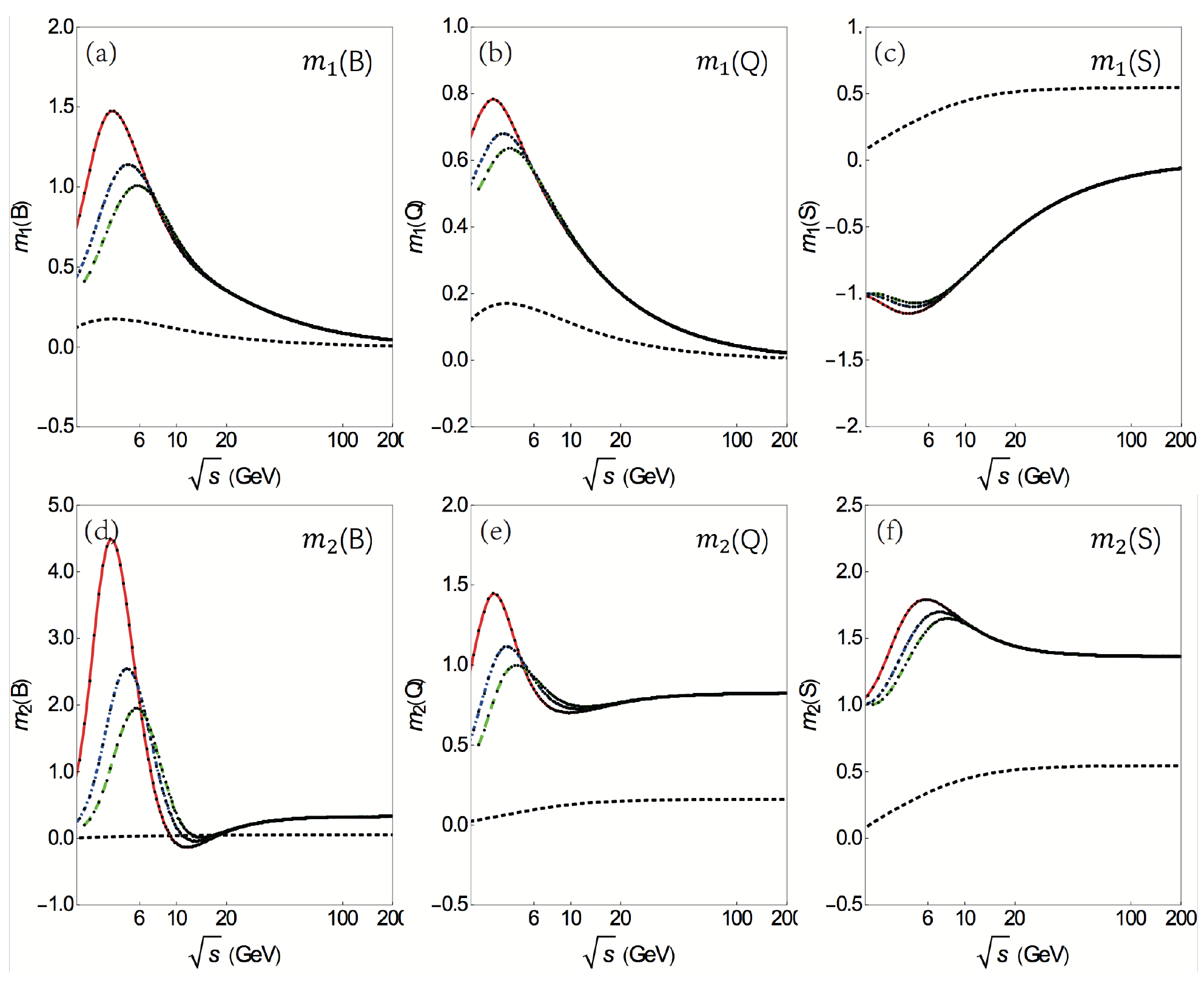}
\hspace{0.5cm}
\includegraphics[width=0.3\textwidth]{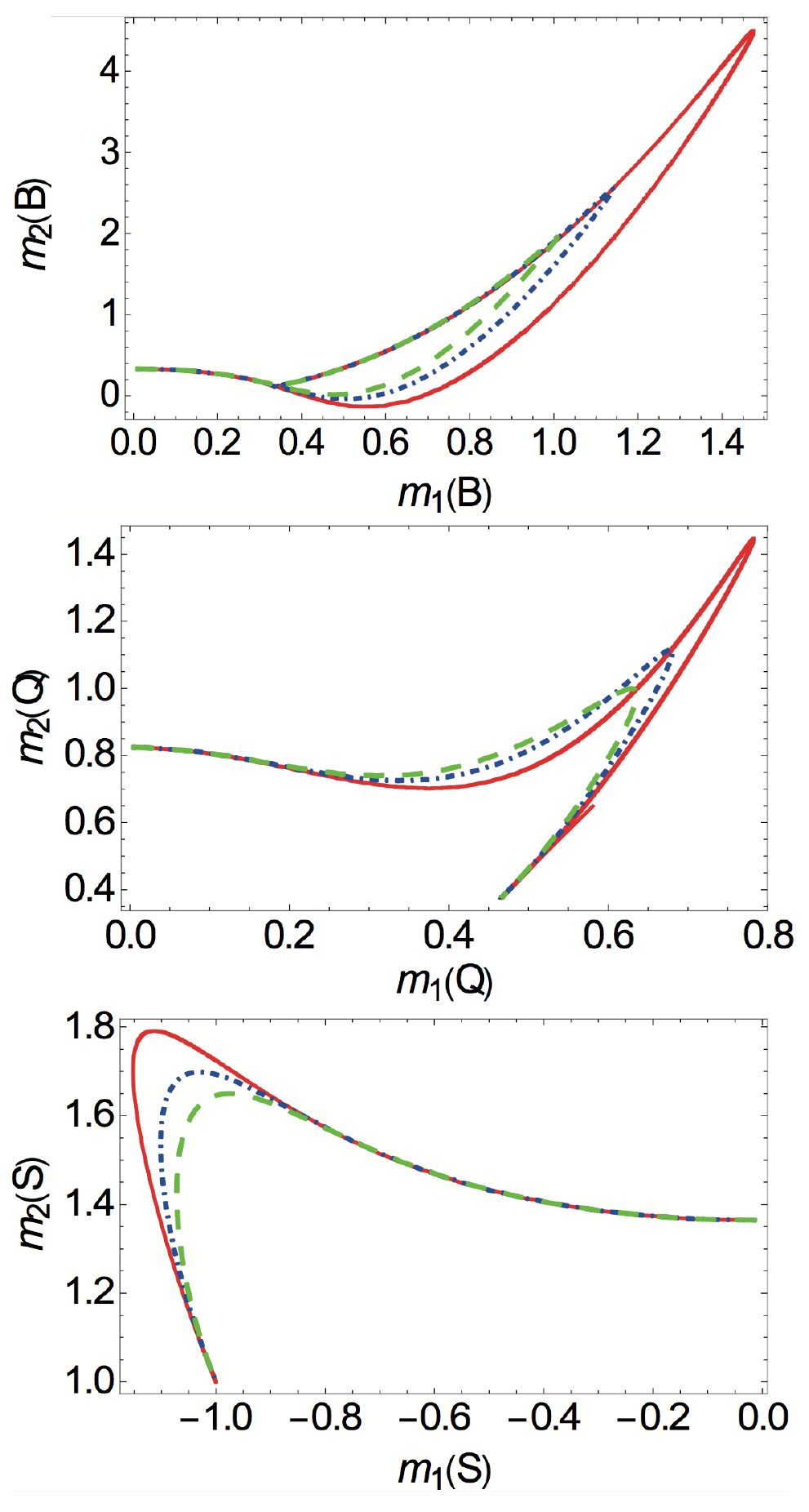}
\caption{(Color online) (Left) The $m_1$ and $m_2$ of baryon ($B$), charge ($Q$) and strangeness ($S$) as a function of colliding energy along the three hypothetical freeze-out lines as plotted in Fig. \ref{fig:freezeout}. The black dashed lines are the results from a free quark gas model. (Right) The correlation plot $m_2$ versus $m_1$ for baryon (top), charge (middle) and strangeness (bottom) along three hypothetical freeze-out lines~\cite{Wenkai_NJL}.}
\label{fig:m1m2all}
\end{figure*}
Figure \ref{fig:sigma_field} left displays the sketch of QCD phase diagram with critical contributions to the $\sigma$ field. When the chemical freeze-out lines (green dashed line) pass by the critical point from the crossover side, the probability distributions of the $\sigma$ field change from gaussian to the double-peak non-gaussian distribution and the corresponding fourth order cumulant change from zero to negative (red region) and to positive (blue region). When this $\sigma$ field couples with the particles, it leads to a non-monotonic energy dependence of the normalized fourth order cumulants of multiplicity distributions ($\omega_{4}=\la(\delta N)^4\ra_c/\la N\ra$) along the chemical freeze-out line, as shown in the right of the Fig.\ref{fig:sigma_field}, where the baseline is unity, the Poisson baseline. However, one has to keep in mind that here we only consider the critical point and statistical fluctuation contributions.  Other dynamical effects in heavy-ion collisions, such as the effects of baryon number conservations, hadronic scattering and resonance decay, are not taken into account. Furthermore, the finite size and finite time effects, non-equilibrium memory effects are also important and need to be carefully studied. 

This critical point induced non-monotonic energy dependence of the fourth order cumulants along the chemical freeze-out line has been confirmed by many other model calculations, such as NJL~\cite{2015_JianDeng_fluctuation,Wenkai_NJL}, PQM~\cite{BFriman_EPJC,2014_Bengt_flu}, chiral hydrodynamics~\cite{Nahrgang_EPJA_2016,herold2016dynamical} and other calculations~\cite{VDW_flu,Marcus_2016,Mukherjee_2016}. It indicates the $\sigma$ field calculations capture the main feature of the critical point. However, it is still a crude model. Here, the $\sigma$ field model only considers the critical fluctuations in static and infinite medium without taking account for the off-equilibrium effects in the dynamical expanding of the fireball created  in heavy-ion collisions. Recently, a theoretical paper discussed critical fluctuations considering the off-equilibrium effects within Kibble-Zurek framework and observed a universal scaling of critical cumulants~\cite{Swagato_2016_PRL}.

\subsubsection{NJL Model}  \label{sec:NJL}
A QCD based effective model-the so called Nambu-Jona-Lasinio (NJL) model is also widely used to study the conserved charge fluctuations near the QCD critical point. In this model, quark and gluon are the basic degree of freedom. Although, there is no mechanism of the quark confinement implemented in the NJL model, it is still a simple and useful way to study the qualitative behavior of the susceptibility of the conserved charges near the QCD critical point. Here we just show the results of two susceptibility ratios calculated from NJL model :
\begin{equation}
{m_1}(q) = \frac{{\chi _3^q}}{{\chi _2^q}},\begin{array}{*{20}{c}}{}&{}
\end{array}{m_2}(q) = \frac{{\chi _4^q}}{{\chi _2^q}}
\end{equation}
where $q=B,Q,S$, $\chi^q_n$ is the $n^{th}$ order susceptibility. Fig.~\ref{fig:m2m1} shows the sign of the $m_1$ and $m_2$ of baryon, charge and strangeness number. The red region are of positive values wheres the blue region represents negative values.  The yellow regions represent the values of  $m_1$ and $m_2$ are very close to zero. One may notice that the signals from baryon number fluctuations are stronger than from charge and strangeness number fluctuations. This is mainly due to the mass effects that the strange quark mass ($m_s$) is much heavier than the mass of light quarks ($m_u,m_d$). Fig.\ref{fig:freezeout} displays three colored hypothetical chemical freeze-out lines. The red solid freeze-out line is fitted to recent experimental data.  The parametrized formula for obtaining the three curves are 
\begin{equation}\label{eq:freezeout}
  T(\mu_B)=a-b\mu_B^2-c\mu_B^4
\end{equation}
where $a=0.158$ GeV, $b=0.14$ $\mathrm{GeV^{-1}}$, and $c=0.04$ (solid), $0.08$ (dot-dashed), $0.12$ (dashed) $\mathrm{GeV^{-3}}$. The relation between baryon chemical potential ($\mu_{B}$) and collision energy can be parametrized as~\cite{Begun_SQM2016}: 
\begin{equation}\label{equ:muBs}
  \mu_B(\sqrt{s})=\frac{1.477}{1+0.343\sqrt{s}}
\end{equation}
With the freeze-out curve and Eq.~\eqref{equ:muBs}, we plot the $m_1,m_2$ of $B,Q,S$ as a function of colliding energy in Fig.~\ref{fig:m1m2all} along the three chemical freeze-out lines as shown in Fig.~\ref{fig:freezeout}. The black dashed lines in Fig.~\ref{fig:m1m2all} left are the results from the free quark gas model. When approaching the critical point at low energies, the NJL model predicts non-monotonic signal of the susceptibility ratios while for the free gas case all moments are close to $0$. Furthermore, we can infer that $m_2(B)$ should be a better probe of the critical behavior due to larger magnitude in signal and also the most important one, having sign changes from negative to positive with respect to collision energy than other cases. As we mentioned, since there has no quark confinement in NJL model, the baselines obtained from NJL model (away from critical point) are different from the ones from hadron resonance gas model, which is unity.  One can also see that the behavior near QCD critical point is very much different from the results of weakly interacting quark gas. The behavior of these two quantities $m_1(B)$ and $m_2(B)$ at colliding energies at few GeV where experiments have not covered yet are of great importance as some other models predict opposite slope of these two quantities compared to the NJL prediction.  Fig. \ref{fig:m1m2all} right shows correlations between the $m_2$ and $m_1$ for baryon, charge and strangeness, respectively. 
We can see that the $m_2$ and $m_1$ correlation along the three chemical freeze-out lines for baryon shows a closed loop with sign changes and looks like a banana shape. This is very different behavior comparing with the charge and strangeness sector. 

\subsection{Baselines and Background Effects in Heavy-ion Collisions} \label{sec:baselines}
In this section, we discuss the statistical baselines and some of the non-CP physics background effects for the fluctuations measurements in heavy-ion collisions. 
The discussion of thermal blurring, diffusion and resonance decay effects can be found in~\cite{Asakawa_review,Asakawa_thermal,2015_Kitazawa_Rapidity} and~\cite{HRG_baseline,HRG_Nahrgang,Marcus_2016}, respectively. 
\subsubsection{Expectations from Poisson,  Binomial and Negative Binomial Statistics}\label{sec:Poisson}
In the following, we discuss some expectations for cumulants of net-proton multiplicity distributions from some basic distributions~\cite{QM2014_baseline}. 
\begin{enumerate}
 \item {\bf Poisson Distributions :}
If the particle and anti-particle are independently distributed as Poissonian distributions. Then the net-proton multiplicity will follow the Skellam distribution, which is expressed as: \\ $P(N) = {(\frac{{{M_p}}}{{{M_{\overline p}}}})^{N/2}}{I_N}(2\sqrt {{M_p}{M_{\overline p}}} )\exp [ - ({M_p} + {M_{\overline p}})],$ where the $N$ is the net-proton number,  $I_{N}(x)$ is a modified Bessel function, $M_{p}$ and $M_{\overline p}$ are the mean number of particles and anti-particles, as shown in Fig. 1. The various order cumulants ($C_{n}$) are closely connected with the moments, e.g., $C_{1}=\la N\ra=M,C_{2}=\la(\delta N)^2\ra=\sigma^{2}, C_{3}=\la(\delta N)^3\ra=S \sigma^{3}, C_{4}=\la(\delta N)^4\ra-3\la(\delta N)^2\ra^2=\kappa \sigma^{4}$, where the $\delta N$ =$N-\la N\ra$, the $\sigma^2$, $S$ and $\kappa$ are variance, skewness and kurtosis, respectively. Then we construct,  $S\sigma  = {C_3}/{C_2} = ({M_p} - {M_{\overline p }})/({M_p} + {M_{\overline p }})$ and $\kappa {\sigma ^2} = {C_4}/{C_2} = 1$, which provides the Poisson expectations for the 
 various order cumulants/moments of net-particle distributions. The only input parameters of the Poisson baseline for cumulants of net-particle distributions are the mean values of the particle and anti-particle distributions.

 \item {\bf Binomial and Negative Binomial Distributions:}
If the particle and anti-particle are independently distributed as Binomial or Negative Binomial distributions (BD/NBD). Then various order cumulants of the net-particle distributions can be expressed in term of cumulants of the particle and anti-particle distributions: $C_n^{net - p} = C_n^p + {( - 1)^n}C_n^{\bar p}$.
The first four order cumulants can be written as: 
$C_2^x = \sigma _x^2 = {\varepsilon _x}{\mu _x},C_3^x = {S_x}\sigma _x^3 = {\varepsilon _x}{\mu _x}(2{\varepsilon _x} - 1),C_4^x = {\kappa _x}\sigma _x^4 = {\varepsilon _x}{\mu _x}(6\varepsilon _x^2 - 6{\varepsilon _x} + 1)$
, where $\varepsilon_x=\sigma _x^2/\mu_x$, $\mu_x=M_x$, $M_{x}$ is the mean values of particles or anti-particles distributions, $x$=particle or anti-particle. $\varepsilon_x<1$ means the underlying distributions of particles or anti-particles are Binomial distributions, 
while $\varepsilon_x>1$ gives Negative Binomial distributions~\cite{westfall}. The input parameters for BD/NBD expectations are the measured mean and variance of the particle and anti-particle distributions.

 \end{enumerate}
 
\begin{figure*}[htb]
		\includegraphics[height=2.6in]{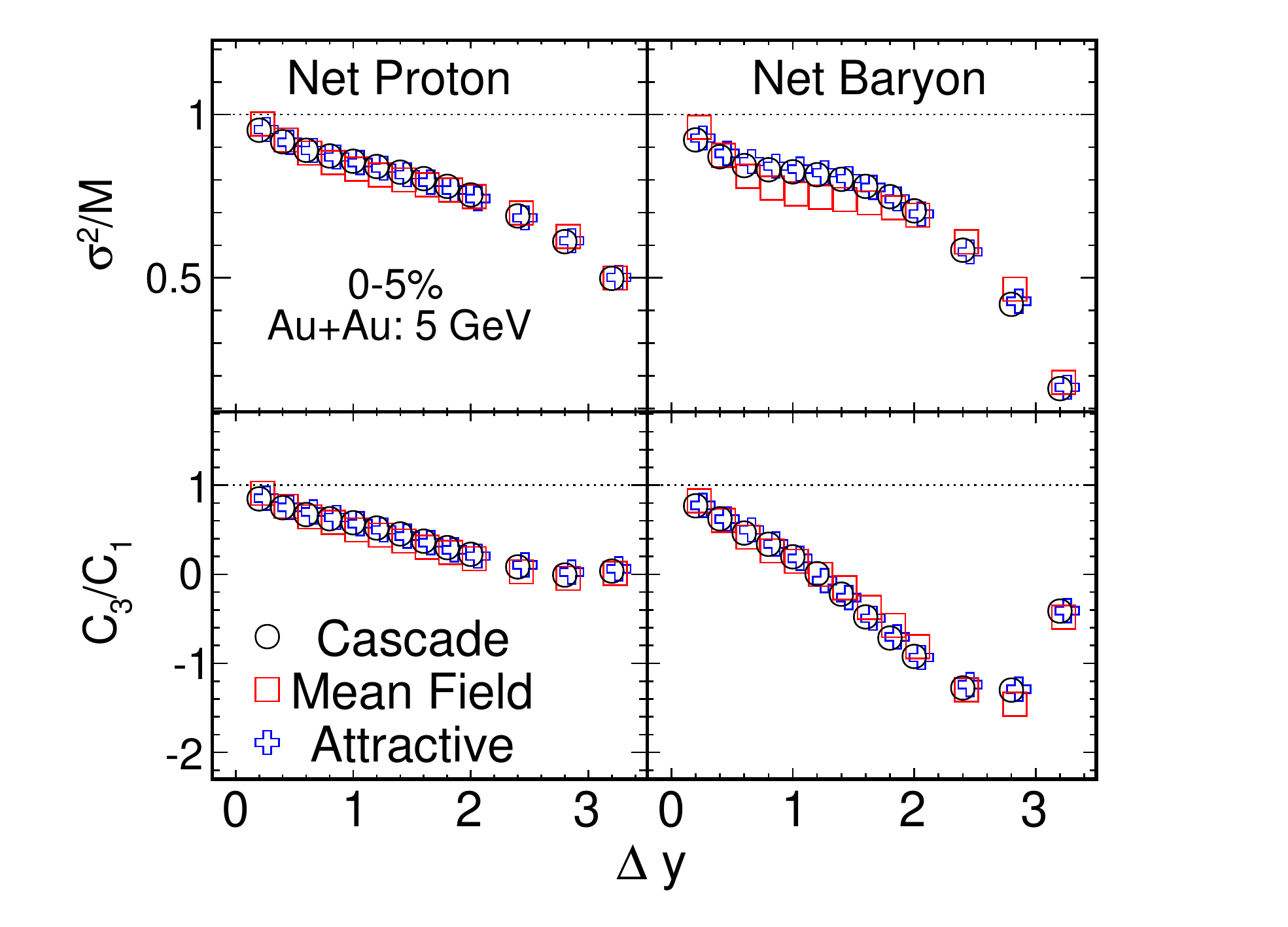}
		\hspace{-1.cm}
		\includegraphics[height=2.6in]{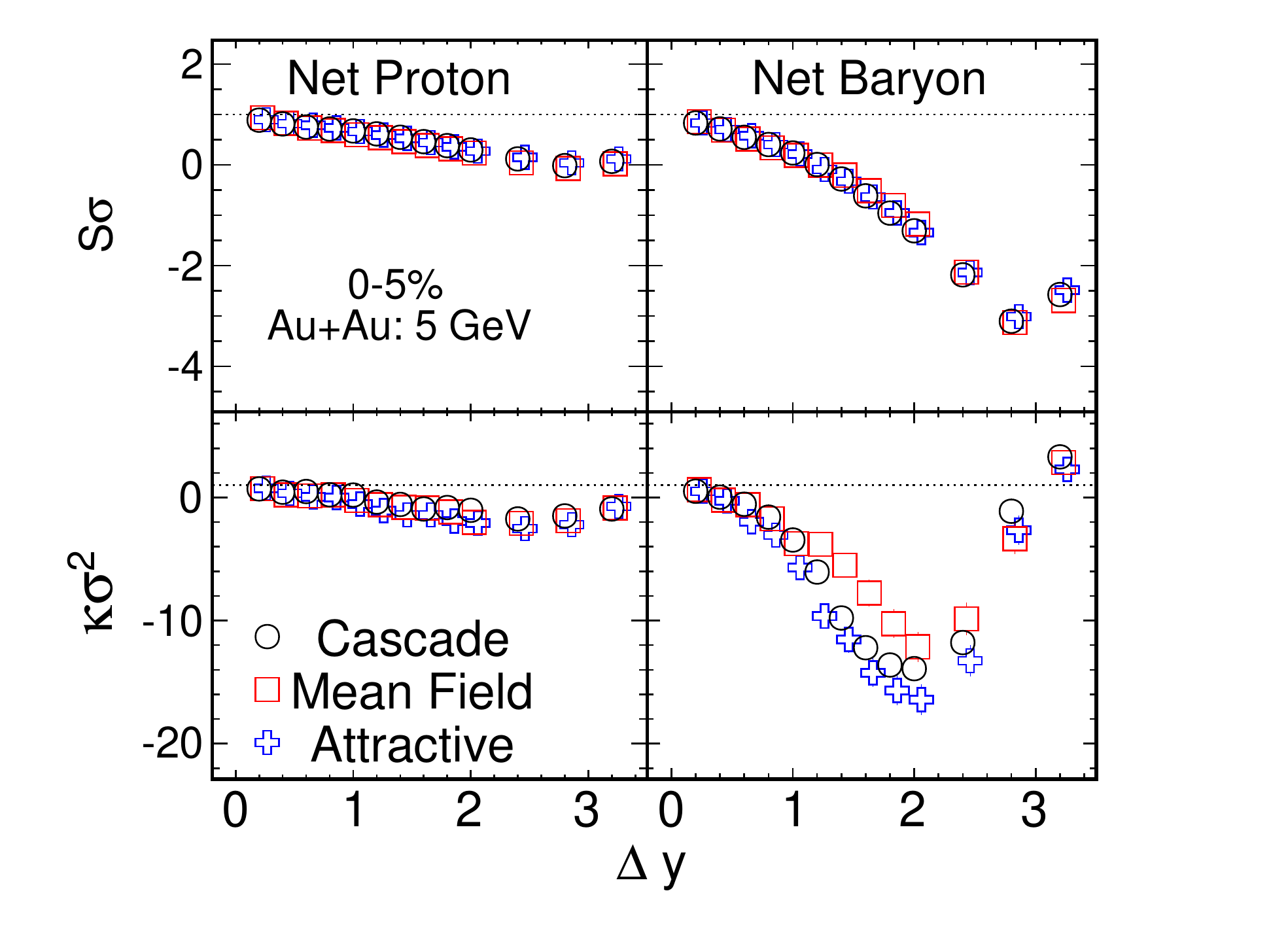}
	\caption{Rapidity dependence for the Cumulants ratios of net proton and net baryon multiplicity distributions in Au+Au collisions at $\sqrt{s_{\text{NN}}} = 5\,\text{GeV}$ GeV from JAM model computed in the three different modes~\cite{ShuHe_PLB}. In the left shows $\sigma^{2}/M$ ($C_2/C_1$) and $C_3/C_1$. The figure in the right shows $S\sigma$ ($C_3/C_2$) and $\kappa\sigma^2$($C_4/C_2$) . The dashed horizontal lines are with the value of unity.} \label{fig:ratios}
\end{figure*}

\begin{figure*}[htb]
\hspace{-0.5cm}
		\includegraphics[height=4in]{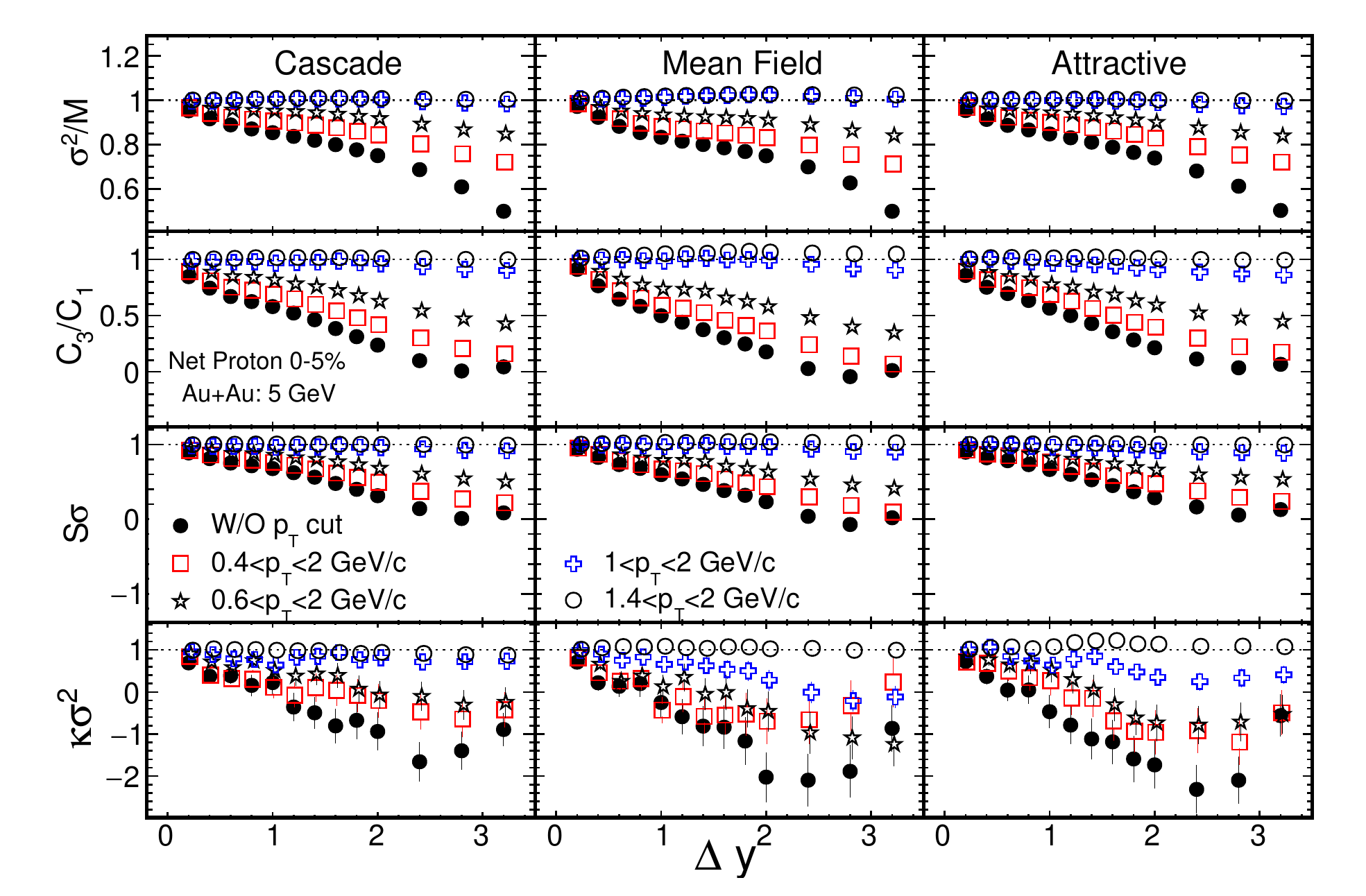}
	\caption{Rapidity dependence for the cumulants ratios of net-proton distributions in Au+Au collisions at $\sqrt{s_{NN}}=5$ GeV from JAM model computed in the three different modes and various transverse momentum ranges~\cite{ShuHe_PLB}. From top to bottom are  $\sigma^{2}/M$ ($C_2/C_1$),  $C_3/C_1$, $S\sigma$ ($C_3/C_2$) and $\kappa\sigma^2$($C_4/C_2$), respectively. The dashed horizontal lines are with the value of unity.}
	\label{fig:pt}
\end{figure*}

\subsubsection{Effects of Baryon Number Conservation and Nuclear Potential on Net-Proton (Baryon) Cumulants}  \label{sec:JAM}
The effects of baryon number conservations (BNS) and mean field potential are more and more important at low energies. To study those effects on the fluctuations of net-proton (baryon) number, the rapidity and transverse momentum dependence for the cumulants of the net-proton (baryon) multiplicity distributions in Au+Au collisions at $\sqrt{s_{\text{NN}}} = 5\,\text{GeV}$ have been studied within a microscopic hadronic transport (JAM) model~\cite{ShuHe_PLB}. The simulations were done with two different modes, which are the mean field and the softening of equation of state (EoS) mode, respectively. The softening of EoS is simulated by introducing attractive orbits in the two-body scattering to realize a smaller pressure of the system. It was found that the mean field potential and softening of EoS have strong effects on the rapidity distributions ($\text{d}N/\text{d}y$) and the shape of the net-proton (baryon)  distributions.  By comparing the results from the two modes with the results from default cascade, one found that the net-proton (baryon) cumulants and ratios from the three modes have similar trends and show strong suppression with respect to unity, which is attributed to the effects of baryon number conservations~\cite{BNS_2013_Koch,HRG_FJH2}. It means that the effects of mean field potential and softening of EoS might be not responsible for the observed strong enhancement in the most central  (0-5\%) Au+Au collisions at 7.7 GeV measured by the STAR experiment at RHIC. 
\begin{figure*}[htb]
\hspace{-1cm}
\begin{minipage}[c]{0.5\linewidth}
\centering 
    \includegraphics[scale=0.4]{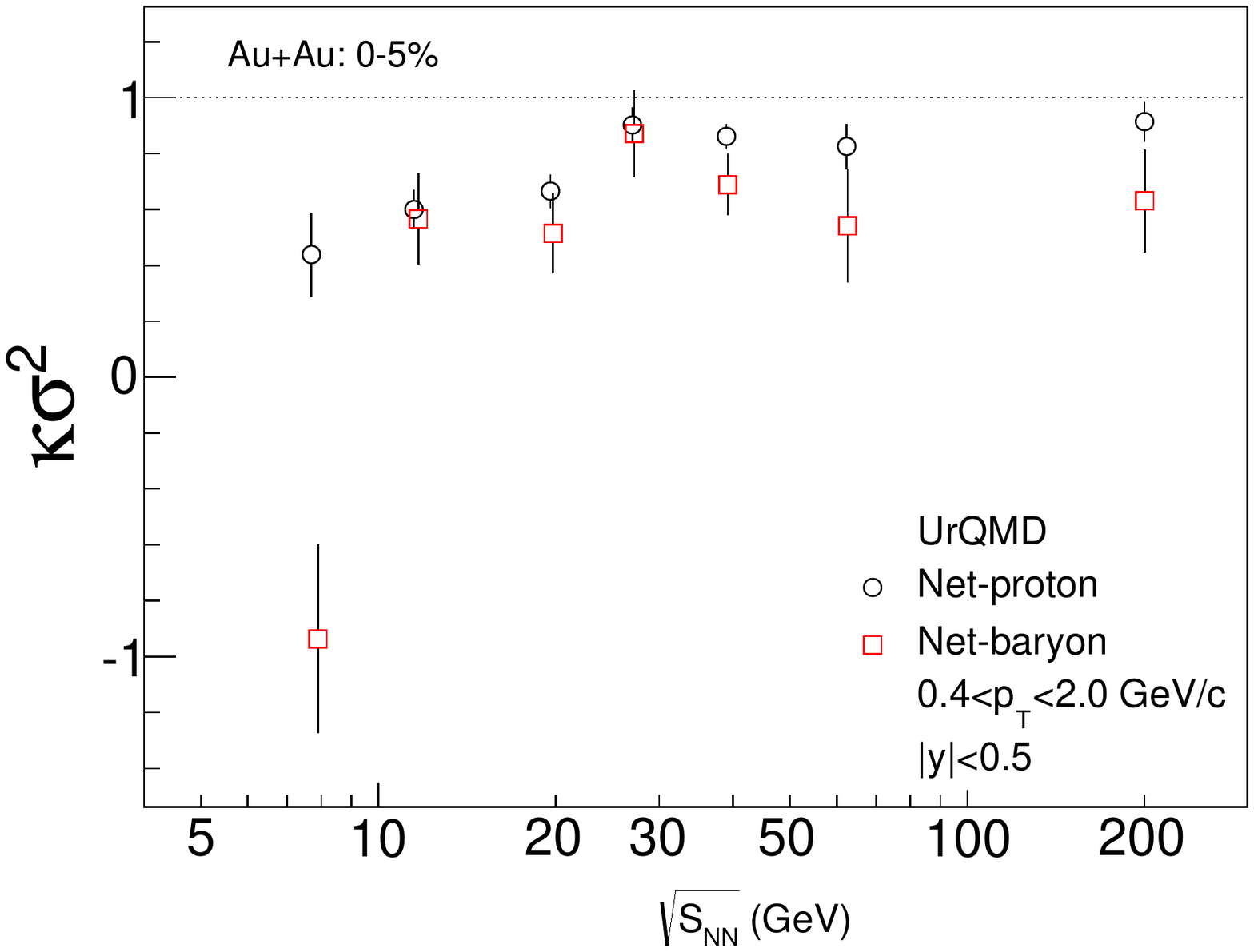}
    \end{minipage}
  \hspace{-0.2in}
  \begin{minipage}[c]{0.5\linewidth}
  \centering 
   \includegraphics[scale=0.5]{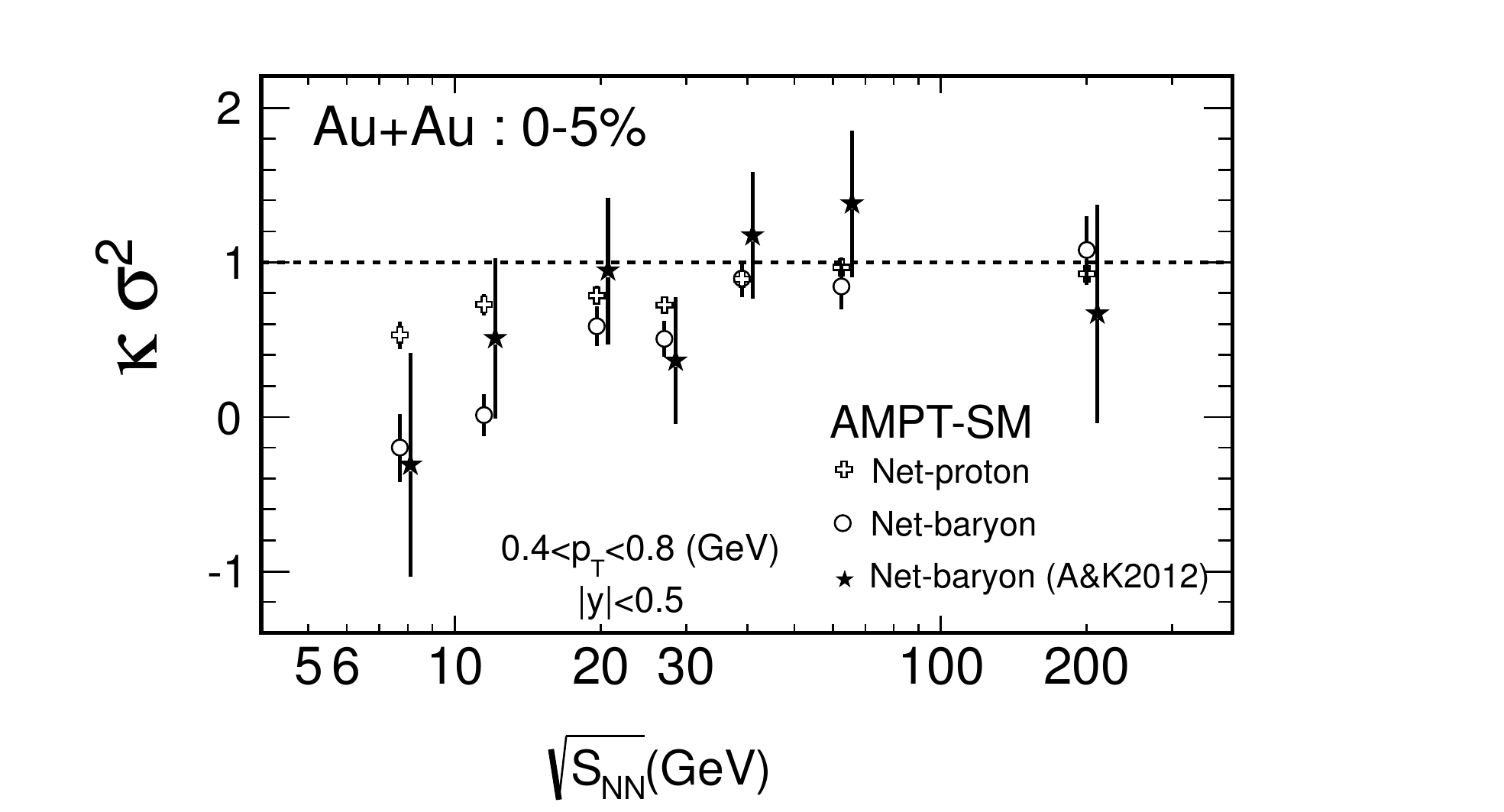}
      \end{minipage} 
            \caption{(Color online) Energy dependence of {\KV} of net-proton and net-baryon distributions for 0-5\% Au+Au collisions from the UrQMD (left) and
    the AMPT string melting model (right). The results marked as solid black stars are based on theoretical calculations using Asakawa 
    and Kitazawa's formula. The error calculation is based on the Bootstrap method. } \label{fig:MP_energy}
\end{figure*}

Figures~\ref{fig:ratios} shows cumulant ratios of net-proton (baryon) distributions in Au+Au collisions at $\sqrt{s_{\text{NN}}} = 5 \,\text{GeV}$ from JAM model. When increasing the rapidity acceptance ($\Delta y$), the net-proton (baryon) cumulant ratios will decrease,  reach a minimum and then increase, which is the typical effects of baryon number conservation~\cite{BNS_2013_Koch,HRG_FJH2}. For different net-proton (baryon) cumulant ratios, the position of the minimum are different. It indicates the mean field potential and softening of EoS will not lead to large increase above unity for the net-proton (baryon) cumulants ratios.  Instead, due to the baryon number conservation, large suppression for the fluctuations of net-proton (baryon) are observed. The rapidity dependence for the cumulants ratios calculated from the three modes are with the similar trend. It suggests that the observed similar trends obtained by JAM model without implementing critical physics are dominated by the effects of baryon number conservation. On the other hand, one observes that the net-baryon cumulant ratios show larger suppression with respect to unity than the net-proton and the higher order cumulant ratios also show larger suppression than the lower order. On the other hand, as the mean field potential implemented in the JAM model is momentum dependent, it is also important to study the momentum dependence for the cumulants of net-proton distributions. In Fig. \ref{fig:pt},  for different transverse momentum range, we plot the cumulant ratios of net-proton distributions as a function of rapidity window, which are calculated with the three different modes. The results computed from different modes are with the similar trends. When the $p_{T}$ coverage is enlarged, the cumulant ratios are suppressed with respected to unity, the Poisson expectations. When the $p_{T}$ range is small, the fluctuations are dominated by Poisson statistics and the cumulant ratios are very close to unity.  Another study for the effect of mean field on baryon number fluctuations done with a Relativistic Mean Field (RMF) approach can be found in~\cite{Kenji_MF}.

\subsubsection{Net-Proton versus Net-Baryon Kurtosis from UrQMD and AMPT Model}\label{sec:UrQMD_AMPT}
The STAR experiment measures net-proton fluctuations instead of net-baryon fluctuations and one may want to know to what extend they can reflect the net-baryon fluctuations in heavy-ion collisions. Therefore, fig. 3 demonstrates the comparison between moments of net-proton and net-baryon distributions from AMPT~\cite{ampt} and UrQMD~\cite{urqmd} model calculations. We can find that the {\KV} of net-baryon distributions are systematically lower than the net-proton results.  The differences are even bigger for low energies than high energies. There are two possible effects for the difference between net-proton and net-baryon fluctuations, one is the non inclusion of neutrons in the net-proton fluctuations, and the other one is the nucleon isospin exchanging process due to $\Delta$ resonance formation via $p\pi$ and $n\pi$ interaction, the so called isospin randomization, which will modify the net-proton fluctuations after the chemical freezeout. A set of formulas have been derived to convert the measured net-proton cumulants to the net-baryon cumulants by taking into account the above two effects~\cite{Asakawa_formula}. The converting formulas for various order net-baryon cumulants can be written as:
\begin{widetext}
\begin{eqnarray}
C_1^{net - B} &=& 2C_1^{net - p} \\
C_2^{net - B} &=& 4C_2^{net - p} - 2C_1^{tot - p} \\
C_3^{net - B} &=& 8C_3^{net - p} - 12(C_2^p - C_2^{\bar p}) + 6C_1^{net - p} \\
C_4^{net - B} &=& 16C_4^{net - p} + 16C_3^{tot - p} - 64(C_3^p + C_3^{\bar p}) + 48C_2^{net - p} + 12C_2^{tot - p} - 26C_1^{tot - p}
\end{eqnarray}
\end{widetext}
where $tot$-$p$ means proton number plus anti-proton number. The right side of fig. 3 shows the net-baryon {\KV} ($C_{4}/C_{2}$) results, converted from the net-protons fluctuations. Within large uncertainties, they are consistent with the net-baryon results directly calculated with the AMPT model.

\begin{figure*}[htb]
\hspace{-0.5cm}
		\includegraphics[height=3.2in]{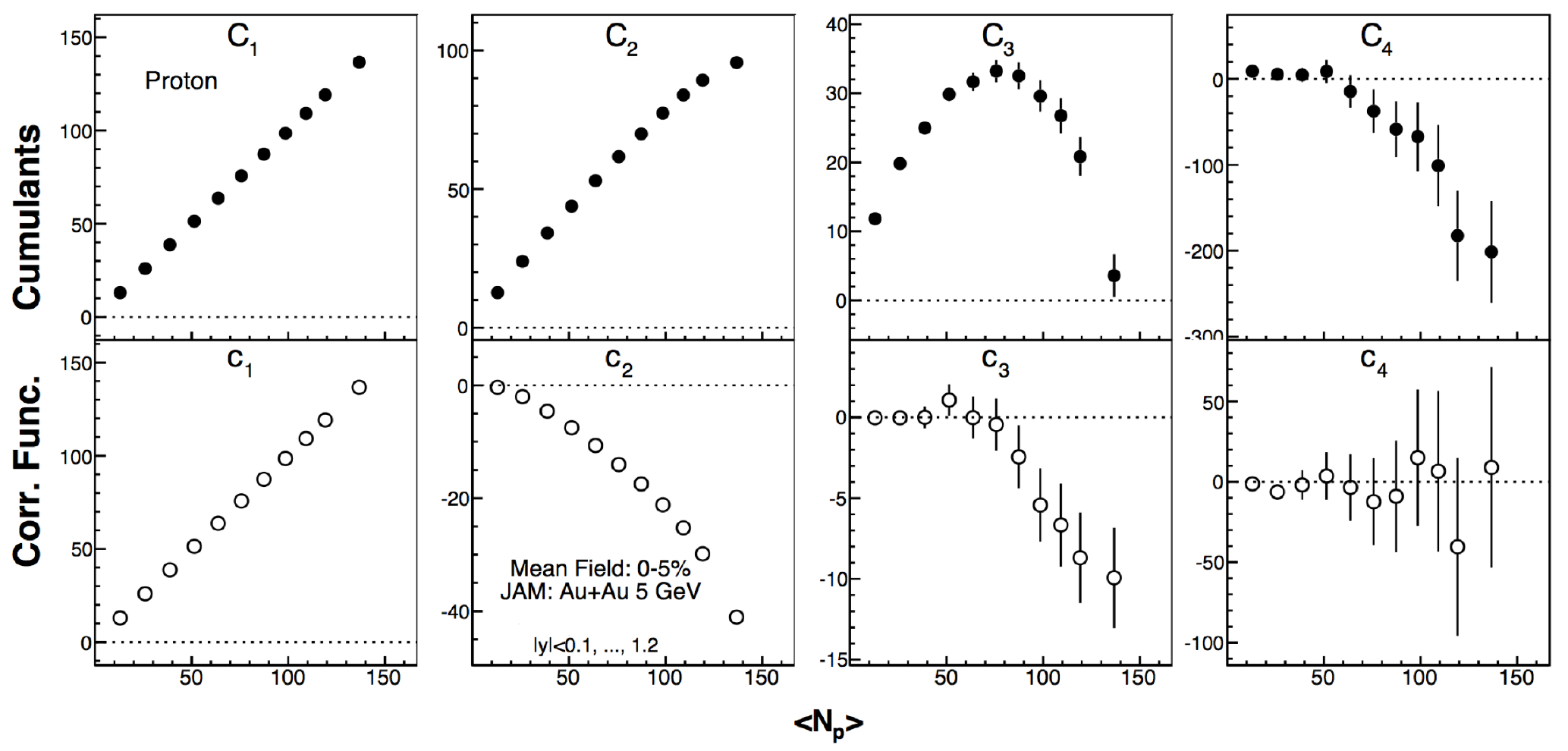}
	\caption{(Color online) Proton cumulants (top panels) and correlation functions (bottom panels) as a function of mean proton number ($\la N_p\ra$) in Au+Au collisions at {\sNN}=5 GeV from JAM model. The mean proton number is varied by changing the rapidity coverage. }
	\label{fig:corr_func}
\end{figure*}
\subsubsection{Cumulants and Correlation Functions} \label{sec:corr_fun}
Fluctuations and correlations are closely related to each other and they are two sides of coins. The various order cumulants can be expressed into the linear combinations of 
the multi-particle correlation functions~\cite{2015_Acceptance_Misha,Corr_func_Koch_2016}, which are directly related to the correlation length ($\xi$) of system. The multi-particle density are related to factorial moments as:
\begin{equation}
{F_n} =  \la N(N - 1)...(N - n + 1) \ra  = \int {d{p_1}} ...d{p_n}\rho ({p_1},...,{p_n})
\end{equation}
where $F_n$ is the $n^{th}$ order factorial moment and $\rho ({p_1},...,{p_n})$ is the $n$ particle density distributions. The integral sums over the interested phase space. 
The generation function of the factorial cumulants is :
\begin{equation}
g(t) = \ln  \la {(1 + t)^N} \ra  = \sum\limits_{k = 1}^\infty  {{c_k}\frac{{{t^k}}}{{k!}}} 
\end{equation}
\begin{equation}{c_k} = {\left. {\frac{{{\partial ^k}g(t)}}{{\partial {t^k}}}} \right|_{t = 0}} \end{equation}
where $c_k$ is the $k^{th}$ order factorial cumulant, $N$ is the random variable. We have the relation between factorial moments and correlation function as:
\begin{eqnarray}
{F_1} &=& \int {dp\rho (p) =  \la N \ra } \\
{F_2} &=& \int {d{p_1}} d{p_2}\rho ({p_1},{p_2}) = {F_1}^2 + {c_2}\\
{F_3} &=& \int {d{p_1}} d{p_2}d{p_3}\rho ({p_1},{p_2},{p_3}) \nonumber \\
&=& {F_1}^3 + 3{c_2}{F_1} + {c_3} \\
{F_4} &=& \int {d{p_1}} d{p_2}d{p_3}d{p_4}\rho ({p_1},{p_2},{p_3},{p_4}) \nonumber \\
&=& {F_1}^4 + 6{c_2}F_1^2 + 4{F_1}{c_3} + 3c_2^2 + {c_4}
\end{eqnarray}
On the other hand, the relation between factorial cumulant ($c_k$) and cumulants can be expressed as:
\begin{eqnarray}
{c_k} = \sum\limits_{i = 0}^k {{s_1}(k,i){C_i}}  \\
{C_k} = \sum\limits_{i = 0}^k {{s_2}(k,i){c_i}} 
\end{eqnarray}
where the $s_1$ is the sterling number of the first kind, $C_i$ is $i^{th}$ order cumulant. 
Then, we have the following equations:
\begin{eqnarray}
{c_1} &=& {C_1} =  \la N \ra \\
{c_2} &=& {C_2} - \la N \ra \\
{c_3} &=& {C_3} - 3{C_2} + 2 \la N \ra \\
{c_4} &=& {C_4} - 6{C_3} + 11{C_2} - 6 \la N \ra
\end{eqnarray}
\begin{eqnarray}
{C_1} &=& {c_1} =  \la N \ra \\
{C_2} &=&  \la N \ra  + {c_2}\\
{C_3} &=&  \la N \ra  + 3{c_2} + {c_3}\\
{C_4} &=&  \la N \ra  + 7{c_2} + 6{c_3} + {c_4}
\end{eqnarray}

It is well known that high order cumulants ($C_n,n>2$) are zero for the gaussian distribution and thus these are ideal probe of the Non-Gaussianity.  For correlation function $c_n$ ($n>1$), they are zero for Poisson distributions, thus can be used to measure the deviation from Poisson fluctuations.  If we define the correlation strength parameter $\hat{c}_{k}$ as:
\begin{equation}
\hat{c}_k=\frac{{c}_k}{\la N \ra^k}
\end{equation}
where $k=2,3,4...,n$. The different order correlation strength parameter $\hat{c}_k$ can reflect different physics process of the system. If the system consists of many independent sources, 
the correlation strength will be diluted and it scales with the multiplicity as :
\begin{equation}
\hat{c}_k \propto \frac{1}{\la N \ra^{k-1}}
\end{equation}
For eg., it is the case that the A+A system is superposed by many p+p collisions. However, if the particle sources are strongly correlated with each other,  which is the case near the critical point, then we have:
\begin{equation}
\hat{c}_k \propto const
\end{equation}
The long range correlation become dominated near the critical point and the cumulant are dominated by the highest order correlation function as: 
$C_k \approx c_k \propto \la N\ra^k$. If the thermal statistical fluctuations dominated in the system, we have $C_k \approx c_k \propto \la N\ra$. 
Thus, to search for the critical point in heavy-ion collisions, it is also important to study the centrality, energy and the rapidity dependence of the multi-particle correlation functions. 
This is effective way to look for the pattern of long range correlations near the critical point in heavy-ion collisions and it is also very useful to study the contributions of the non-critical 
backgrounds, such as the baryon number conservations, resonance decay, hadronic scattering. 

Figure~\ref{fig:corr_func} shows the rapidity dependence of the proton cumulants and correlation functions in Au+Au collisions at {\sNN}=5 GeV from JAM transport model calculation~\cite{JAM_model}. It can be found that the second and third order
correlation functions show negative values when enlarging the rapidity acceptance. The large negative values of the second and third order proton correlation functions also leads to strong suppression of the fourth order proton cumulant. Those observations can be understood in terms of the baryon number conservations. As one can see in Fig. \ref{fig:corr_func}, it seems that the fourth order proton correlation function is consistent with zero, which is due to the absence of long range correlations in this model calculations. It means the baryon number conservations, which is a large background effect for searching for the critical point with fluctuations of conserved quantities in heavy-ion collisions, has negligible effects on the fourth order proton correlation functions.  In other words, due to insensitive to the baryon number conservations, the fourth order proton correlation function is an ideal probe of the long range correlations induced by the critical point. 

\section{Data Analysis Methods} \label{sec:DataAnalysis}
In the data analysis,  we applied a series of analysis techniques to suppress backgrounds and make precise measurements of the event-by-event fluctuation analysis in heavy-ion collisions.  Those include : (1) Centrality bin width correction~\cite{WWND2011,technique}. This is to remove centrality bin width effect, which is caused by volume variation within a finite centrality bin size.  (2) Carefully define the collision centrality to suppress volume fluctuations and auto-correlations~\cite{technique}.  (3) Efficiency correction for the cumulants. (4) Estimate the statistical error with Delta theorem and/or Bootstrap methods~\cite{Delta_theory,voker_eff1,voker_eff2,Unified_Errors}. Those techniques are very crucial to precisely measure the dynamical fluctuation signals from heavy-ion collisions.   Let's discuss those techniques one by one. 

\begin{figure*}[htbp]
\begin{center}
\hspace{-1cm}
\includegraphics[width=0.7\textwidth]{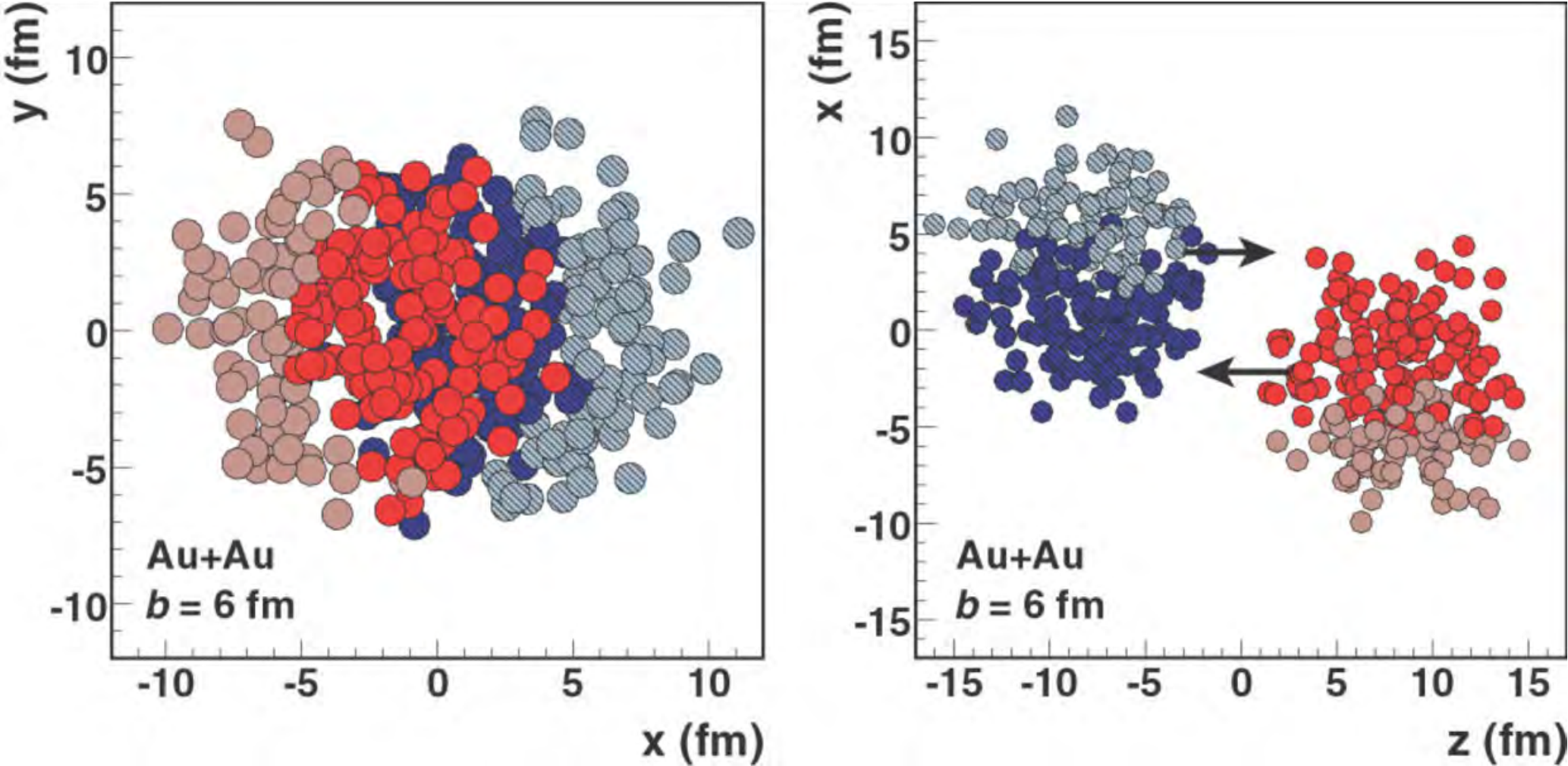}
\caption{ Illustrationof a Glauber Monte Carlo event for Au+Au at $\sqrt{s_\mathrm{NN}} =
200$~GeV with impact parameter $b = 6$~fm in the transverse plane
(left panel) and along the beam axis (right panel)~\cite{miller2007glauber}. The nucleons are
drawn with a radius $\sqrt{\sigma^\mathrm{NN}_\mathrm{inel}/\pi}/2$.
Darker disks represent participating nucleons.} \label{fig:glauber_mc_event} 
\end{center}
\end{figure*}

\begin{figure}[tbp]
\begin{center}
\hspace{-1.cm}
\includegraphics[width=85mm]{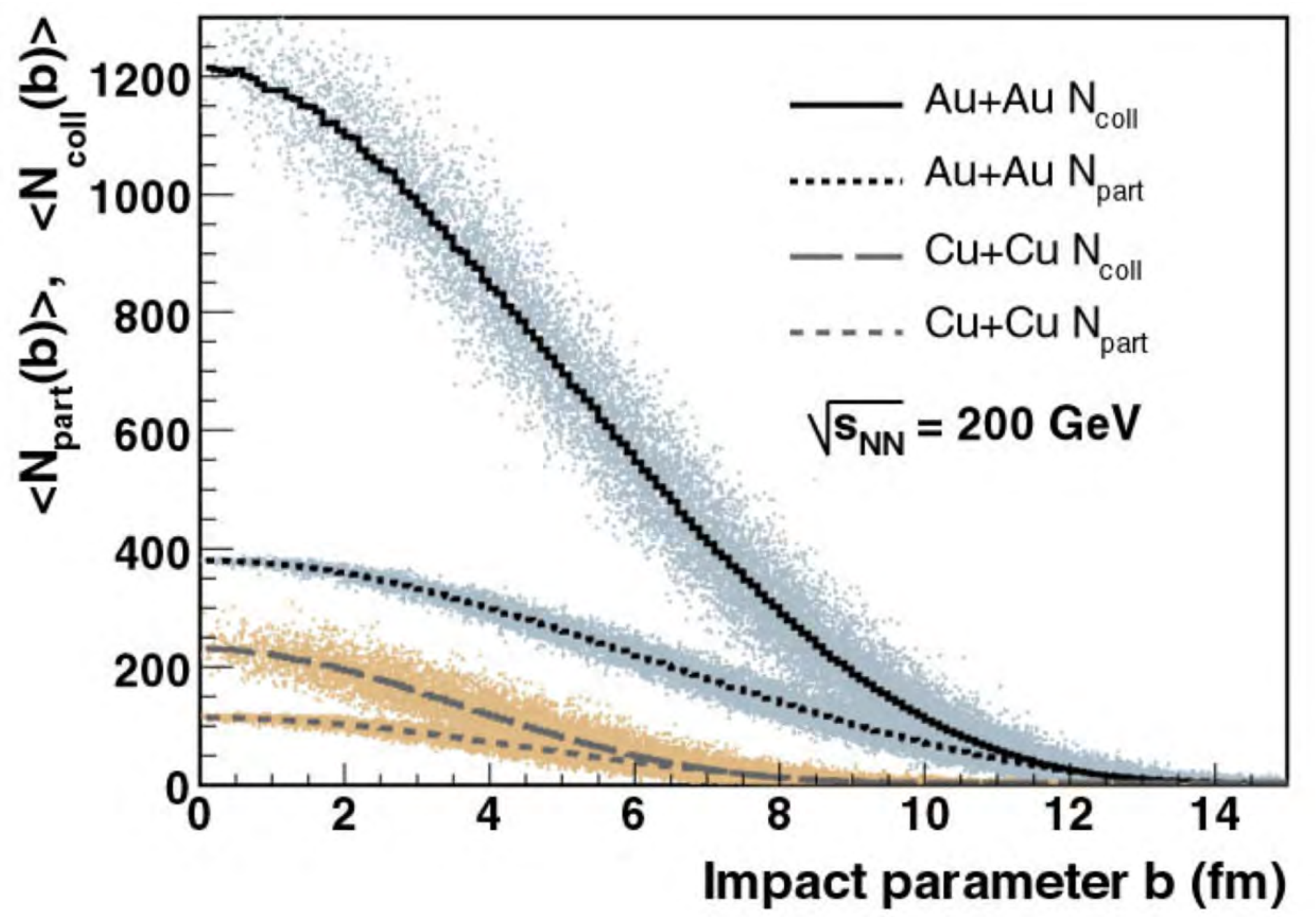}
\caption{ Average number of participants ($\langle N_\mathrm{part}
  \rangle$) and binary nucleon-nucleon collisions ($\langle
  N_\mathrm{coll} \rangle$) along with event-by-event fluctuation of
  these quantities in the Glauber Monte Carlo calculation as a
  function of the impact parameter $b$~\cite{miller2007glauber}. \label{fig:npart_ncoll_vs_b} }
\end{center}
\end{figure}

\begin{figure}[tbp]
\centering\mbox{
\includegraphics[width=0.4\textwidth]{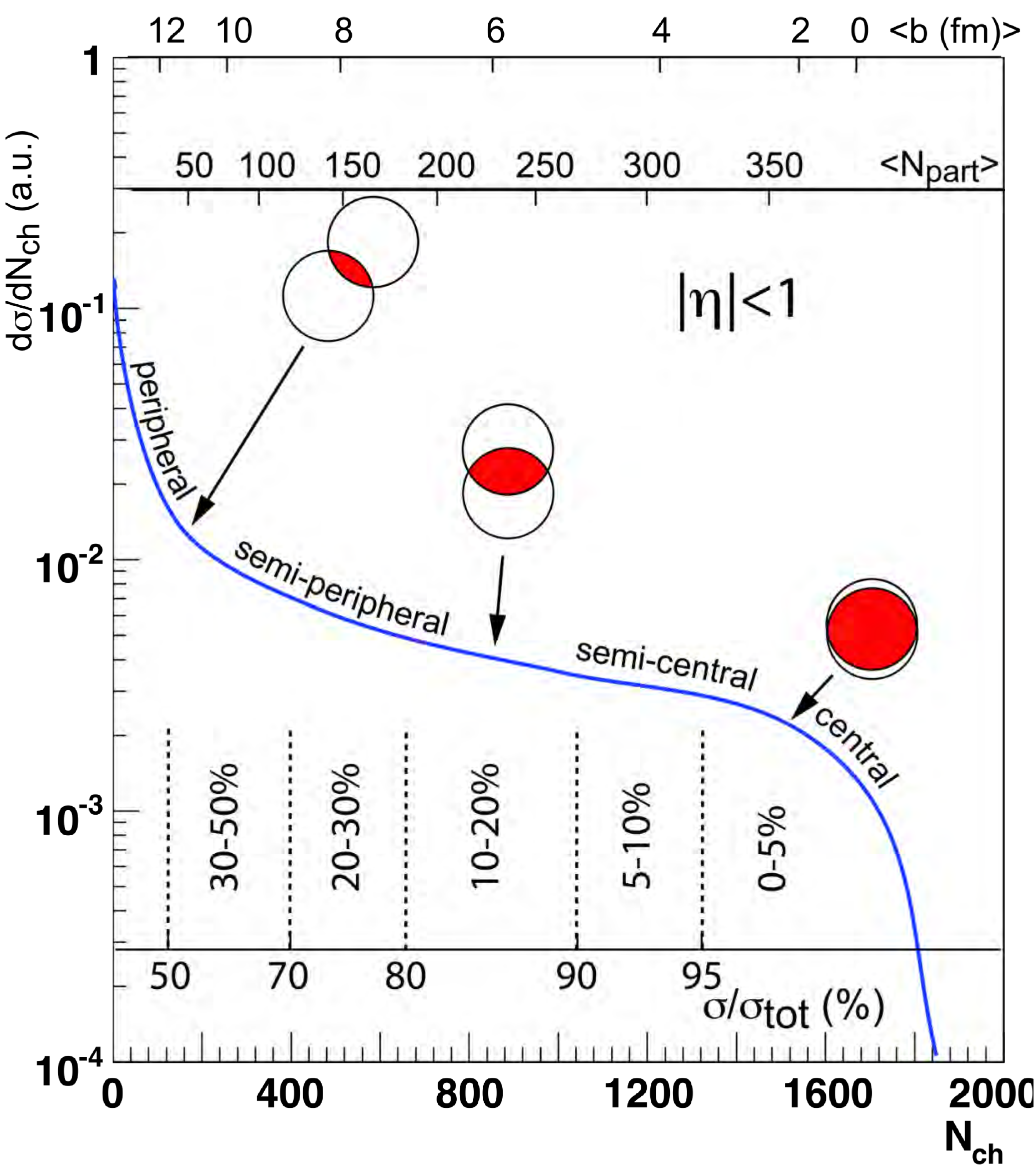}}
\caption{An
illustrated example of the correlation of the total inclusive
charged-particle multiplicity $N_{ch}$ with Glauber-calculated
quantities($b, N_{part}$)~\cite{miller2007glauber}. The plotted distribution and various
values are illustrative and not actual measurements. . } \label{fig:Centrality} 
\end{figure}

\begin{figure*}[hptb]
\centering
\renewcommand{\figurename}{FIG.}
\hspace{-0.4cm}
\includegraphics[width=0.46\textwidth]{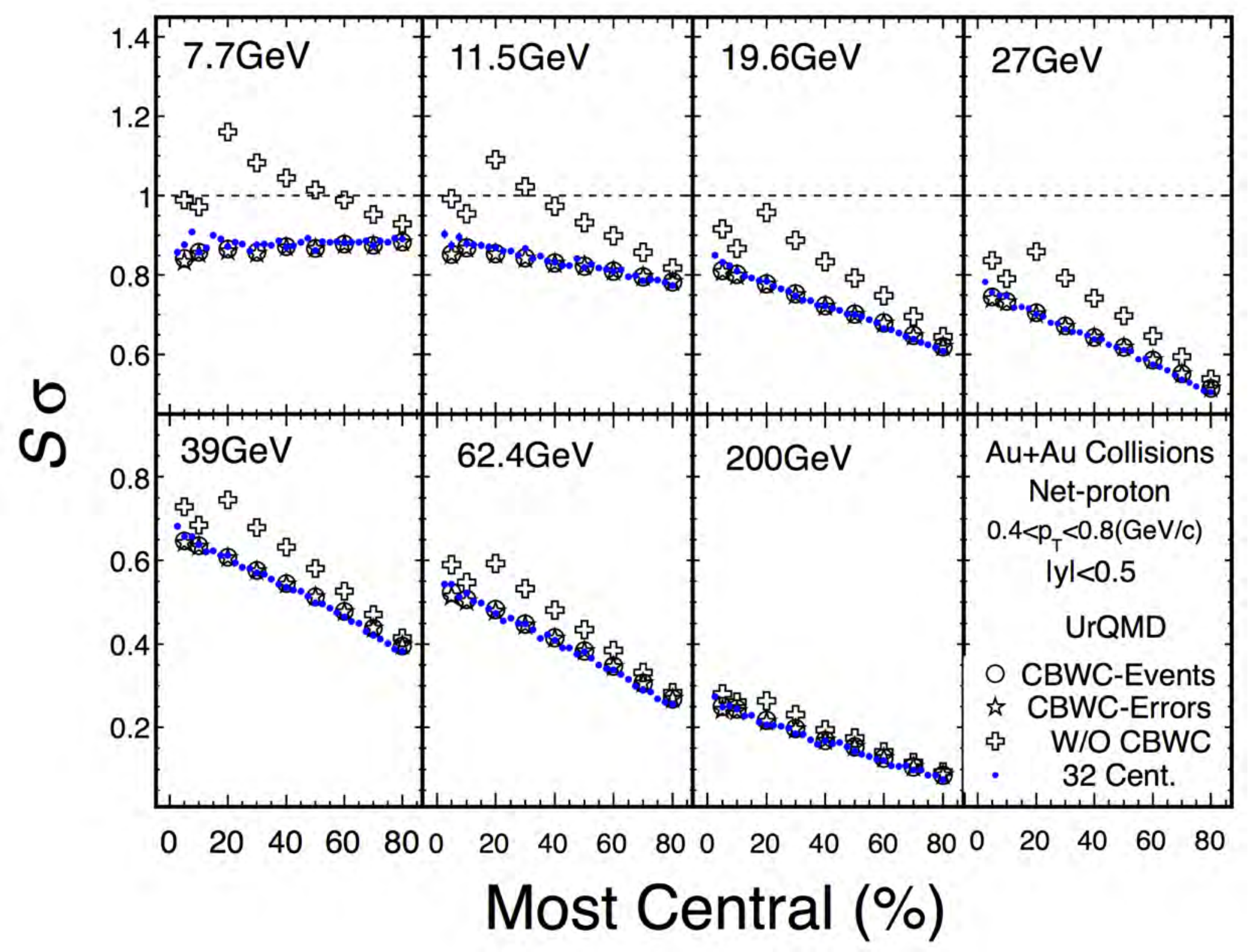}
\hspace{0.2cm}
\includegraphics[width=0.45\textwidth]{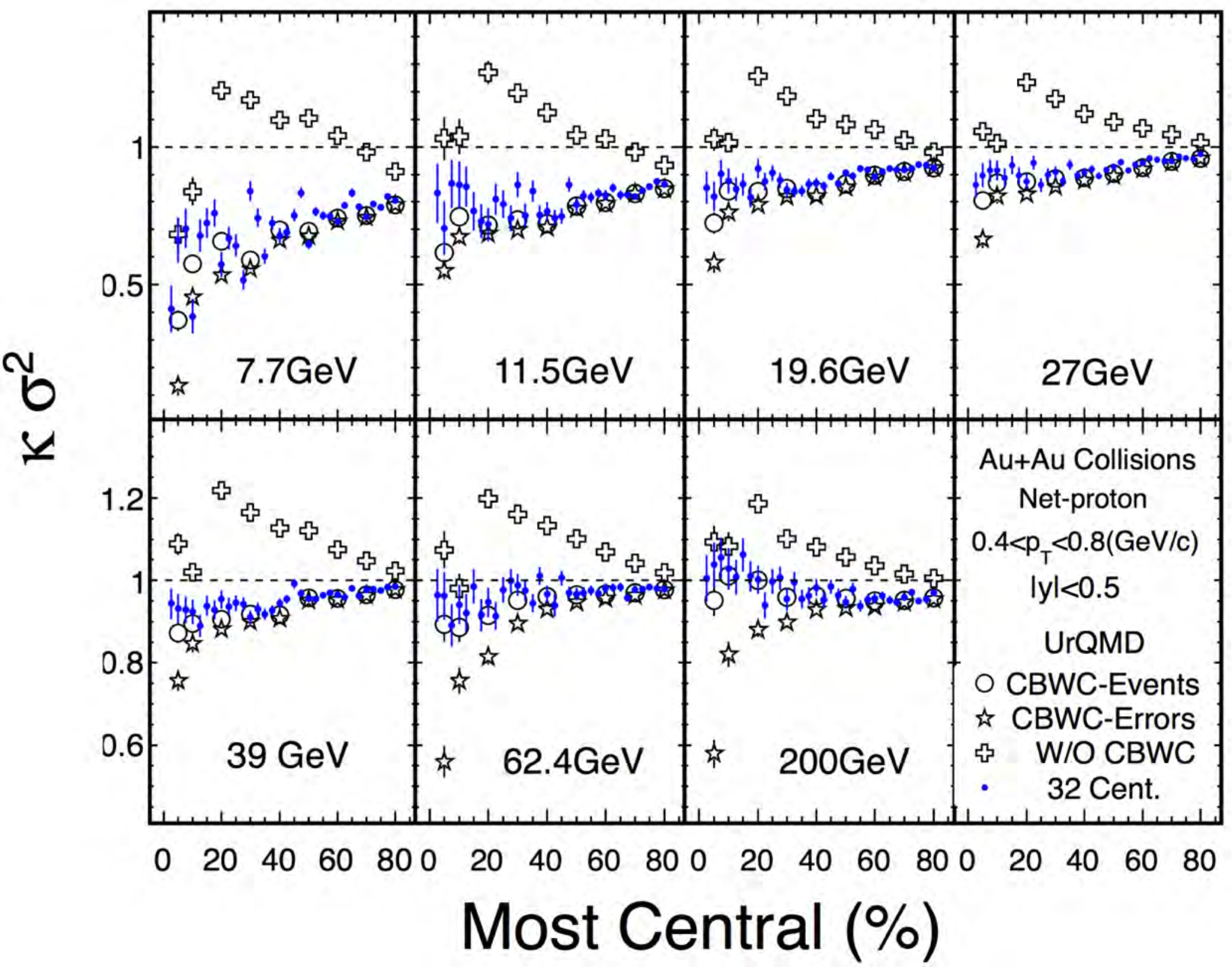}
\caption{(Color online) The centrality dependence of the moments
products $S\sigma$ (Left) and $\kappa\sigma^2$  (Right)of net-proton multiplicity distributions for Au+Au collisions at \sNN=7.7, 11.5, 19.6, 27, 39, 62.4, 200GeV in UrQMD model~\cite{technique}. The solid dots represent the results calculated from 32 centrality bins.}
\label{Plot::CBWC}
\end{figure*}

\subsection{Collision Geometry and Centrality Definition} \label{sec:Centrality}
Before introducing the background suppression methods, we would like to firstly discuss about the centrality definition used in the heavy-ion collisions. 
The definition of the collision centralities for two colliding nuclei is not unique and can be defined by different quantities. A commonly used quantity is the so called impact parameter $b$, defined as the distance between the geometrical centers of the colliding nuclei in the plane transverse to their direction. Other quantities, such as the number of participant nucleons, $N_{part}$ and the number of binary collisions, $N_{coll}$, can be also used. Fig. \ref{fig:glauber_mc_event} shows a Glauber Monte Carlo event of Au+Au collision at \sNN\ = 200 GeV with impact parameter $b=6$ fm~\cite{miller2007glauber}. The blue and red solid circles represent the participant nucleons from the two colliding gold nuclei. Fig. \ref{fig:npart_ncoll_vs_b} shows the average number of participant nucleons ($\la N_{part} \ra$ ) and average number of binary collisions ($\la N_{coll}\ra$) as a function of impact parameter $b$. One can see that there is no one-to-one correspond between $N_{part}$, $N_{coll}$ and impact parameter $b$. Unfortunately, all of those geometrical variables can't be directly measured in the heavy-ion collision experiment. Since the particle multiplicity can be easily measured and also can reflect the initial geometry of heavy-ion collision. The centrality in heavy-ion collisions is usually determined by a comparison between experimental measured particle multiplicity and Glauber Monte-Carlo simulations~~\cite{miller2007glauber}. It is denoted as a percentage value (for e.g. 0-5\%, 5-10\%,...) for a collection of events to represent the fraction of the total cross section. Fig. \ref{fig:Centrality} illustrates how to define a  collision centrality in heavy-ion collisions by comparing the particle multiplicities with Glauber Monte Carlo simulation
and the correlation between the particle multiplicities and the Glauber calculated quantities $b$ and $N_{part}$. However, the relation between measured particle multiplicities and collision geometry is not one-to-one correspondence and there are fluctuations in the particle multiplicity even for a fixed collision geometry.

\subsection{Centrality Bin Width Correction}  \label{sec:CBWC}
The centrality bin width effect is caused by the volume variation within a wide centrality bin and will cause an artificial centrality dependence for the fluctuation observables~\cite{WWND2011,technique}. 
The centrality bin width correction (CBWC) is to suppress the volume fluctuations effects in the event-by-event fluctuation analysis within finite centrality bin width.
Experimentally, measurements are usually reported for a wide centrality bin (a range of particle multiplicity),
such as $0-5\%$,$5-10\%$,...etc., to reduce statistical errors. We know that the smallest centrality bin is determined by a single value of particle multiplicity.
To suppress the centrality bin width effect in a wide centrality bin, we calculate the cumulants ($C_{n}$) for each single particle multiplicity bin ($N_{ch}$).  Then, the results reported for this wide centrality bin ($N_{ch}$) is to take the weighted average. The weight is the corresponding number of events in the particle multiplicity bin divided by the total events of the wide centrality bin. The method can be expressed as:
\begin{equation} \label{eq:cbwc}
{C_n} = \frac{{\sum\limits_{r = {N_1}}^{{N_2}} {{n_r}C_n^r} }}{{\sum\limits_{r = {N_1}}^{{N_2}} {{n_r}} }} = \sum\limits_{r = {N_1}}^{{N_2}} {{\omega _r}C_n^r} 
\end{equation}
where the $n_r$ is the number of events for multiplicity bin $r$ and
the corresponding weight for the multiplicity $r$, ${\omega _r} = {n_r}/\sum\limits_{r = {N_1}}^{{N_2}} {{n_r}}$. $N_1$ and $N_2$ are the lowest and highest multiplicity values for one centrality bin. Once having the centrality bin width corrected cumulants via Eq.(\ref{eq:cbwc}), we can calculate the various order cumulant ratios, for e.g. {\KV}$=C_{4}/C_{2}$ and {\SD}$=C_{3}/C_{2}$, where the $\kappa$ and $S$ are kurtosis and skewness, respectively. The final statistical error of cumulants and cumulant ratios for wide centrality bin can be calculated by standard error propagation. 

To demonstrate the centrality bin width effect and test the method of centrality bin width correction, we have calculated the cumulants of net-proton distributions in Au+Au collisions from UrQMD model in different ways. Figure~\ref{Plot::CBWC} show the centrality dependence of the cumulant ratios $(S\sigma , \kappa\sigma^2)$ of net-proton multiplicity distributions for Au+Au collisions at \sNN=7.7, 11.5, 19.6, 27, 39, 62.4 and 200 GeV from the UrQMD model calculations. The open circle and open cross in Fig.~\ref{Plot::CBWC} represent the results obtained with and without applying the CBWC in the nine centralities (0-5\%, 5-10\%, 10-20\%, 20-30\%...70-80\%), respectively. For the nine centralities, we clearly observe that the results
with CBWC are very different from those without CBWC. This indicates the volume fluctuations in one wide centrality bin do have a significant impacts on the value of cumulants and the CBWC will make the values of the cumulants systematically lower by reducing the effects of volume fluctuations in one wide centrality bin. The solid circles show the centrality dependence for 32 finer centrality bins (0-2.5\%, 2.5-5\%, 5-7.5\%...77.5\%-80\%) without CBWC. In the case of 32 centrality bin, due to the finer bin width, the centrality bin width effects are expected to be very small. Interestingly, we found that the results calculated from 32 centrality bins show good agreement with the results from nine centralities with CBWC. This further confirms the effectiveness of the CBWC method described above.  On the other hand, we also tried to use the statistical errors ($error$) as weight to perform the CBWC by replacing the weight factor $n_{r}$ in Eq. (\ref{eq:cbwc}) with $1/error^{2}$ for each single multiplicity bin. The statistical error can be obtained by the Delta theorem and/or bootstrap methods at each multiplicity bin. It is found that the \SD\ with CBWC using the statistical error as weight are consistent with the results with events number weighted CBWC, but not for $\kappa\sigma^2$. It means the error weighted method can not be used for CBWC, which may be due to the statistical error is not only related to the number of events but also the cumulants itself.

\subsection{Volume Fluctuations Effects}   \label{sec:volume}
Volume fluctuations are long standing notorious background for the event-by-event fluctuation analysis in heavy-ion collisions~\cite{technique,Haojie_vol,PBM_vol,Skokov_vol,haojie_plb,bzdak2016correlated}. This is originated from that one cannot directly measure the collision centrality and/or initial collision geometry of the system of two nuclei. It is difficult to completely eliminated as it is usually convoluted with the real fluctuation signals. Consequently, this will lead to undesirable volume fluctuations in the event-by-event fluctuation analysis of particle multiplicity in heavy-ion collisions. The volume fluctuations will enhance the values of cumulants of the the event-by-event multiplicity distributions. However, the model calculations in the paper~\cite{bzdak2016correlated} conclude that the effects of volume fluctuations is too small to explain the large increase found in the preliminary result of 0-5\% most central net-proton {\KV} in Au+Au collisions at {\sNN}=7.7 GeV measured by the STAR experiment.
\begin{figure}[hptb]
\centering
\renewcommand{\figurename}{FIG.}
\hspace{-0.5cm}
\includegraphics[width=0.5\textwidth]{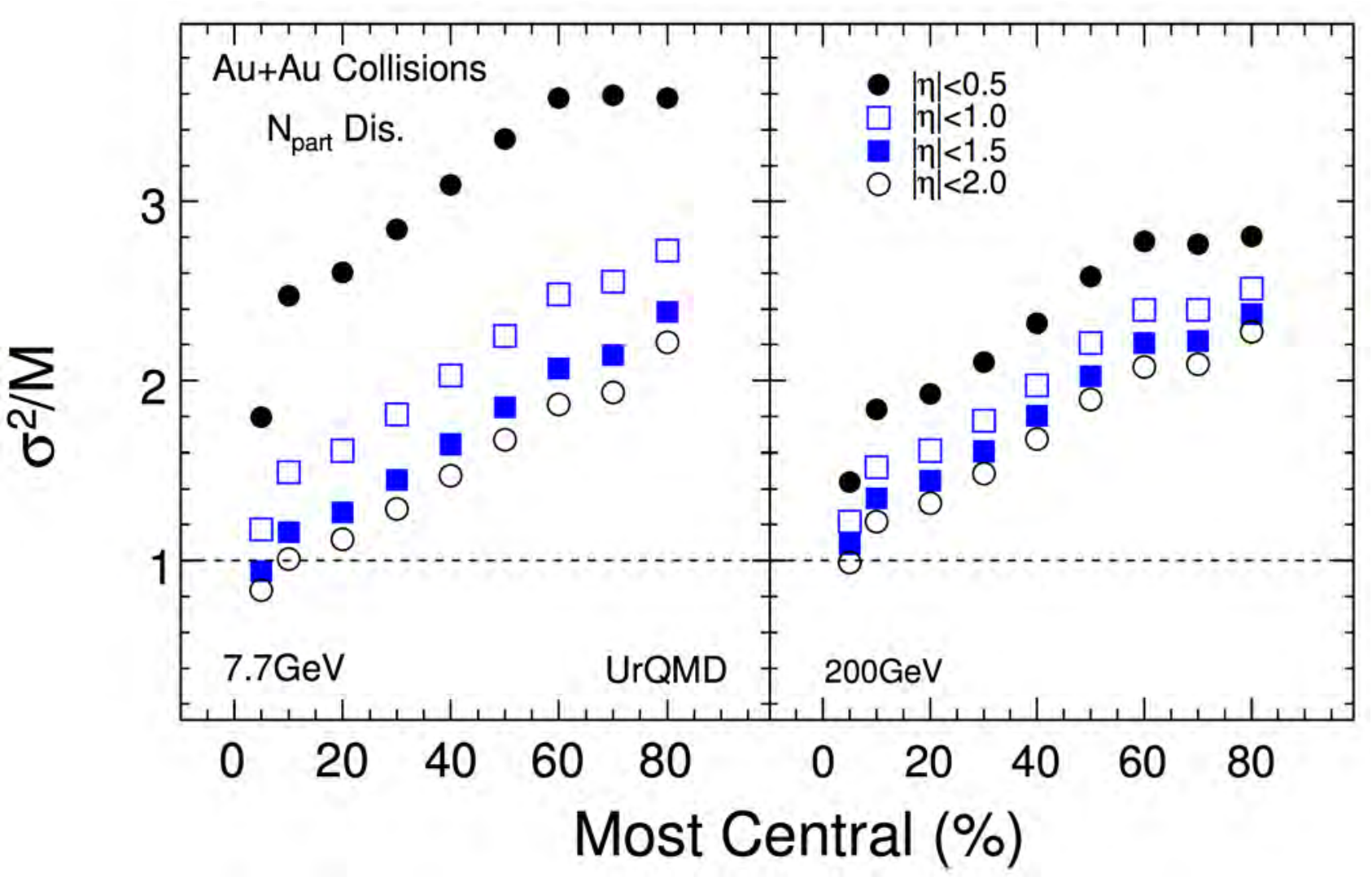}
\caption{(Color online) The centrality dependence of the $\sigma^2 / M$ of $N_{part}$ distributions for Au+Au collisions at \sNN=7.7 and 200 GeV in UrQMD model~\cite{technique}. Four different centrality definitions are corresponding to the charged particles with different $\eta$ coverage ($|\eta|<0.5,1.0,1.5,2.0$). }
\label{Plot::R21}
\end{figure}
\begin{figure}[hptb]
\centering
\hspace{-0.5cm}
\includegraphics[width=0.5\textwidth]{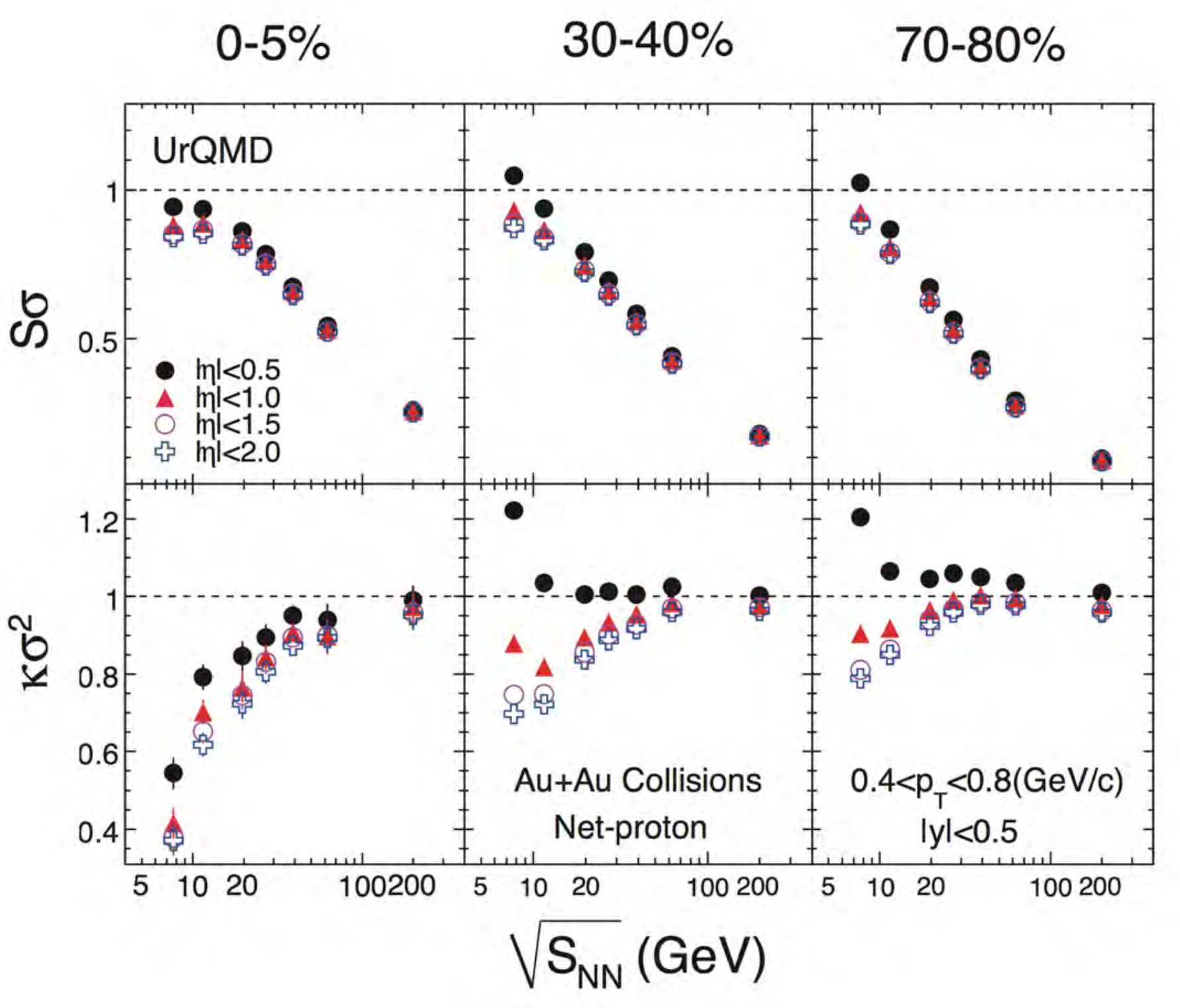}
\caption{(Color online) The energy dependence of the moments products $(S\sigma , \kappa\sigma^2)$ of net-proton multiplicity distributions for Au+Au collisions at \sNN=7.7, 11.5, 19.6, 27, 39, 62.4, 200 GeV in UrQMD model~\cite{technique}. Four different centrality definitions are corresponding to the charged particles with different $\eta$ coverage ($|\eta|<0.5,1.0,1.5,2.0$). }
\label{Plot::energy}
\end{figure}
\begin{figure*}[hptb]
\centering
\hspace{-0.5cm}
\includegraphics[width=0.45\textwidth]{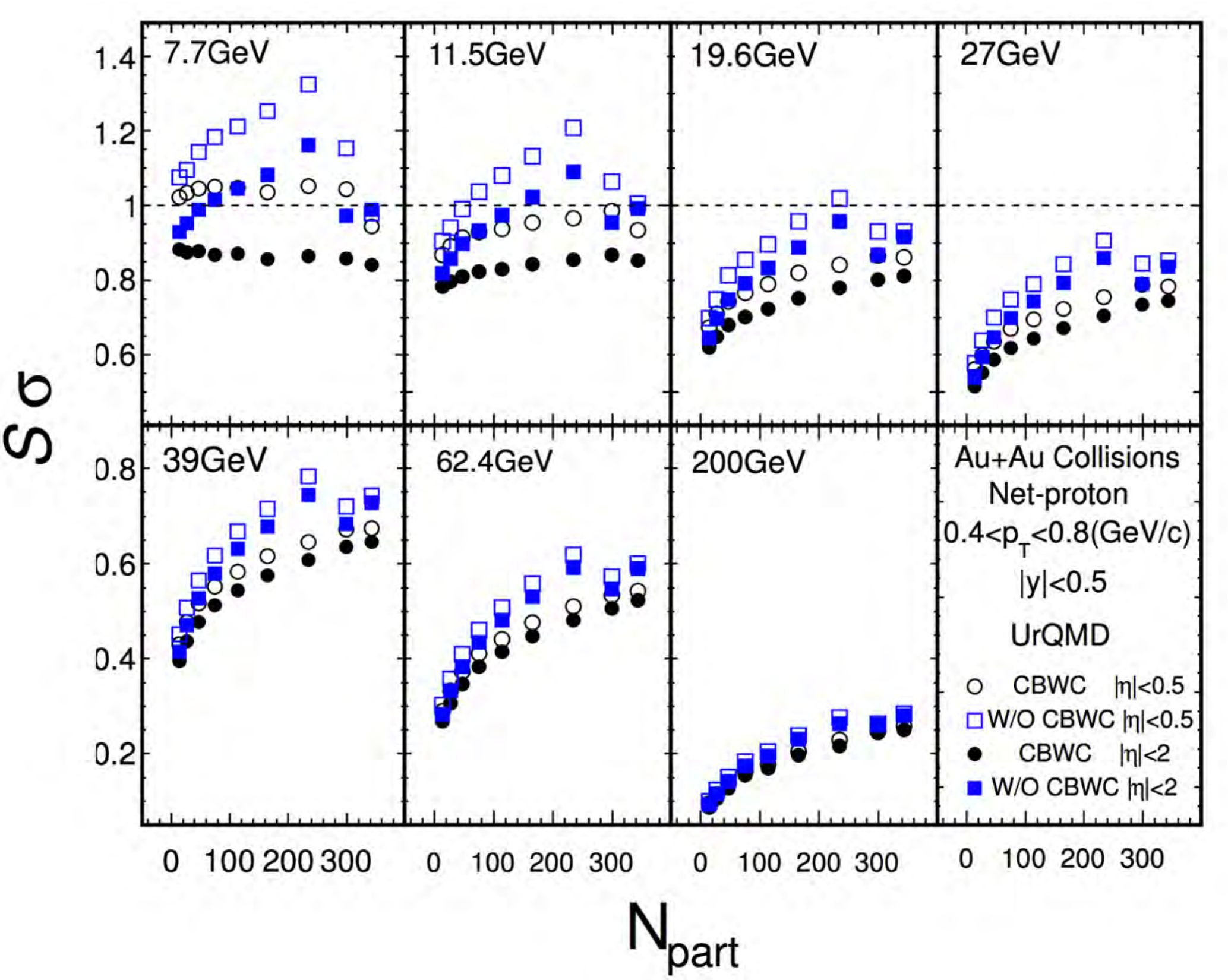}
\hspace{0.3cm}
\includegraphics[width=0.46\textwidth]{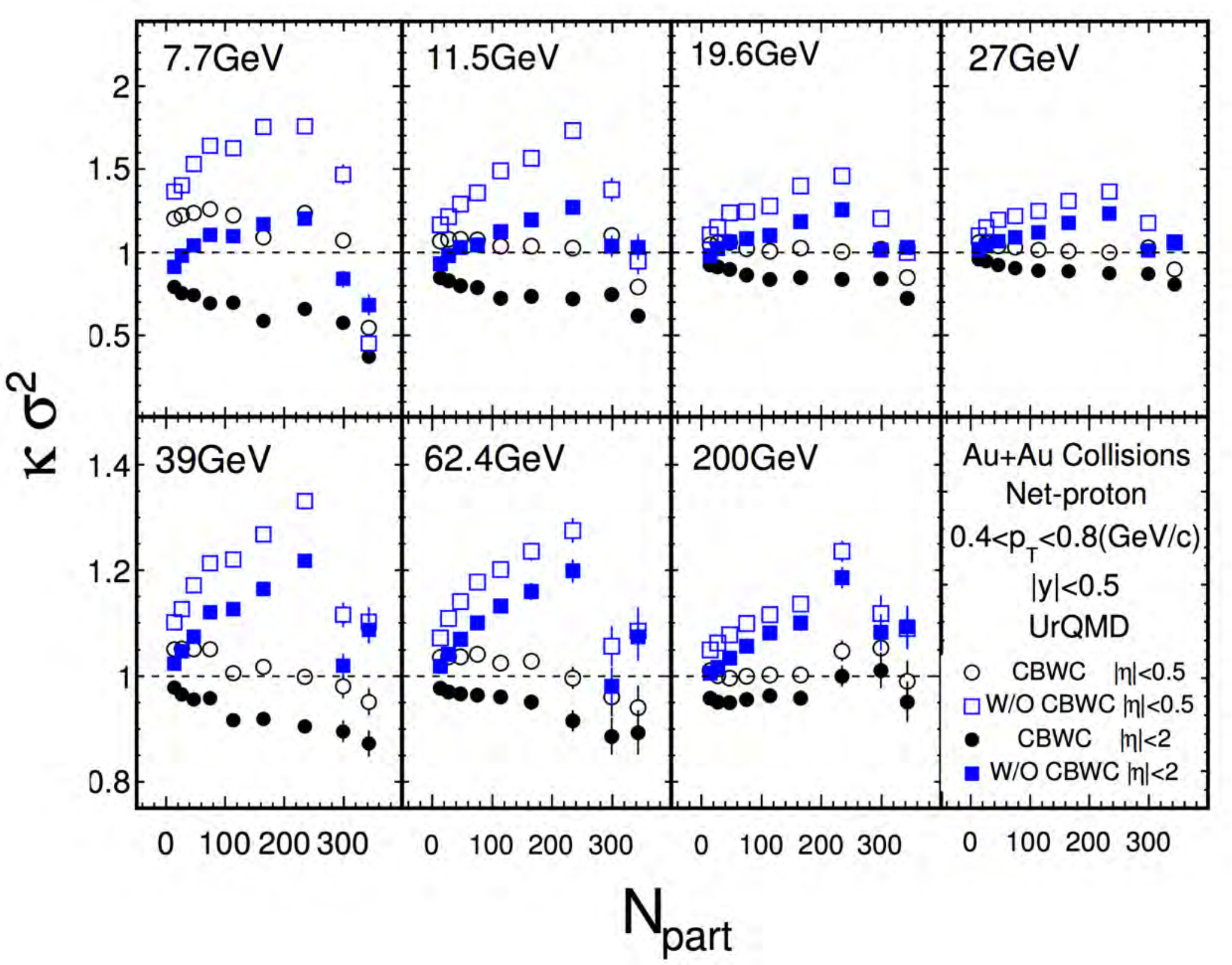}
\caption{(Color online) The centrality dependence of the moments
products $S\sigma$ (Left) and $\kappa\sigma^2$  (Right)of net-proton multiplicity distributions for Au+Au collisions at \sNN=7.7, 11.5, 19.6, 27, 39, 62.4, 200GeV in UrQMD model~\cite{technique}. The open circles and squares are the results with CBWC and without CBWC at $|\eta| < 0.5$ for the centrality definition, respectively. The solid circles and squares are the results with at $|\eta| < 2$ in the centrality definition, respectively. }
\label{fig:CBWC_vol}
\end{figure*}

\begin{figure*}[hptb]
\centering
\hspace{-0.5cm}
\includegraphics[width=0.46\textwidth]{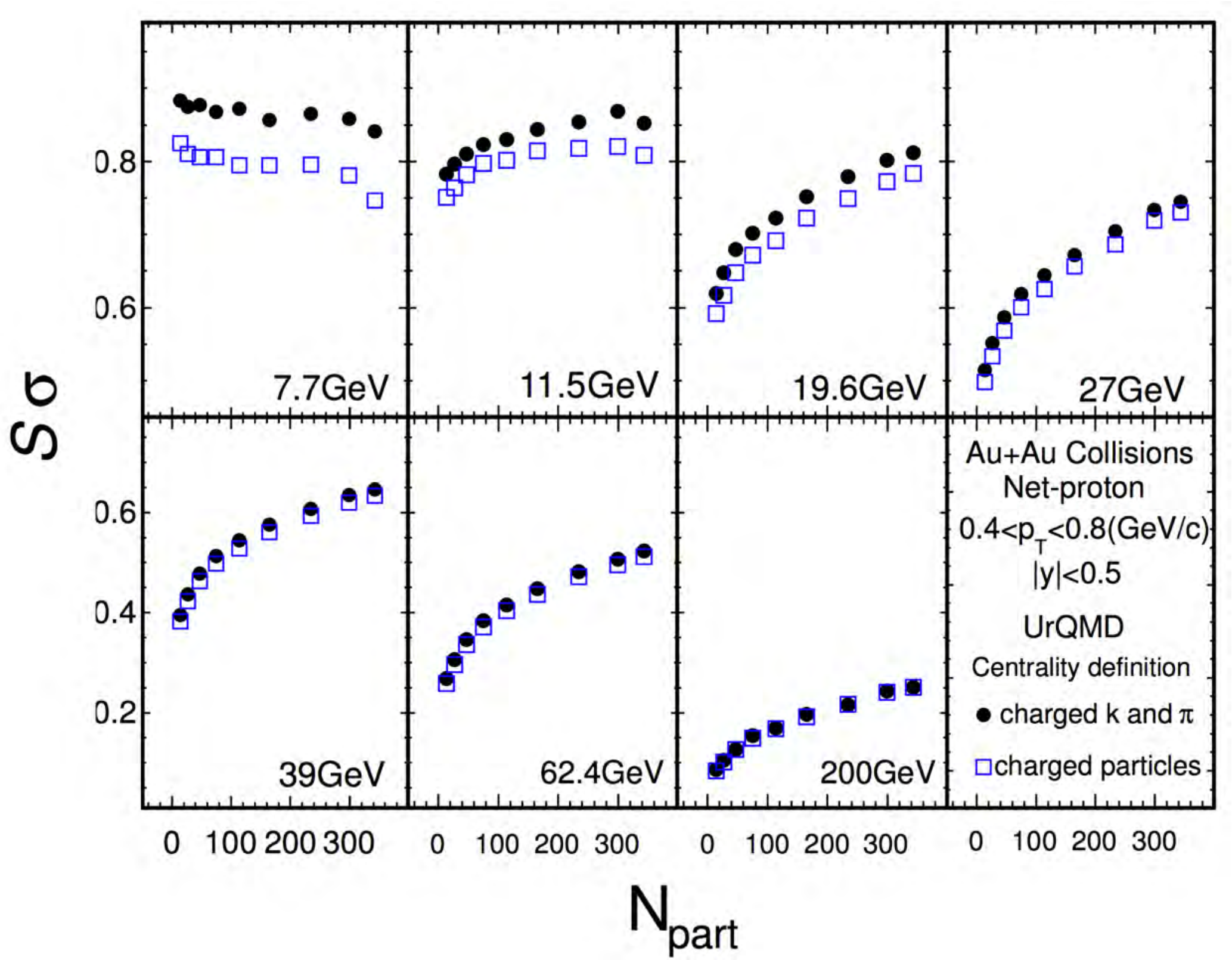}
\hspace{0.3cm}
\includegraphics[width=0.46\textwidth]{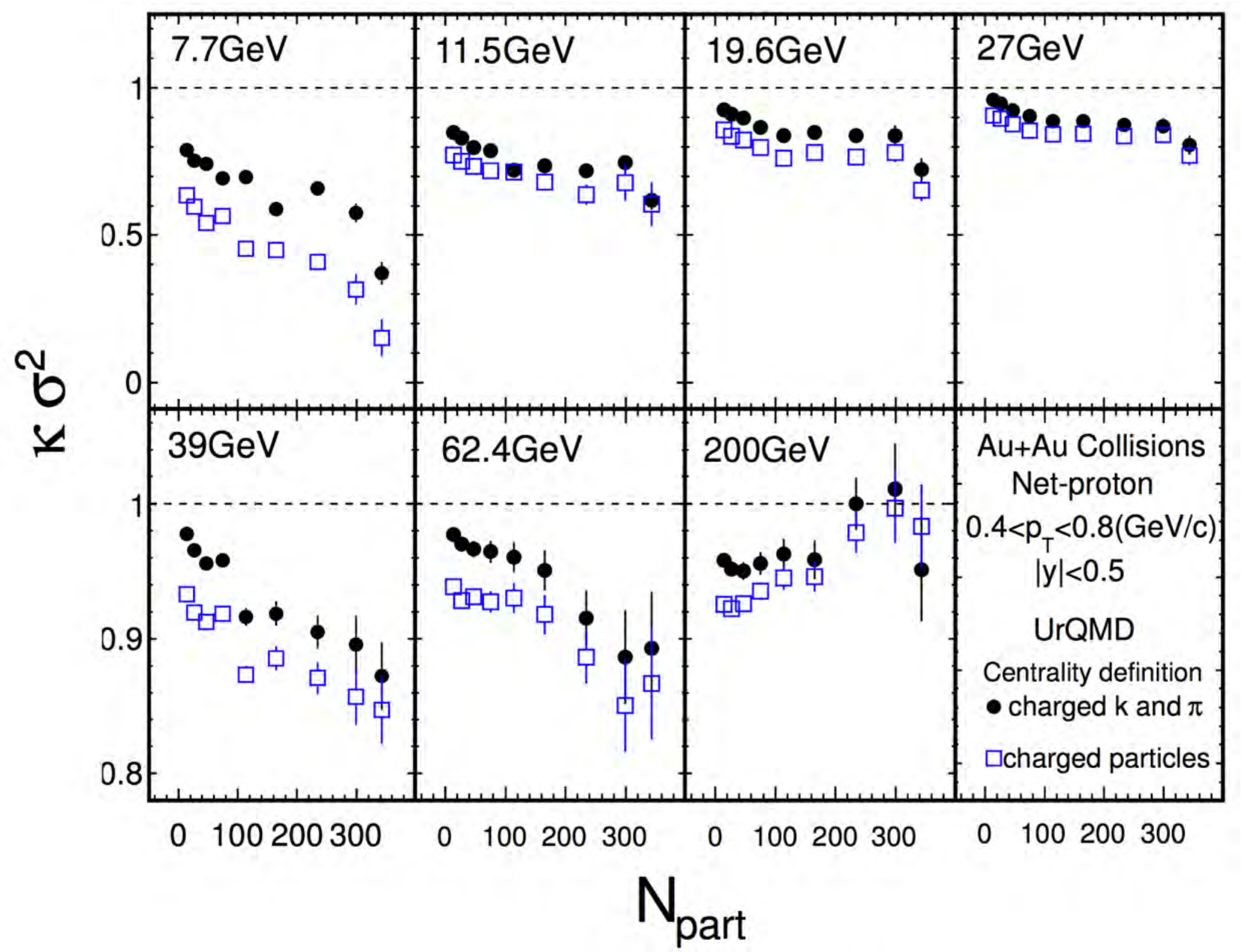}
\caption{(Color online) The centrality dependence of the moments
products $S\sigma$ (Left) and $\kappa\sigma^2$  (Right)of net-proton multiplicity distributions for Au+Au collisions at \sNN=7.7, 11.5, 19.6, 27, 39, 62.4, 200GeV in UrQMD model with the two different centrality definitions~\cite{technique}. }
\label{fig:CBWC_auto}
\end{figure*}

In the following, we will demonstrate the volume fluctuations in net-proton multiplicity fluctuations from Au+Au collisions by using UrQMD model simulations and discuss the method to suppress the volume fluctuations. To avoid auto-correlation, the centrality are defined with charged particle multiplicities by excluding the protons and anti-protons used in the analysis. 
The relation between measured particle multiplicity and impact parameter is not one-to-one correspond and there are fluctuations in the particle multiplicity even for a fixed impact parameter. Thus, we could obtain a finite resolution of initial collision geometry by using particle multiplicity to determine the centrality. As the $N_{part}$ can reflect the initial geometry (volume) of the colliding nuclei, the $\sigma^2/M$ of $N_{part}$  distributions can be regarded as the centrality resolution for a certain centrality definition. Fig.~\ref{Plot::R21} shows the centrality dependence of $\sigma^2/M$ of number of participant nucleons ($N_{part}$) distributions for Au+Au collisions at \sNN =7.7 and 200 GeV in UrQMD calculations with four different centrality definitions.  The different centrality definitions are corresponding to the charged particles with four different $\eta$ coverage ($|\eta|<0.5,1.0,1.5,2.0$). It shows that when we define the centrality with $|\eta|<2$, the $\sigma^2/M$ of $N_{part}$ distributions for 7.7 and 200 GeV are similar.  We can find that more particles are used in the centrality determination, the better centrality resolution and smaller fluctuation of the initial geometry (volume fluctuation) we obtain. Fig.~\ref{Plot::energy} shows the energy dependence of moment product ($S\sigma, \kappa\sigma^2$) of net-proton multiplicity distributions in Au+Au collisions from UrQMD calculations for three different centralities (0-5\%, 30-40\%, 70-80\%) with four different $\eta$ ranges for centrality definitions. The centrality bin width corrections have been applied for all of the cases. Different centrality definitions could cause different results due to changing of magnitude of the volume fluctuations.  By extending the $\eta$ coverage of the charged particles used in centrality definition, we find the volume fluctuations are strongly suppressed. The $\kappa\sigma^2$ (fourth order fluctuation) is more sensitive to the volume fluctuations than the {\SD} (third order fluctuation). On the other hand, the volume fluctuations have much smaller effects in the most central collisions (0-5\%) than in peripheral and mid-central collisions.  With those studies, we conclude that having more particles in the centrality definition is an effective way to improve the centrality resolution and suppress the effects of volume fluctuations on the event-by-event fluctuations observables in heavy-ion collisions. 

In principle, both the centrality bin width effects and centrality resolution effects are originated from volume fluctuations. The former is the volume variation within one wide centrality bin, and the latter is due to the initial volume fluctuations. These are two different effects and should be treated separately. The centrality bin width effects not only depend on the bin size but also dependent on the centrality resolutions (or the way to define the centrality). In this sense, these two effects are related with each other and both depend on the centrality definition. Fig.\ref{fig:CBWC_vol} shows the {\SD} and {\KV} of net-proton distributions in Au+Au collisions from UrQMD model calculations with centrality definitions at different $\eta$ coverage ($|\eta|<$ 0.5 and 2).  The larger $\eta$ coverage means more particles are included in the centrality definition and with better centrality resolution. Indeed, it shows that the results from wider $\eta$ coverage centrality definition are get suppressed comparing with the results from narrower $\eta$ coverage centrality definition. At fixed centrality definition,  the results with CBWC are always smaller than the results without CBWC.

\subsection{Auto-correlation Effects}   \label{sec:autocorrelation}
The auto-correlation effect is a background effect in the fluctuation analysis and will suppress the magnitude of the signals. For eg., in net-proton fluctuation analysis, to avoid the auto-correlation, we should exclude the corresponding protons and anti-protons from the centrality definition. For net-kaon fluctuations, we need to exclude $K^{+}$ and $K^{-}$ in the centrality definition. To illustrate this effects, we calculate the net-proton fluctuations in Au+Au collisions from UrQMD model with two different centrality definitions. One is using all charged particles and the other use the multiplicity of only charged kaon and pion to define the collision centrality. Fig.~\ref{fig:CBWC_auto} shows that for {\SD}  and {\KV} of net-proton distributions, the results with auto-correlation are smaller than the ones without auto-correlation. Meanwhile, the auto-correlation effects are stronger at lower energies. This is because the overlap fraction of proton/anti-protons used in the fluctuation analysis and in the centrality definition increase when decreasing the energies. In the data analysis, to avoid auto-correlation, we have to exclude the particles used in the fluctuation analysis from the centrality definition.

\subsection{Efficiency Correction for Cumulants}   \label{sec:EfficiencyCorrection}
The detector always have a finite particle detection efficiency. The observed event-by-event particle multiplicity distributions are the convolution between the original distributions and the efficiency response function. We need to correct this efficiency effect and a deconvolution operation is needed to recover the true fluctuations signals.  However,  it is not straightforward to get the efficiency corrected results for the cumulants of particle multiplicity distributions, especially for the higher order fluctuations. 

It is well know that the detection efficiency response function is binomial distribution for a detector with good performance. Based on binomial efficiency response function, there has many discussions about the efficiency correction methods for moment analysis~\cite{voker_eff1,voker_eff2}.  Here, we provide a unified description of efficiency correction and error estimation for cumulants of multiplicity distributions~\cite{Unified_Errors}. The principle idea is to express the moments and cumulants in terms of the factorial moments, which can be easily corrected for efficiency effect. By knowing the covariance between factorial moments, we use the standard error propagation based on the Delta theorem in statistics to derive the error formulas for efficiency corrected cumulants. More important, this method can be also applied to the phase space dependent efficiency case, where the efficiency of proton or anti-proton are not constant within studied phase space. One needs to note that the efficiency correction and error estimation should be done for each single particle multiplicity bin in each centrality and just before the centrality bin width correction. 

In the STAR experiment, the particle detection efficiency can be obtained from the so called Monte Carlo (MC) embedding techniques~\cite{spectra_STAR}.  The Monte Carlo tracks are blended into real events at the raw data level. The tracks are propagated through the full simulation chain of the detector geometry with a realistic simulation of the detector response. The efficiency can be obtained by the ratio of matched MC tracks to input MC tracks. It contains the net effects of tracking efficiency, detector acceptance, decays, and interaction losses. 
For illustration purpose, we discuss the application of the efficiency correction on the net-proton fluctuation analysis in heavy-ion collisions. Experimentally, we measure net-proton number event-by-event wise, $n=n_{p}-n_{\bar{p}}$, which is proton number minus anti-proton number.  The average value over the whole event ensemble is denoted by $\la n \ra$, where the single angle brackets are used to indicate ensemble average of an event-by-event distributions. For simplify, let us discuss constant efficiency case for (anti-)proton within the entire phase space. The probability distribution function of 
measured proton number $n_{p}$ and anti-proton number $n_{\bar{p}}$ can be expressed as~\cite{voker_eff1}:
\begin{widetext}
\begin{equation}  \label{eq:conv} 
\begin{split}
 p({n_p},{n_{\bar p}}) &= \sum\limits_{{N_p} = n_p}^\infty  {\sum\limits_{{N_{\bar p}} = n_{\bar p}}^\infty  {P({N_p},{N_{\bar p}}) \times \frac{{{N_p}!}}{{{n_p}!\left( {{N_p} - {n_p}} \right)!}}{{({\varepsilon _p})}^{{n_p}}}{{(1 - {\varepsilon _p})}^{{N_p} - {n_p}}}} } \\
& \times  \frac{{{N_{\bar p}}!}}{{{n_{\bar p}}!\left( {{N_{\bar p}} - {n_{\bar p}}} \right)!}}{({\varepsilon _{\bar p}})^{{n_{\bar p}}}}{(1 - {\varepsilon _{\bar p}})^{{N_{\bar p}} - {n_{\bar p}}}} 
\end{split}
\end{equation}
\end{widetext}
where the $P({N_p},{N_{\bar p}})$ is the original joint probability distribution of number of proton ($N_p$) and anti-proton ($N_{\bar p}$), $\varepsilon _p$ and $\varepsilon _{\bar p}$ are the efficiency of proton and anti-proton, respectively. To derive the efficiency correction formula for moments and cumulants, let us introduce the bivariate factorial moments:
\begin{widetext}
\begin{align}
& {F_{i,k}(N_p, N_{\bar p})} = \left \langle \frac{{{N_p}!}}{{\left( {{N_p} - i} \right)!}}\frac{{{N_{\bar p}}!}}{{\left( {{N_{\bar p}} - k} \right)!}}\right \rangle = \sum\limits_{{N_p} = i}^\infty  {\sum\limits_{{N_{\bar p}} = k}^\infty  {P({N_p},{N_{\bar p}})\frac{{{N_p}!}}{{\left( {{N_p} - i} \right)!}}\frac{{{N_{\bar p}}!}}{{\left( {{N_{\bar p}} - k} \right)!}}} }   \label{eq:fact1} \\ 
&  {f_{i,k}(n_p, n_{\bar p})} =\left \langle \frac{{{n_p}!}}{{\left( {{n_p} - i} \right)!}}\frac{{{n_{\bar p}}!}}{{\left( {{n_{\bar p}} - k} \right)!}}\right\rangle  = \sum\limits_{{n_p} = i}^\infty  {\sum\limits_{{n_{\bar p}} = k}^\infty  {p({n_p},{n_{\bar p}})\frac{{{n_p}!}}{{\left( {{n_p} - i} \right)!}}\frac{{{n_{\bar p}}!}}{{\left( {{n_{\bar p}} - k} \right)!}}} }  \label{eq:fact2}
\end{align}
\end{widetext}
With the Eq. (\ref{eq:conv}), (\ref{eq:fact1}) and (\ref{eq:fact2}), one can obtain a useful relation between the efficiency corrected and uncorrected factorial moments as:
\begin{equation}  \label{eq:relation} 
{F_{i,k}(N_p, N_{\bar p})} = \frac{{{f_{i,k}(n_p, n_{\bar p})}}}{{{{({\varepsilon _p})}^i}{{({\varepsilon _{\bar p}})}^k}}}
\end{equation}

Then, the various order moments and cumulants can be expressed in terms of the factorial moments. Before deriving the formulas for the moments and cumulants of net-proton distributions, we need some mathematical relationships between moments, central moments, cumulants and factorial moments. Let us define a multivariate random vector $\boldsymbol{X}=(X_1,X_2,...,X_k)^{'}$ and a set of number $\boldsymbol{r}=(r_1,r_2,...,r_k)^{'}$. The multivariate moments, central moments and factorial moments can be written as:
\begin{align} \label{eq:momdefinition}
& {m_{\boldsymbol{r}}}(\boldsymbol{X}) = E\left[\prod\limits_{i = 1}^k {X_i^{{r_i}}} \right] \\
& {\mu _{\boldsymbol{r}}}(\boldsymbol{X}) = E\left[\prod\limits_{i = 1}^k {(X_i  - E[{X_i}])^{{r_i}}}\right] \\
& {F_{\boldsymbol{r}}}(\boldsymbol{X}) = E\left[\prod\limits_{i = 1}^k {\frac{{X_i^{}!}}{{(X_i^{} - {r_i})!}}}\right]
\end{align}
where $E$ denotes the expectation value operator, and the ${m_{\boldsymbol{r}}}(\boldsymbol{X})$, $ {\mu _{\boldsymbol{r}}}(\boldsymbol{X})$ and  $ {F_{\boldsymbol{r}}}(\boldsymbol{X}) $ are multivariate moments, central moments and factorial moments, respectively. Then, we have the relation between the moments and central moments by using binomial expansions: 
\begin{equation} 
\begin{array}{l}
{\mu _{\boldsymbol{r}}}({\boldsymbol{X}}) = \begin{array}{*{20}{c}}
{\sum\limits_{{i_1} = 0}^{{r_1}} {} }& \cdots &{\sum\limits_{{i_k} = 0}^{{r_k}} {{{( - 1)}^{{i_1} + {i_2} \cdots  + {i_k}}}(\begin{array}{*{20}{c}}
{{r_1}}\\
{{i_1}}
\end{array}) \cdots (\begin{array}{*{20}{c}}
{{r_k}}\\
{{i_k}}
\end{array})} }
\end{array}\\
\begin{array}{*{20}{c}}
{}&{ \times {{(E[{X_1}])}^{{i_1}}} \cdots }&{{{(E[{X_k}])}^{{i_k}}}}
\end{array}m_{{\boldsymbol{r-i}}}^{}({\boldsymbol{X}})
\end{array}
\end{equation} 
where  $\boldsymbol{i}=(i_1,i_2,...,i_k)^{'}$. To get the relation between moments and factorial moments, one needs the Stirling numbers of the first ($s_{1}(n,i)$) and second kind ($s_{2}(n,i)$), which are defined as: 
 \begin{align}
&\frac{{N!}}{{(N - n)!}} = \sum\limits_{i = 0}^n {{s_1}(n,i)} {N^i}\\
&{N^n} = \sum\limits_{i = 0}^n {{s_2}(n,i)} \frac{{N!}}{{(N - i)!}}
\end{align}
 where $N$, $n$ and $i$ are non-negative integer number.
The recursion equations for the Stirling numbers of the first and second kind are:
\begin{equation}
\begin{split}
&{s_1}(n,i) = {s_1}(n - 1,i - 1) - (n - 1)\times{s_1}(n - 1,i)\\
&{\left. {{s_1}(n,i)} \right|_{n < i}} = 0,{\left. {{s_1}(n,i)} \right|_{n = i}} = 1,{\left. {{s_1}(n,0)} \right|_{n > 0}} = 0
\end{split}
\end{equation}
and
\begin{equation}
\begin{split}
&{s_2}(n,i) = {s_2}(n - 1,i - 1) + i\times{s_2}(n - 1,i)\\
&{\left. {{s_2}(n,i)} \right|_{n < i}} = 0,{\left. {{s_2}(n,i)} \right|_{n = i}} = 1,{\left. {{s_2}(n,0)} \right|_{n >0}} = 0 
\end{split}
\end{equation}
The Stirling number of the first kind may have the negative value while the value of the second kind is always non-negative.
With the two kinds of Stirling numbers, one can write down the relations between moments and factorial moments as:
\begin{equation} \label{eq:mtof}
{m_{\boldsymbol{r}}}({\boldsymbol{X}}) = \begin{array}{*{20}{c}}
{\sum\limits_{{i_1} = 0}^{{r_1}} {} }& \cdots &{\sum\limits_{{i_k} = 0}^{{r_k}} {{s_2}({r_1},{i_1})} }
\end{array} \cdots {s_2}({r_k},{i_k}){F_{\boldsymbol{r}}}({\boldsymbol{X}})
\end{equation}

\begin{equation} \label{eq:ftom}
{F_{\boldsymbol{r}}}({\boldsymbol{X}}) = \begin{array}{*{20}{c}}
{\sum\limits_{{i_1} = 0}^{{r_1}} {} }& \cdots &{\sum\limits_{{i_k} = 0}^{{r_k}} {{s_1}({r_1},{i_1})} }
\end{array} \cdots {s_1}({r_k},{i_k}){m_{\boldsymbol{r}}}({\boldsymbol{X}})
\end{equation}

With Eq. (\ref{eq:momdefinition}) to (\ref{eq:ftom}), one can express the moments
of net-proton distributions in terms of the factorial moments. There are two variables in net-proton number calculation, the number of protons ($N_p$) and anti-protons ($N_{\bar{p}}$).
The $n^{th}$ order moments of net-proton distributions can be expressed in term of factorial moments: 
\begin{widetext}
\begin{equation} \label{eq:mtof2}
\begin{array}{l}
{m_n}({N_p} - {N_{\bar p}}) = <{({N_p} - {N_{\bar p}})^n}> = \sum\limits_{i = 0}^n {{{( - 1)}^i}\left( {\begin{array}{*{20}{c}}
n\\
i
\end{array}} \right)} <N_p^{n - i}N_{\bar p}^i>\\
 = \sum\limits_{i = 0}^n {{{( - 1)}^i}\left( {\begin{array}{*{20}{c}}
n\\
i
\end{array}} \right)} \left[ {\sum\limits_{{r_1} = 0}^{n - i} {\sum\limits_{{r_2} = 0}^i {{s_2}(n - i,{r_1}){s_2}(i,{r_2}){F_{{r_1},{r_2}}}({N_p},{N_{\bar p}})} } } \right]\\
 = \sum\limits_{i = 0}^n {\sum\limits_{{r_1} = 0}^{n - i} {\sum\limits_{{r_2} = 0}^i {{{( - 1)}^i}\left( {\begin{array}{*{20}{c}}
n\\
i
\end{array}} \right){s_2}(n - i,{r_1}){s_2}(i,{r_2}){F_{{r_1},{r_2}}}({N_p},{N_{\bar p}})} } } 
\end{array}
\end{equation}
\end{widetext}
Actually, two steps are needed to obtain this equation, the first step is to expand the moments of net-proton to the bivariate moments by using binomial expansion, and the other one is 
to express the bivariate moments in term of the factorial moments using the Eq. (\ref{eq:mtof}). Now, one can easily calculate the efficiency corrected moments of net-proton distributions in heavy-ion collisions by using the Eq. (\ref{eq:relation}) and (\ref{eq:mtof2}). Finally, we can express the efficiency corrected cumulants of net-proton distribution with 
the efficiency corrected moments by using the recursion relation:
\begin{equation} \label{eq:cumulants}
\begin{split}
&{C _r}({N_p} - {N_{\bar p}}) = {m_r}({N_p} - {N_{\bar p}}) \\
 &- \sum\limits_{s = 1}^{r - 1} {\left( \begin{array}{c}
r - 1\\
s - 1
\end{array} \right)} {C _s}({N_p} - {N_{\bar p}}){m_{r - s}}({N_p} - {N_{\bar p}})
\end{split}
\end{equation}
where the $C_r$ denotes the $r^{th}$ order cumulants of net-proton distributions. 
In principle, one can also express the factorial moments in Eq. (\ref{eq:mtof2}) in terms of the cumulants and the various order efficiency corrected cumulants can be expressed by the measured cumulants and efficiency as :
\begin{widetext}
\begin{equation} \label{eq:effcorrPRL}
\begin{split}
C_1^{X-Y}&=\frac{\la x\ra -\la y\ra}{\varepsilon} \\
C_2^{X - Y} &= \frac{{C_2^{x - y} + (\varepsilon  - 1)( \la x \ra  +  \la y \ra )}}{{{\varepsilon ^2}}}\\
C_3^{X - Y} &= \frac{{C_3^{x - y} + 3(\varepsilon  - 1)(C_2^x - C_2^y) + (\varepsilon  - 1)(\varepsilon  - 2)( \la x \ra  -  \la y \ra )}}{{{\varepsilon ^3}}}\\
C_4^{X - Y} &= \frac{{C_4^{x - y} - 2(\varepsilon  - 1)C_3^{x + y} + 8(\varepsilon  - 1)(C_3^x + C_3^y) + (5 - \varepsilon )(\varepsilon  - 1)C_2^{x + y}}}{{{\varepsilon ^4}}}\\
 &+ \frac{{8(\varepsilon  - 1)(\varepsilon  - 2)(C_2^x + C_2^y) + ({\varepsilon ^2} - 6\varepsilon  + 6)(\varepsilon  - 1)( \la x \ra  +  \la y \ra )}}{{{\varepsilon ^4}}}
\end{split}
\end{equation}
\end{widetext}
where the $(X,Y)$ and $(x,y)$ are the numbers of $(p, \bar{p})$ produced and measured, respectively. $\varepsilon=\varepsilon_p=\varepsilon_{\bar{p}}$ is the $p(\bar{p})$ efficiency. 
Obviously, the efficiency corrected cumulants are sensitive to the efficiency and depend on the lower order measured cumulants. For more detail discussion of this method, one can also refer to~\cite{Kitazawa_Eff}. 
\begin{figure*}[hptb]
\centering
\includegraphics[width=0.45\textwidth]{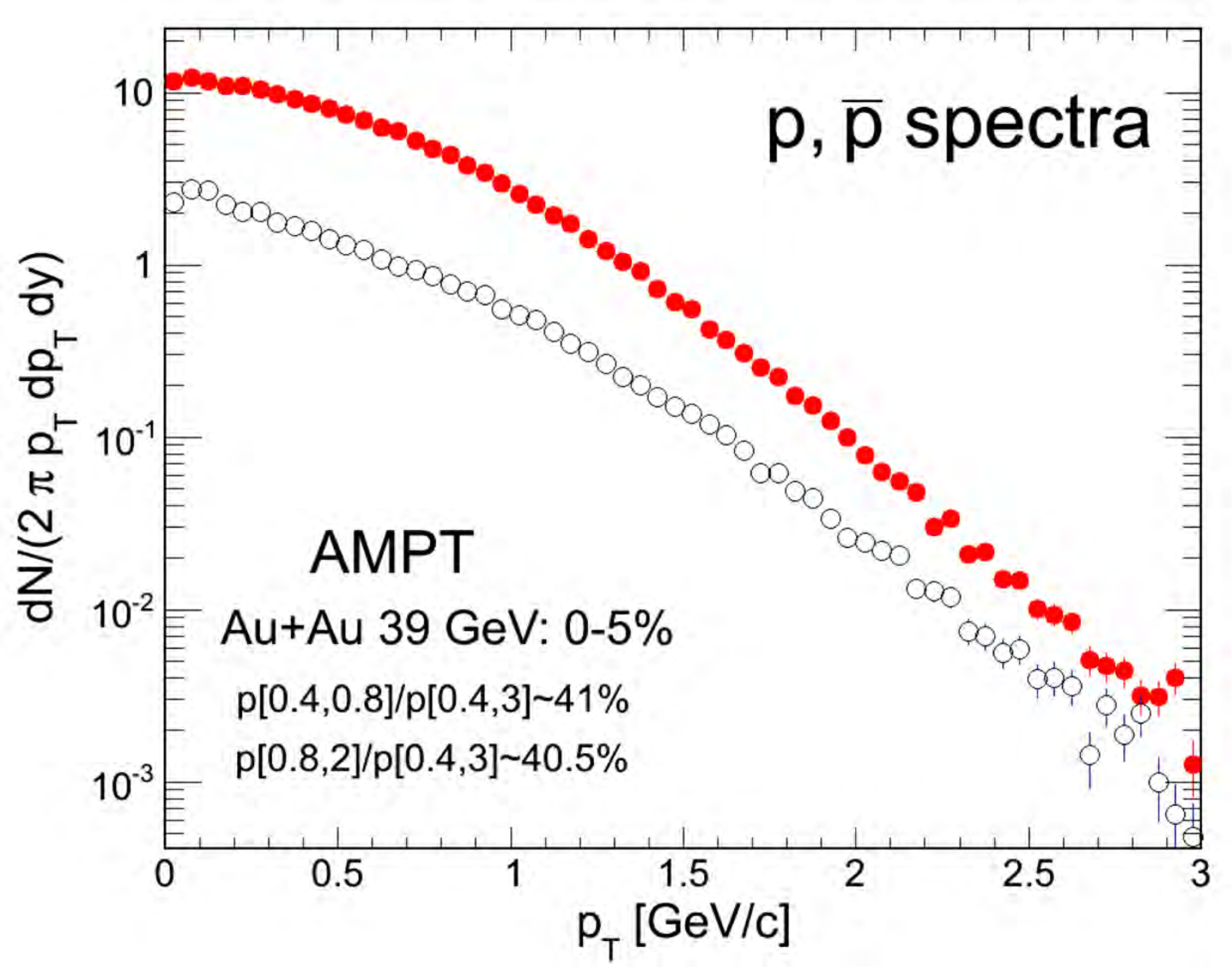}
\hspace{1cm}
\includegraphics[width=0.47\textwidth]{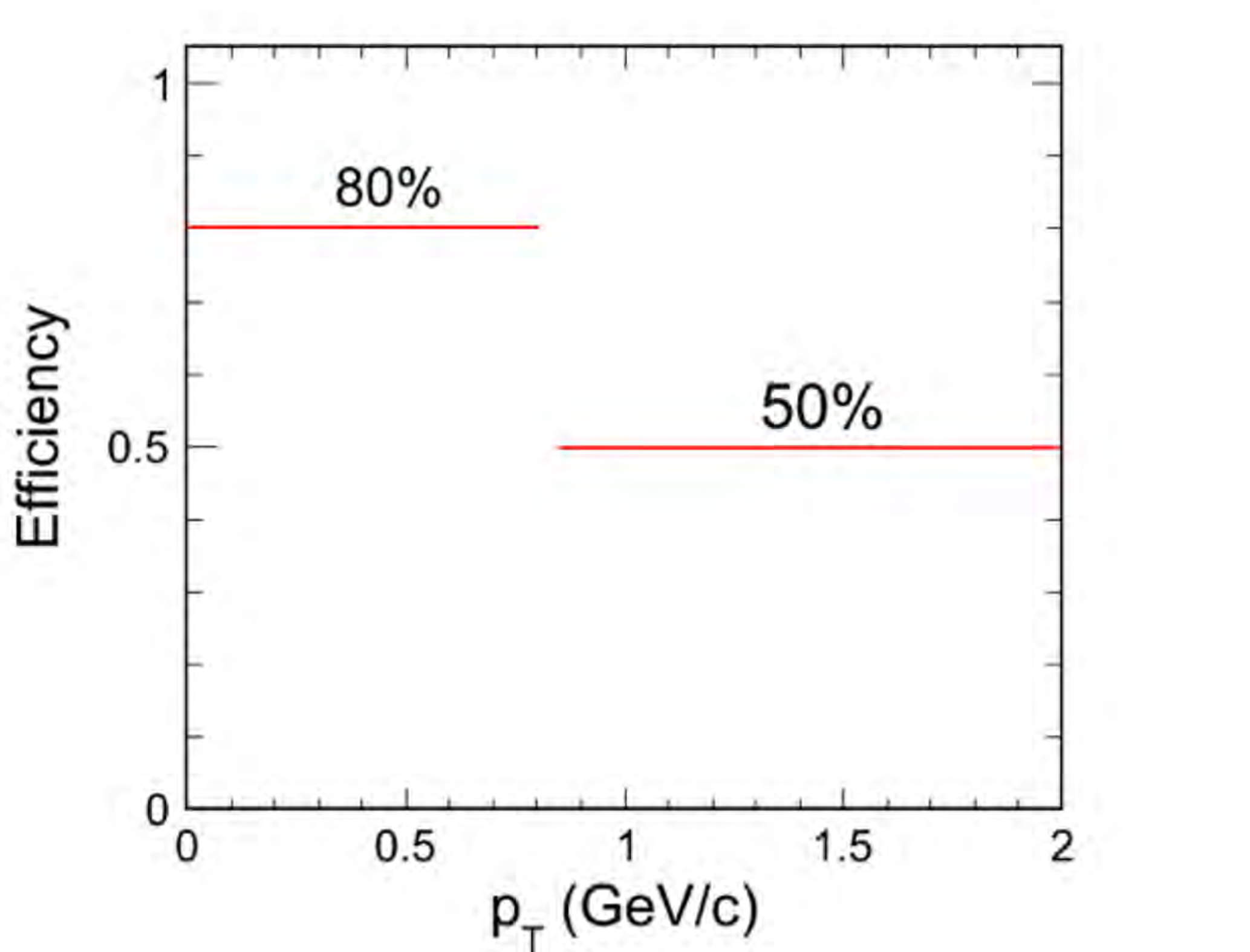}
\caption{(Color online) (Left) The invariant $p_{T}$ spectra of protons and anti-protons in Au+Au collisions at \sNN\ = 39 GeV from AMPT string melting calculation.  
(Right) Illustration of $p_{T}$ dependent detection efficiency for protons and anti-protons input by hand with low $p_{T}$ ($0.4<p_{T}<0.8$ GeV/c) : 80\% and  
high $p_{T}$ ($0.8<p_{T}<2$ GeV/c): 50\%.} \label{fig:efficiency_pt}
\end{figure*}

\begin{figure*}[hptb]
\centering
\hspace{-0.5cm}
\includegraphics[width=0.6\textwidth]{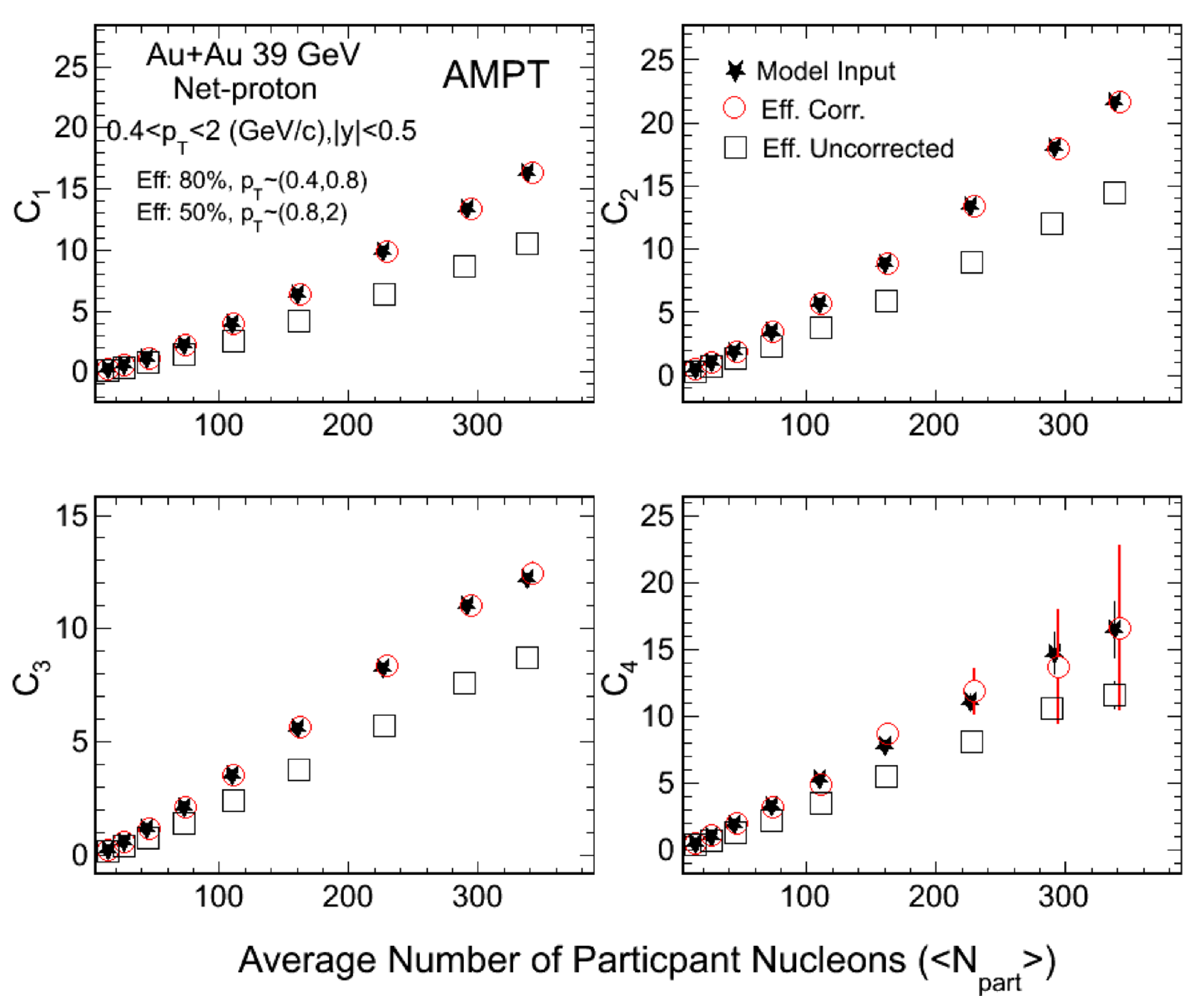}
\caption{(Color online) The cumulants of net-proton distributions in Au+Au collisions at \sNN\ = 39 GeV from AMPT model calculations. The black stars denote the 
results obtained from original model results without any efficiency effects.  The black empty squares represent the measured cumulants by applying the phase space dependent  efficiency effects ( low $p_{T}$ ($0.4<p_{T}<0.8$ GeV/c) : 80\% and  high $p_{T}$ ($0.8<p_{T}<2$ GeV/c): 50\%.). The red circles are efficiency corrected cumulants by using the phase space dependent efficiency correction formulas. } \label{fig:AMPT_test}
\end{figure*}

In the previous discussion,  the detection efficiency of proton and anti-proton are considered to be constant within the entire phase space. In many cases, the efficiency of proton and anti-proton will depend on the phase space (transverse momentum ($p_{T}$), rapidity (y), azimuthal angle ($\phi$)). In this sense, one has to re-consider  the efficiency correction method.  In the paper~\cite{voker_eff2}, a new method for dealing with this case has been discussed, but the formulae for efficiency correction are rather involved and difficult to understand.  In the following, we will provide an alternative efficiency correction method for the phase space dependent efficiency, which is straightforward and easier to understand.  For simplify, we only consider the phase space of the proton and anti-proton are decomposed into two sub-phase spaces (1 and 2), within which the efficiency of proton and anti-proton are constant. We use the symbol $\varepsilon _{{p_1}}^{},\varepsilon _{{p_2}}^{}$
and $\varepsilon _{{{\bar p}_1}}^{},\varepsilon _{{{\bar p}_2}}^{}$ to denote the efficiency of proton and anti-proton in the two sub-phase spaces, and the corresponding number of proton and anti-proton in the two sub-phase spaces are $N_{p_1}$, $N_{p_2}$ and $N_{\bar{p}_1}$,  $N_{\bar{p}_2}$, respectively. Using the relations in Eq. (\ref{eq:mtof}) and (\ref{eq:ftom}), one has:
\begin{widetext}
\begin{equation} \label{eq:4Fto2F}
\begin{split}
{F_{{r_1},{r_2}}}({N_p},{N_{\bar p}}) &= {F_{{r_1},{r_2}}}({N_{{p_1}}} + {N_{{p_2}}},{N_{{{\bar p}_1}}} + {N_{{{\bar p}_2}}})= \sum\limits_{{i_1} = 0}^{{r_1}} {\sum\limits_{{i_2} = 0}^{{r_2}} {{s_1}({r_1},{i_1})} } {s_1}({r_2},{i_2})\la {({N_{{p_1}}} + {N_{{p_2}}})^{{i_1}}}{({N_{{{\bar p}_1}}} + {N_{{{\bar p}_2}}})^{{i_2}}} \ra\\
& = \sum\limits_{{i_1} = 0}^{{r_1}} {\sum\limits_{{i_2} = 0}^{{r_2}} {{s_1}({r_1},{i_1})} } {s_1}({r_2},{i_2})\la {\sum\limits_{s = 0}^{{i_1}} {\left( {\begin{array}{*{20}{c}}
{{i_1}}\\
s
\end{array}} \right)N_{{p_1}}^{{i_1} - s}N_{{p_2}}^s\sum\limits_{t = 0}^{{i_2}} {\left( {\begin{array}{*{20}{c}}
{{i_2}}\\
t
\end{array}} \right)N_{{{\bar p}_1}}^{{i_2} - t}N_{{{\bar p}_2}}^t} } } \ra \\
& = \sum\limits_{{i_1} = 0}^{{r_1}} {\sum\limits_{{i_2} = 0}^{{r_2}} {\sum\limits_{s = 0}^{{i_1}} {\sum\limits_{t = 0}^{{i_2}} {{s_1}({r_1},{i_1}){s_1}({r_2},{i_2})\left( {\begin{array}{*{20}{c}}
{{i_1}}\\
s
\end{array}} \right)\left( {\begin{array}{*{20}{c}}
{{i_2}}\\
t
\end{array}} \right)} } } } \la N_{{p_1}}^{{i_1} - s}N_{{p_2}}^sN_{{{\bar p}_1}}^{{i_2} - t}N_{{{\bar p}_2}}^t \ra \\
&= \sum\limits_{{i_1} = 0}^{{r_1}} {\sum\limits_{{i_2} = 0}^{{r_2}} {\sum\limits_{s = 0}^{{i_1}} {\sum\limits_{t = 0}^{{i_2}} {\sum\limits_{u = 0}^{{i_1} - s} {\sum\limits_{v = 0}^s {\sum\limits_{j = 0}^{{i_2} - t} {\sum\limits_{k = 0}^t {{s_1}({r_1},{i_1}){s_1}({r_2},{i_2})\left( {\begin{array}{*{20}{c}}
{{i_1}}\\
s
\end{array}} \right)\left( {\begin{array}{*{20}{c}}
{{i_2}}\\
t
\end{array}} \right)} } } } } } } } \\
 &\times {s_2}({i_1} - s,u){s_2}(s,v){s_2}({i_2} - t,j){s_2}(t,k) \times {F_{u,v,j,k}}(N_{{p_1}}^{},N_{{p_2}}^{},N_{{{\bar p}_1}}^{},N_{{{\bar p}_2}}^{})
\end{split}
\end{equation}
\end{widetext}
Based on the Eq. (\ref{eq:4Fto2F}), we build up a relation between the bivariate factorial moments of proton and anti-proton distributions in the entire phase space and
the multivariate factorial moments of proton and anti-proton distributions in the two sub-phase spaces. As a direct extension of Eq. (\ref{eq:relation}) for multivariate case, the efficiency corrected multivariate factorial moments of proton and anti-proton distributions in the sub-phase spaces can be obtained as:
 \begin{equation} \label{eq:relation2}
{F_{u,v,j,k}}(N_{{p_1}}^{},N_{{p_2}}^{},N_{{{\bar p}_1}}^{},N_{{{\bar p}_2}}^{}) = \frac{{{f_{u,v,j,k}}(n_{{p_1}}^{},n_{{p_2}}^{},n_{{{\bar p}_1}}^{},n_{{{\bar p}_2}}^{})}}{{{{({\varepsilon _{{p_1}}})}^u}{{({\varepsilon _{{p_2}}})}^v}{{({\varepsilon _{{{\bar p}_1}}})}^j}{{({\varepsilon _{{{\bar p}_2}}})}^k}}}\end{equation}
where  ${{f_{u,v,j,k}}(N_{{p_1}}^{},N_{{p_2}}^{},N_{{{\bar p}_1}}^{},N_{{{\bar p}_2}}^{})}$ is the measured multivariate factorial moments of proton and anti-proton distributions. 
By using  Eq. (\ref{eq:mtof2}), (\ref{eq:cumulants}), (\ref{eq:4Fto2F}) and (\ref{eq:relation2}), one can obtain the efficiency corrected moments of net-proton distributions for the case, where
the proton (anti-proton) are with different efficiency in two sub-phase spaces. If the efficiency of the proton (anti-proton) have large variations within the phase space, one needs to further divide the phase space into small ones. It is easy and straightforward to do this, but it is time consuming and requires more computing resources.

To verify the phase space dependent efficiency correction formulas, we perform a calculation of the net-proton fluctuations with AMPT string melting model. The invariant $p_{T}$ spectra of proton and anti-proton from AMPT can be found in the Fig.\ref{fig:efficiency_pt} left. In Fig.\ref{fig:efficiency_pt} right, we set by hand the $p_{T}$ dependent efficiency for (anti-)protons with the efficiency at low $p_{T}$ ($0.4<p_{T}<0.8$ GeV/c) : 80\% and  high $p_{T}$ ($0.8<p_{T}<2$ GeV/c): 50\%, respectively. The efficiency response function is set to be binomial distribution. Then, the measured net-proton distributions are the convolution between original model inputs and the binomial distributions. By doing this, we can calculate the measured cumulants of net-proton of Au+Au collisions at \sNN\ = 39 GeV from AMPT string melting model with $p_{T}$ dependent efficiency.  With the same procedures as we did in the real data analysis, we apply the phase space dependent efficiency formulas to do the efficiency correction for the measured cumulants. Fig. \ref{fig:AMPT_test} shows that the efficiency corrected cumulants are consistent with the results from original model input within uncertainties. The statistical errors for the efficiency corrected cumulants are calculated from Delta theorem, which will be discussed later. Finally, this test confirms that the phase space dependent efficiency correction formulas we obtained are reliable and work well. On the other hand, if the efficiency response function is non-binomial type, instead of using analytical formulas, the unfolding method with real response matrix should be used~\cite{Koch_eff_mul}. 

\subsection{Error Estimation for the Efficiency Corrected Cumulants}   \label{sec:ErrorEstimation}
Based on the Delta theorem in statistics, we obtained the error formulas for various order cumulants and cumulant ratios~\cite{Delta_theory}. However, those formulas can only be applied to the case, where the efficiency is unity  ($\varepsilon=1$).  It is not straightforward and easy to calculated the statistical errors for efficiency corrected cumulants with $\varepsilon \neq 1$ and one can not directly use the formulas obtained in the paper~\cite{Delta_theory}. In the following, we will derive general error formulas for estimating the statistical errors of efficiency corrected cumulants of conserved quantities in heavy-ion collisions based on the Delta theorem in statistics.  With those analytical formulas,  one can predict the expected errors with the number of events and efficiency numbers.
\begin{figure*}[htb]
\begin{center}
\includegraphics[width=0.65\textwidth]{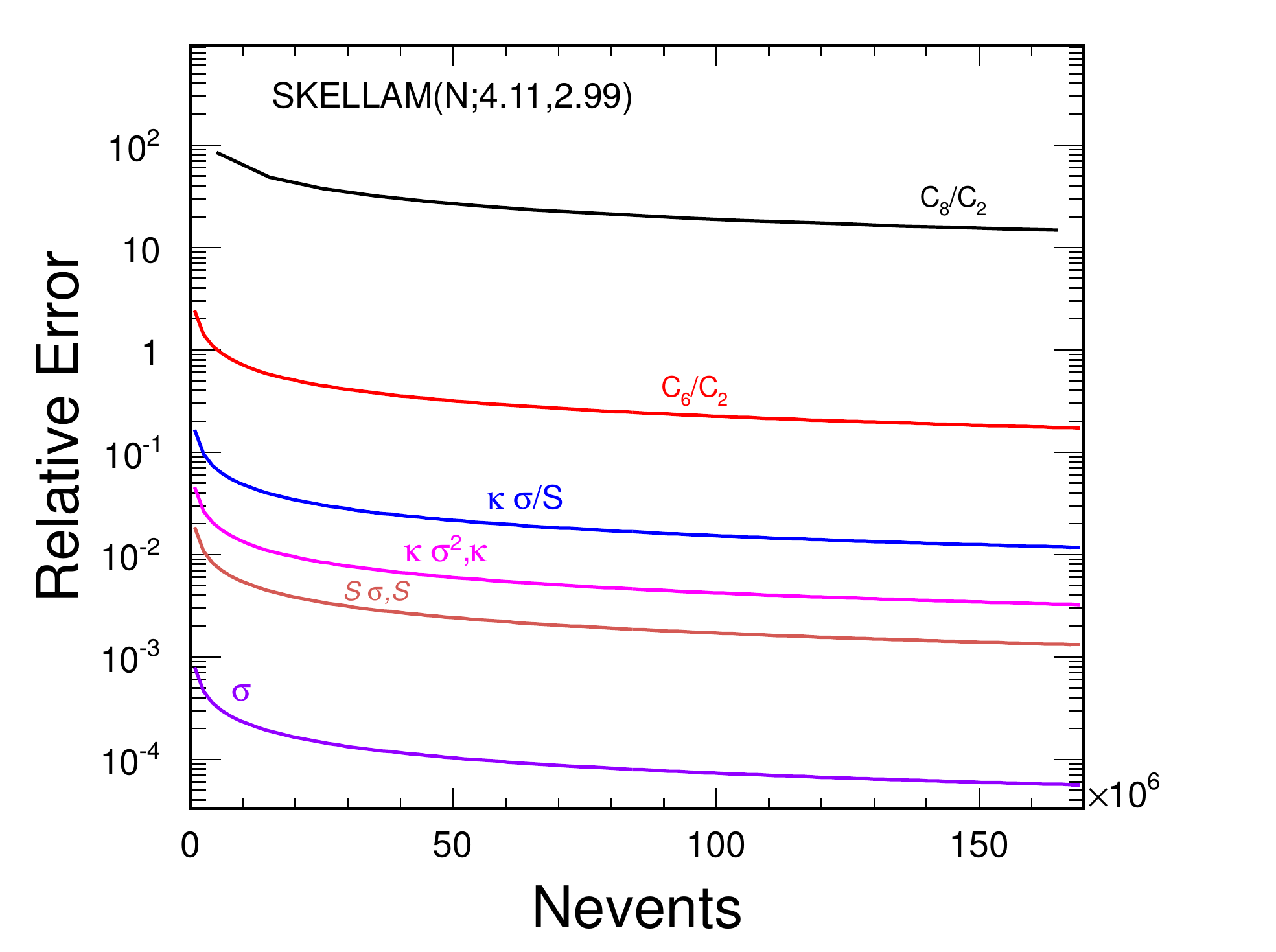}
\caption[]{Relative errors as a function of number of events for various cumulants and cumulant ratios of Skellam distributions based on
the error formulas~\cite{Delta_theory}. } \label{fig:relative_error}
\end{center}
\end{figure*}
The Delta theorem in statistics is a fundamental theorem which is used to approximate the distribution of a transformation
of a statistic in large samples if we can approximate the distribution of the statistic itself. Distributions of transformations of a statistic
are of great importance in applications. We will give the theorem without proofs and one can see~\cite{asytheory, junshan}.

{ \bf {\textit {Delta Theorem}}:} Suppose that
${\bf{ X}}=\{X_{1},X_{2},...,X_{k}\}$ is normally distributed as
$N({\bf{\mu}}, {\bf{\Sigma}}/n)$,  with $\bf \Sigma$ a covariance
matrix. Let ${\bf {g(x)}}=( g_{1}({\bf x}),...,g_{m}({\bf x}))$, ${\bf
{x}}=(x_{1},...x_{k})$, be a vector-valued function for which each
component function $g_{i}({\bf x})$ is real-valued and has a non-zero
differential $g_{i}(\mu)$, at ${\bf{x}}={\bf{\mu}}$. Put
\begin{equation}{\bf{D}} = \left[ {\left. {\frac{{\partial g_i
}}{{\partial x_j }}} \right|_{x = \mu } } \right]_{m \times
k}\end{equation} Then
\begin{equation} \label{eq:limitvar} {\bf{g}}({\bf{X}})\xrightarrow[]{d}N({\bf{g}}(\mu),\frac{{\bf{D\Sigma D^{'}}}}{n})\end{equation}
where $n$ is the number of events. 
\begin{figure*}[htbp]
\begin{center}
\hspace{-1.5cm}
\includegraphics[scale=0.8]{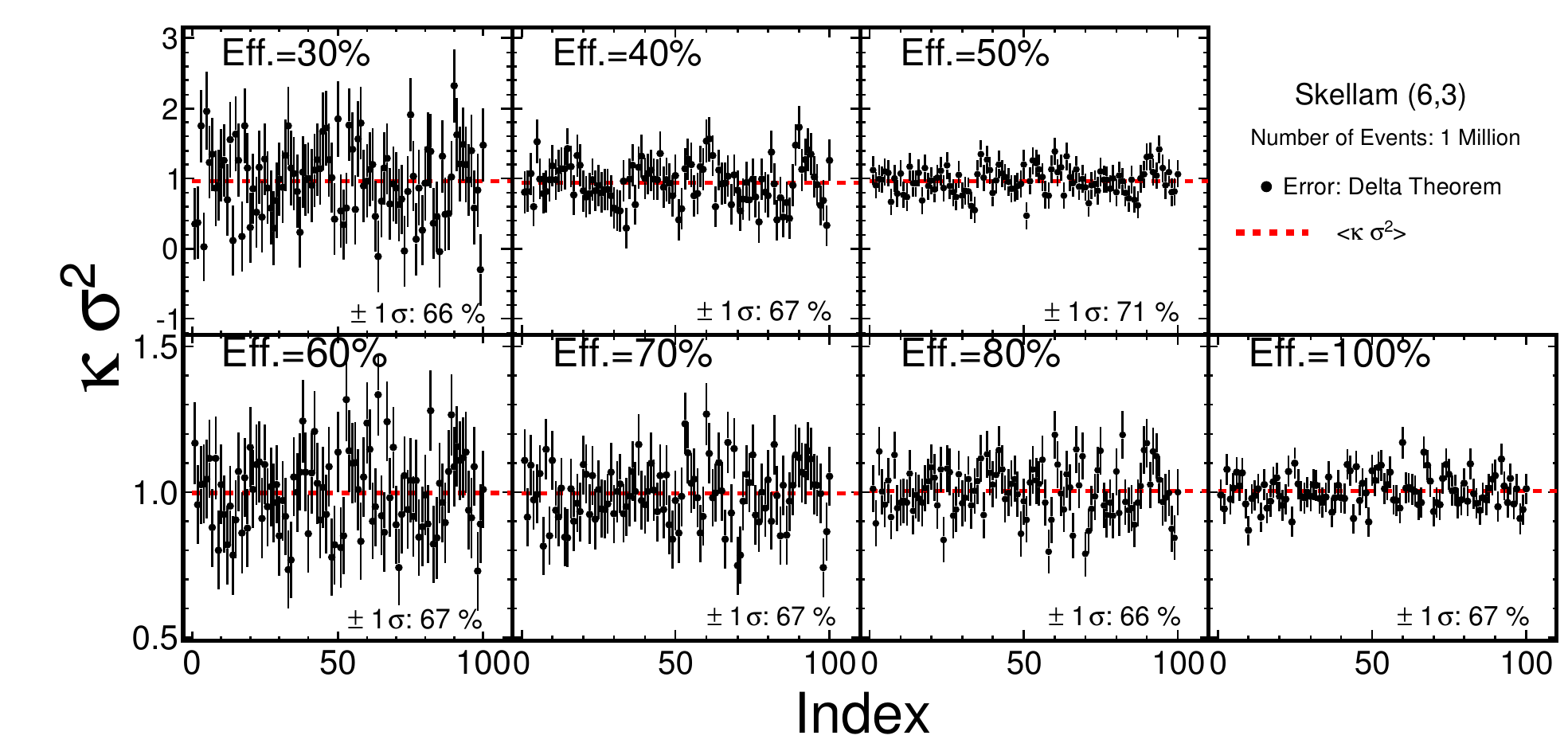}
\caption[]{(Color online) Each data point in each panel represents the efficiency corrected {\KV} and statistical error for an event sample with one million events that independently and randomly generated from the original skellam distribution with efficiency effects. Different panels are with different efficiency varying from 30\% to 100\% The error estimation is based on the Delta theorem. The dashed line in each panel is the average {\KV} value of the 100 samples~\cite{Unified_Errors}.   } \label{fig:KV_eff}
\end{center}
\end{figure*}
\begin{figure*}[htbp]
\begin{center}
\hspace{-1.5cm}
\includegraphics[scale=0.8]{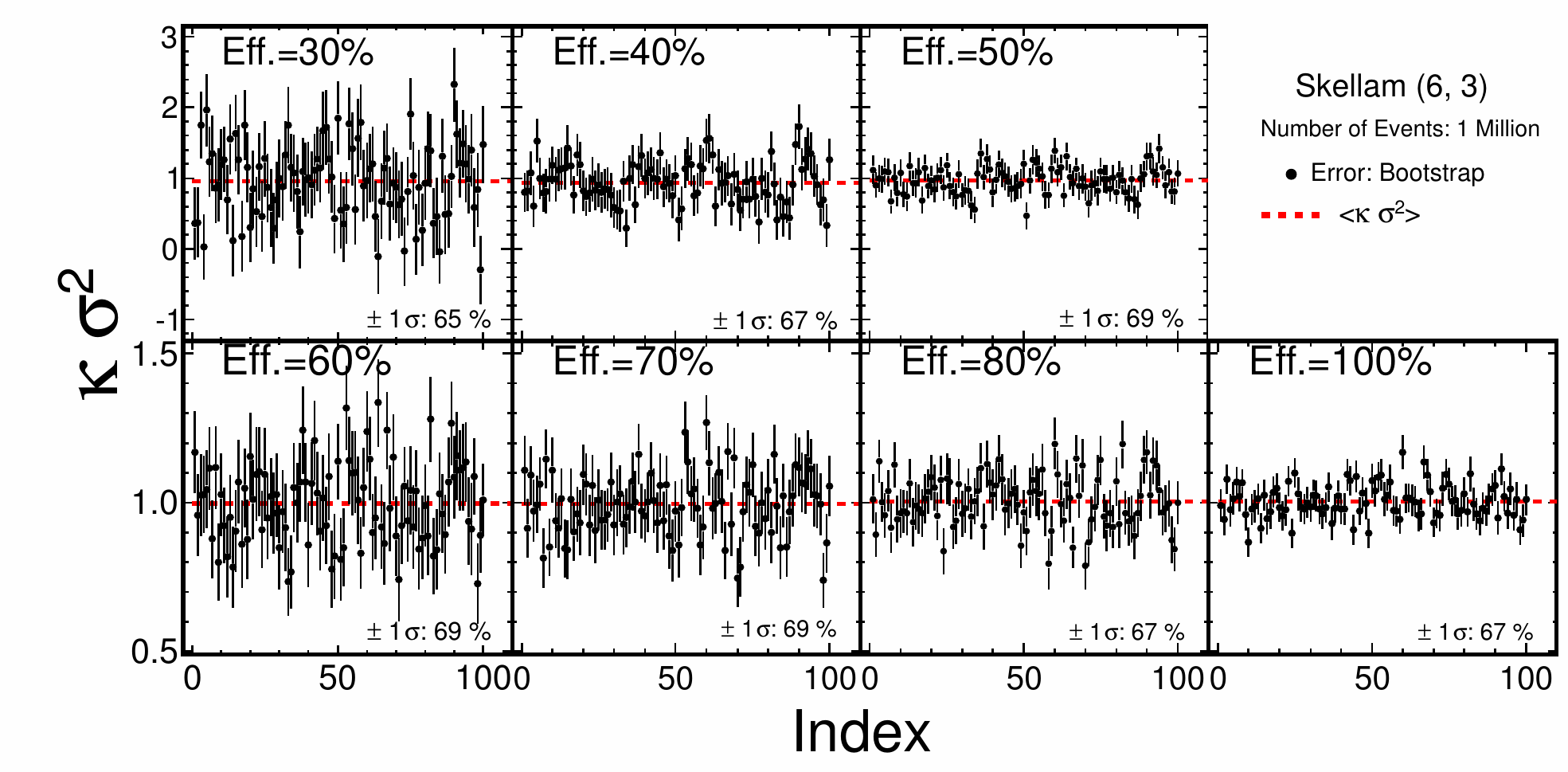}
\caption[]{(Color online) Each data point in each panel represents the efficiency corrected {\KV} and statistical error for an event sample with one million events that independently and randomly generated from the original skellam distribution with efficiency effects. Different panels are with different efficiency varying from 30\% to 100\% The error estimation is based on the Bootstrap. The dashed line in each panel is the average {\KV} value of the 100 samples~\cite{Unified_Errors}.     } \label{fig:KV_eff_Boot}
\end{center}\end{figure*}

Based on the Delta theorem, one can derive the general error formula for a statistic quantity. Suppose,  statistic quantity $\phi$ is as a function of random variables ${\bf{ X}}=\{X_{1},X_{2},...,X_{m}\}$, then the transformation functions ${\bf {g(X)}}=\phi({\bf X})$.  The {\bf D} matrix can be written as: 
\begin{equation}
{\bf{D}} = {\left[ {\frac{{\partial \phi }}{{\partial {\bf X}}}} \right]_{1\times m}}
\end{equation} 
and the covariance matrix $\Sigma$ is:
\begin{equation}
\Sigma  = n \times Cov({X_i},{X_j})
\end{equation} 
Based on Eq. (\ref{eq:limitvar}), the variance of the statistic $\phi$ can be calculated as:

\begin{widetext}
\begin{equation} \label{eq:error}
\begin{split}
V(\phi ) &=\frac{{\bf{D\Sigma D^{'}}}}{n}= \sum\limits_{i = 1,j = 1}^m {\left( {\frac{{\partial \phi }}{{\partial {X_i}}}} \right)} \left( {\frac{{\partial \phi }}{{\partial {X_j}}}} \right)Cov({X_i},{X_j})\\
 &= \sum\limits_{i = 1}^m {{{\left( {\frac{{\partial \phi }}{{\partial {X_i}}}} \right)}^2}} V({X_i}) + \sum\limits_{i = 1,j = 1,i \ne j}^m {\left( {\frac{{\partial \phi }}{{\partial {X_i}}}} \right)} \left( {\frac{{\partial \phi }}{{\partial {X_j}}}} \right)Cov({X_i},{X_j})
\end{split}
\end{equation}
\end{widetext}
where $V(X_i)$ is the variance of variable $X_i$ and $Cov(X_i,X_j)$ is the covariance between $X_i$ and $X_j$. 
To calculate the statistical errors, one needs to know the variance and covariance of the variable $X_i$ and $X_j$ in the Eq. (\ref{eq:error}).  Since the efficiency corrected moments are expressed in terms of the factorial moments, the factorial moments are the random variable $X_i$ in Eq.  (\ref{eq:error}).  Then, we need to know the expression for variance and covariance of the factorial moments. It is known that the covariance of the multivariate moments~\cite{advancetheory} can be written as:
\begin{equation} \label{eq:covm}Cov({m_{r,s}},{m_{u,v}}) = \frac{1}{n}({m_{r + u,s + v}} - {m_{r,s}}{m_{u,v}}) \end{equation}
where $n$ is the number of events, $m_{r,s}=<X_1^{r}X_2^{s}>$ and ${m_{u,v}}=<X_1^{u}X_2^{v}>$ are the multivariate moments, the $X_1$ and $X_2$ are random variables. Then, we can obtain the variance of the cumulants and cumulant ratios as:
\begin{widetext}
\begin{eqnarray}
Var( \la N \ra ) &=& \mu _2^{}/n,\begin{array}{*{20}{c}}
{}&{}
\end{array}Var({C_2}) = ({\mu _4} - \mu _2^2)/n\\
Var({C_3}) &=& ({\mu _6} - \mu _3^2 - 6{\mu _4}{\mu _2} + 9\mu _2^3)/n\\
Var({C_4}) &=& ({\mu _8} - 12{\mu _6}{\mu _2} - 8{\mu _5}{\mu _3} + 48{\mu _4}\mu _2^2 - \mu _4^2 + 64\mu _3^2{\mu _2} -36\mu _2^4)/n \\
Var( S \sigma )&=&[9 - 6m_4  + m_3 ^2 (6 +m_4 ) - 2m_3 m_5  + m_6 ]\sigma ^2 /n \\ 
Var(\kappa  \sigma ^2 ) &=& [ - 9 + 6m_4 ^2  + m_4 ^3  + 8m_3^2 (5 + m_4 ) - 8m_3 m_5  + m_4 (9 - 2m_6 ) - 6m_6  + m_8 ]\sigma ^4/n \\ 
Var(\kappa \sigma/S) &=&[64m_3 ^4  - 8m_3 ^3 m_5  - ( - 3 + m_4 )^2 ( - 9 + 6m_4 - m_6 ) +2m_3 ( - 3 + m_4 )(9m_5  - m_7 ) \nonumber
\\ &+& m_3 ^2 (171 - 48m_4  + 8m_4 ^2  - 12m_6  + m_8 )]\sigma ^2 /(n \times m_3 ^4 ) \\
 Var(C_6/C_2)  &= &[10575 - 30m_{10}  + m_{12}  + 18300m_3 ^2  + 2600m_3 ^4  - 225( - 3 + m_4 )^2  - 7440m_3 m_5 \\ \nonumber
  &- &520m_3 ^3 m_5  + 216m_5 ^2  - 2160m_6  - 200m_3 ^2 m_6  + 52m_3 m_5 m_6  + 33m_6 ^2  \\   \nonumber
 & +& ( - 3 + m_4 )(10(405 - 390m_3 ^2  + 10m_3 ^4  + 24m_3 m_5 ) - 20(6 + m_3 ^2 )m_6  + m_6 ^2 ) \\  \nonumber
  &+ &840m_3 m_7  - 12m_5 m_7  + 345m_8  + 20m_3 ^2 m_8  - 2m_6 m_8  - 40m_3 m_9 ]\sigma ^8/n  \\  \nonumber
\end{eqnarray}
\end{widetext}
where $\mu_r=\la (\delta N)^{r} \ra$ is the $r^{th}$ order central moments, $m_r=\mu_r/\sigma^{r}$ and $n$ is the number of events. For normal distributions with width $\sigma$, the statistical error of the cumulants and cumulant ratios at different orders can be approximated as:
\begin{eqnarray}
error({C_r}) &\propto& \frac{{{\sigma ^r}}}{{\sqrt n }} \label{eq:err_cum} \\
error({C_r}/{C_2}) &\propto& \frac{{{\sigma ^{(r - 2)}}}}{{\sqrt n }}
\end{eqnarray}
Figure \ref{fig:relative_error} shows the relative errors of cumulants and cumulant ratios of Skllellam distribution as a function of number of events $N$. It is found that the higher orders cumulants are with larger relative errors than the low orders at the same number of events $N$. 

Based on Eq. (\ref{eq:ftom}) and (\ref{eq:covm}), one can obtain the 
covariance for the multivariate factorial moments as:
\begin{widetext}
\begin{equation} \label{eq:covF}
\begin{split}
&Cov({f_{r,s}},{f_{u,v}}) = Cov\left(\sum\limits_{i = 0}^r {\sum\limits_{j = 0}^s {{s_1}(r,i){s_1}(s,j){m_{i,j}},} } \sum\limits_{k = 0}^u {\sum\limits_{h = 0}^v {{s_1}(u,k){s_1}(v,h){m_{k,h}}} } \right)\\
 &= \sum\limits_{i = 0}^r {\sum\limits_{j = 0}^s {\sum\limits_{k = 0}^u {\sum\limits_{h = 0}^v {{s_1}(r,i){s_1}(s,j){s_1}(u,k){s_1}(v,h)} }  \times Cov({m_{i,j}},{m_{k,h}})} } \\
 &= \frac{1}{n}\sum\limits_{i = 0}^r {\sum\limits_{j = 0}^s {\sum\limits_{k = 0}^u {\sum\limits_{h = 0}^v {{s_1}(r,i){s_1}(s,j){s_1}(u,k){s_1}(v,h)} }  \times } } ({m_{i + k,j + h}} - {m_{i,j}}{m_{k,h}})\\
&=\frac{1}{n}{\sum\limits_{i = 0}^r {\sum\limits_{j = 0}^s {\sum\limits_{k = 0}^u {\sum\limits_{h = 0}^v {\sum\limits_{\alpha  = 0}^{i + k} {\sum\limits_{\beta  = 0}^{j + h} {{s_1}(r,i){s_1}(s,j){s_1}(u,k){s_1}(v,h){s_2}(i + k,\alpha ){s_2}(j + h,\beta ){f_{\alpha ,\beta }}} } } } } } }  \\
&-\frac{1}{n}{f_{r,s}}{f_{u,v}} \\
 &= \frac{1}{n}({f_{(r,u),(s,v)}} - {f_{r,s}}{f_{u,v}})
\end{split}
\end{equation}
where the $f_{(r,u),(s,v)}$ is defined as:
\begin{equation} \label{eq:fact4}
\begin{split}
&{f_{(r,u),(s,v)}} = \left\langle {\frac{{{X_1}!}}{{({X_1} - r)!}}\frac{{{X_1}!}}{{({X_1} - u)!}}\frac{{{X_2}!}}{{({X_2} - s)!}}\frac{{{X_2}!}}{{({X_2} - v)!}}} \right\rangle \\
&={\sum\limits_{i = 0}^r {\sum\limits_{j = 0}^s {\sum\limits_{k = 0}^u {\sum\limits_{h = 0}^v {\sum\limits_{\alpha  = 0}^{i + k} {\sum\limits_{\beta  = 0}^{j + h} {{s_1}(r,i){s_1}(s,j){s_1}(u,k){s_1}(v,h){s_2}(i + k,\alpha ){s_2}(j + h,\beta ){f_{\alpha ,\beta }}} } } } } } }  
\end{split}
\end{equation}
\end{widetext}
The definition of bivariate factorial moments $f_{r,s}$, $f_{u,v}$ and $f_{\alpha,\beta}$ are the same as Eq. (\ref{eq:fact2}).
The Eq. (\ref{eq:covF}) can be put into the standard error propagation formulae (\ref{eq:error}) to calculate the statistical errors of the efficiency corrected moments.

\begin{figure}[htbp]
\begin{center}
\hspace{-0.5cm}
\includegraphics[scale=0.35]{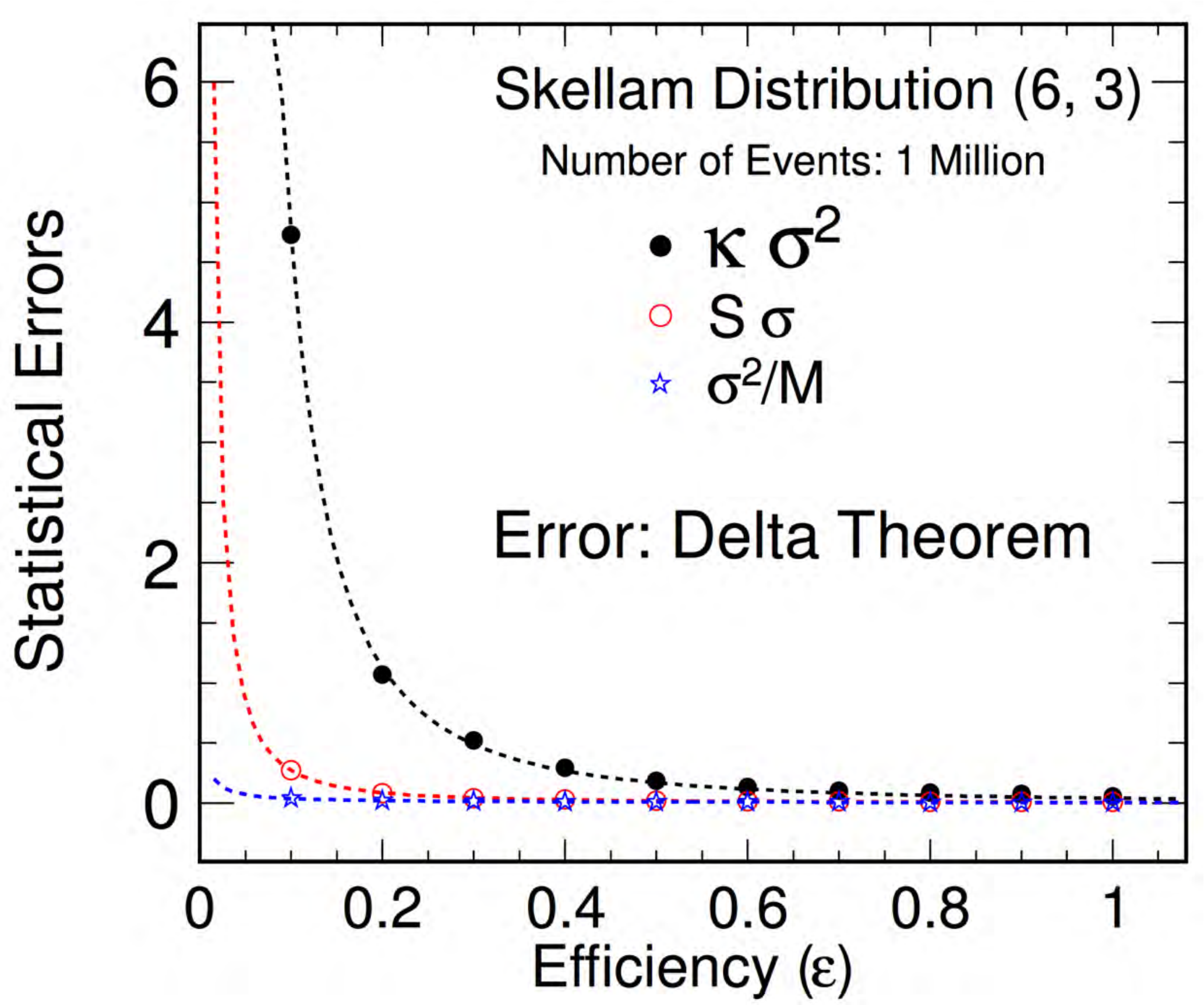}
\caption[]{(Color online) The statistical errors of efficiency corrected {\KV}, {\SD} and {\VM} as a function of efficiency for the original skellam distribution.
The errors are calculated by the Delta theorem~\cite{Unified_Errors}.} \label{fig:error_eff}
\end{center}\end{figure}

\begin{figure*}
\begin{minipage}[c]{0.4\linewidth}
\centering 
    \includegraphics[scale=0.33]{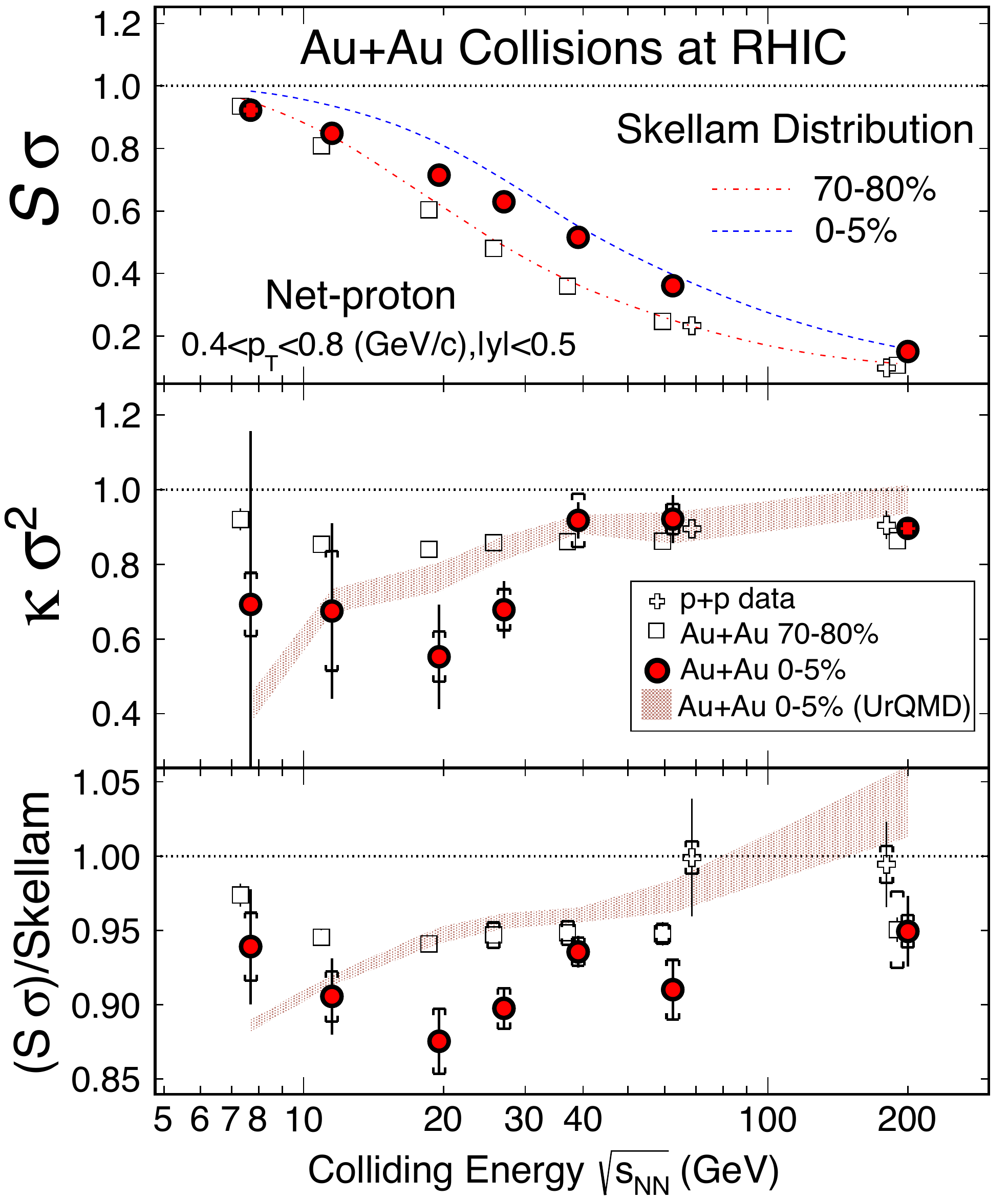}
        \end{minipage}
      \hspace{0.2in}
  \begin{minipage}[c]{0.4\linewidth}
  \centering 
   \includegraphics[scale=0.34]{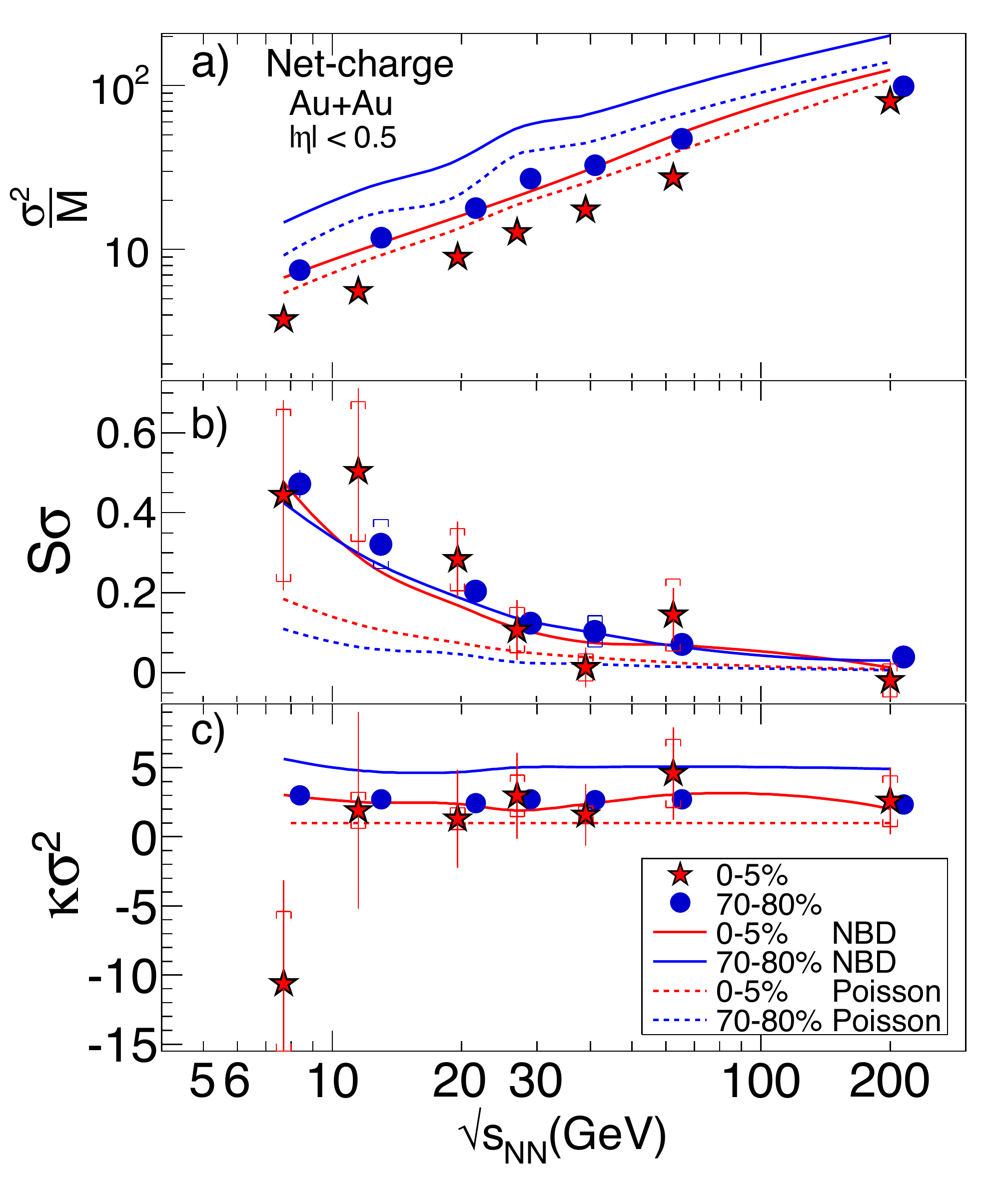}
          \end{minipage} 
         \caption{(Color online) Energy dependence of moments of net-proton (left)~\cite{STAR_BES_PRL} and net-charge (right)~\cite{netcharge_PRL} distributions for Au+Au collisions at RHIC BES energies. The statistical and systematical error are shown in bars and brackets, respectively. } \label{fig:cumulants_energy}
\end{figure*}
Besides the Delta theorem for estimating the statistical errors, another computer intensive one is the so called bootstrap, which is based on resampling method. with the bootstrap method, one needs to prepare $B$ new samples. Every new sample is sampling randomly with replacement from the original sample and are with the same number of events as the original one.  The uncertainty on a statistic quantity is estimated by the root mean square of the $B$ values of the statistic quantity obtained from these samples. In the MC simulation, we set the number of new samples $B=200$. The variance of the statistic quantity $\Phi$ can be given by
\begin{equation}
\begin{split}
V(\Phi ) &= \frac{{\sum\limits_{b = 1}^B {{{\left( {{\Phi _b} - \frac{1}{B}\sum\limits_{b = 1}^B {{\Phi _b}} } \right)}^2}} }}{{B - 1}}\\
 &= \frac{B}{{B - 1}}\left[ {\frac{1}{B}\sum\limits_{b = 1}^B {\Phi _b^2 - {{\left( {\frac{1}{B}\sum\limits_{b = 1}^B {{\Phi _b}} } \right)}^2}} } \right]
\end{split}
\end{equation}
For comparison, we show the error estimation for the efficiency corrected {\KV} of Skellam distributions with Delta theorem and Bootstrap method in the Fig. \ref{fig:KV_eff} and Fig. \ref{fig:KV_eff_Boot}, respectively.  Both the Delta theorem and Bootstrap method can reasonably describe the statistical errors of the efficiency corrected {\KV} with various efficiency numbers ranging from 30\% to 100\%. The probability for the error bars of those data points touching the mean value are very close to the expected value 68\%. Since we concentrate on the comparison of the magnitude of the statistical error calculated from the Delta theorem and Bootstrap methods, the data points are calculated from the same data sets and thus the {\KV} values are identical, while the statistical error bars of the data points in the two figures are not identical. This consistency verifies that the analytical error formulas derived from Delta theorem is correct. However, the calculation speed of Delta theorem method is much faster than that of Bootstrap method. On the other hand, since one cannot obtain events further into the tails than those in the original sample, the  bootstrap method might run into difficulties if the quantity whose variance is being estimated depends heavily on the tails of distributions.

Figure  \ref{fig:error_eff} shows the statistical errors for the efficiency corrected {\KV}, {\SD} and {\VM} as a function of efficiency. In simulation, the efficiency effects are implemented for the original skellam distribution and the number of events is fixed to be one million for each data point. It can be found that the statistical errors are dramatically increase when decreasing the efficiency number, especially for higher order cumulant ratios.  We also fit those data points with the functional form:
\begin{equation}
f(\varepsilon ) = \frac{1}{{\sqrt n }}\frac{a}{{{\varepsilon ^b}}}\end{equation}
where $n$ is the number of events which is fixed to be one million here, $a$ and $b$ are free parameters. The fitting results of $a$ and $b$ are 40.6 and 2.06 for {\KV},  6.02 and 1.65 for {\SD}, 4.96 and 0.89 for {\VM}, respectively. The parameters $a$ and $b$ depend on the original distribution and the studied statistic quantity. We can understand the effects of the efficiency on the statistical errors in an intuitive way.  The efficiency will cause the loss of information of the original distributions, especially at the tails. The smaller the efficiency is, larger uncertainties we will get for the efficiency corrected results and needs more events to recover the original information.
 
\begin{figure*}
\hspace{-0.5cm}
\begin{minipage}[c]{0.4\linewidth}
\centering 
    \includegraphics[scale=0.35]{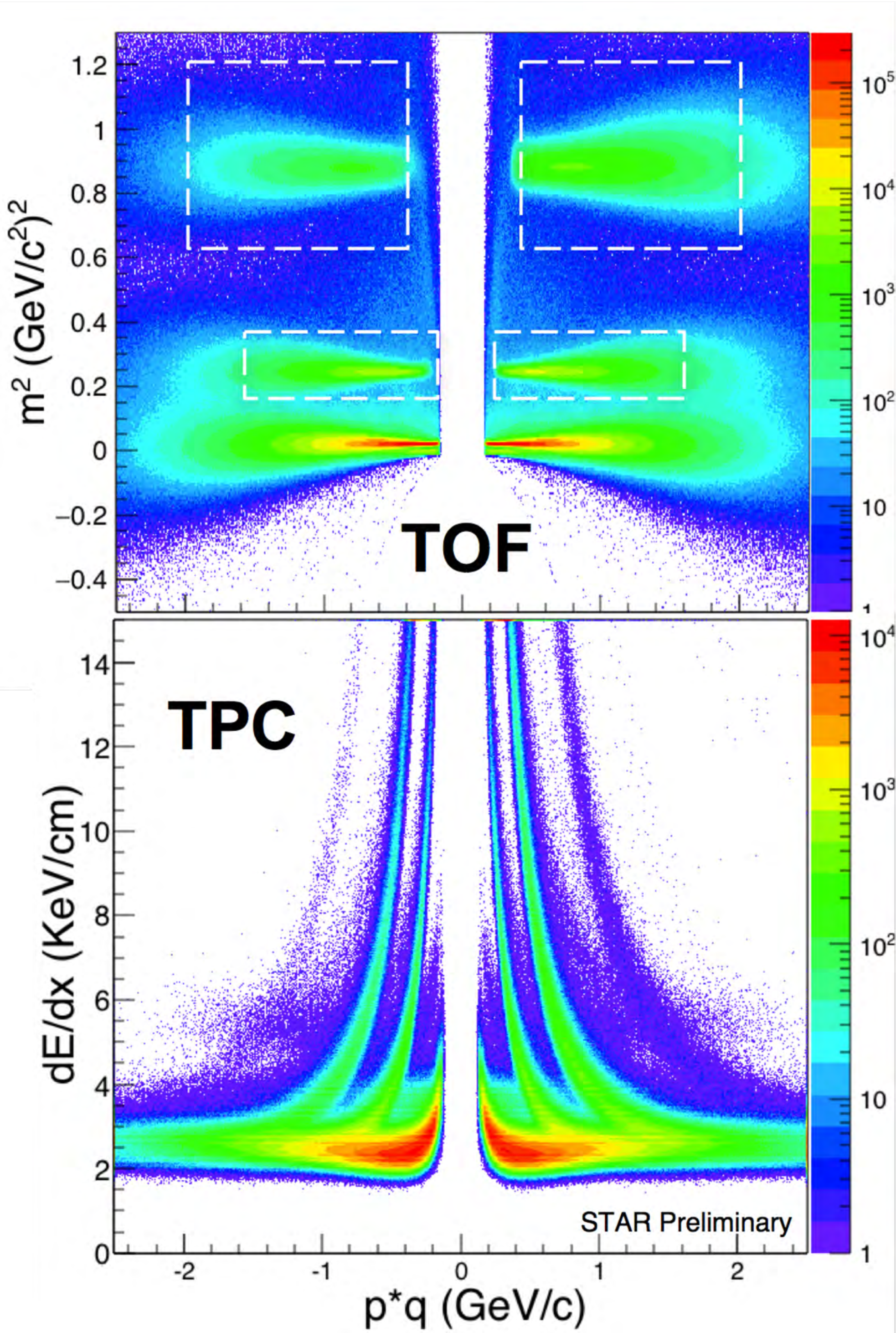}
        \end{minipage}
      \hspace{1cm}
  \begin{minipage}[c]{0.4\linewidth}
  \centering 
  \includegraphics[scale=0.4]{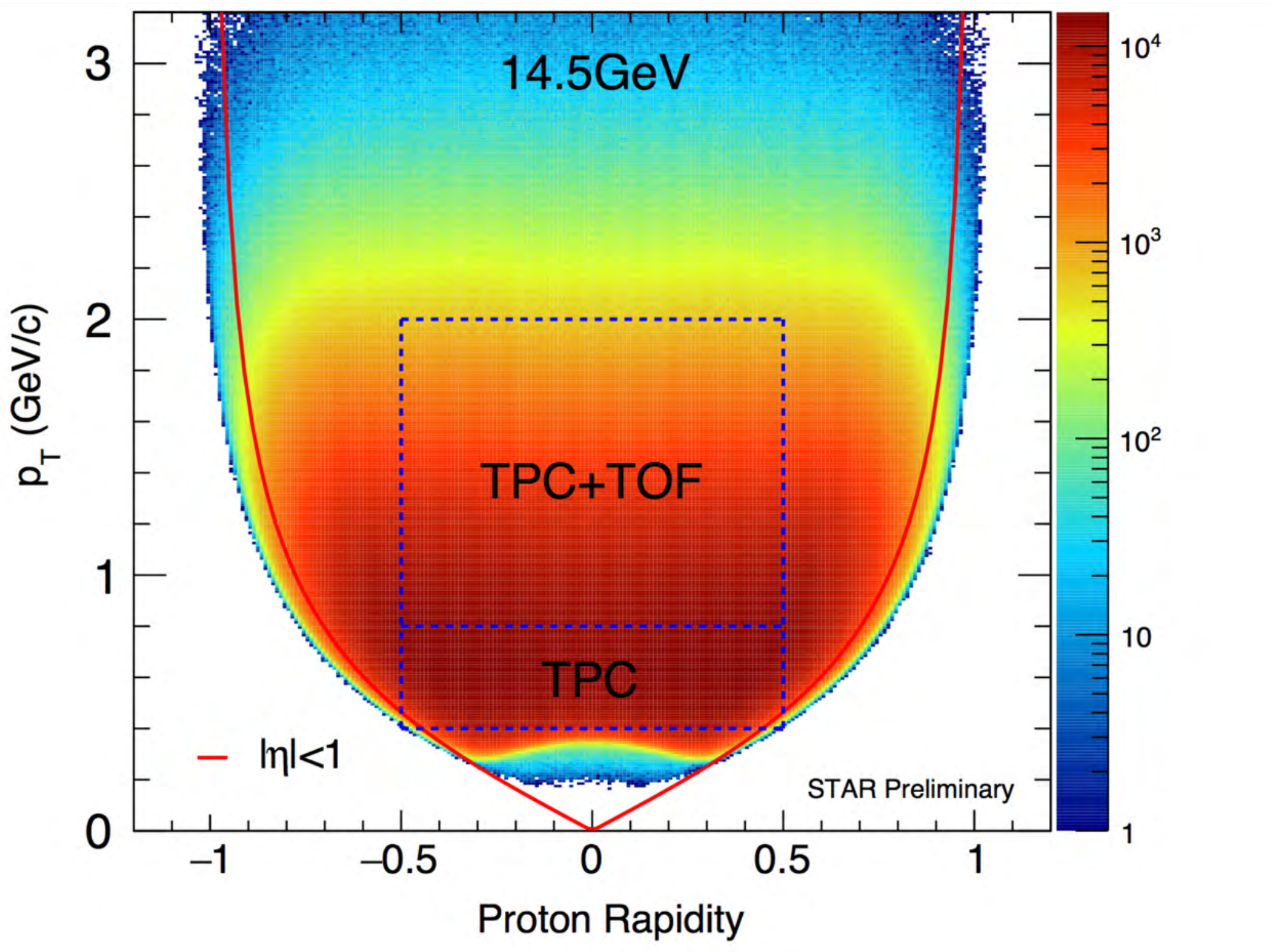}
          \end{minipage} 
         \caption{(Color online) (Left) Particle identification plot for the Time of Flight (ToF) : mass square versus rigidity (momentum times charge) and Time Projection Chamber (TPC): ionization energy loss versus rigidity. (Right) Proton phase space ($p_{T}$ vs. y) in Au+Au collisions at \sNN\ = 14.5 GeV measured by the STAR detector~\cite{Xuji_AGS_2016}.} \label{fig:PID_PS}
\end{figure*}
\begin{figure*}
\includegraphics[scale=0.65]{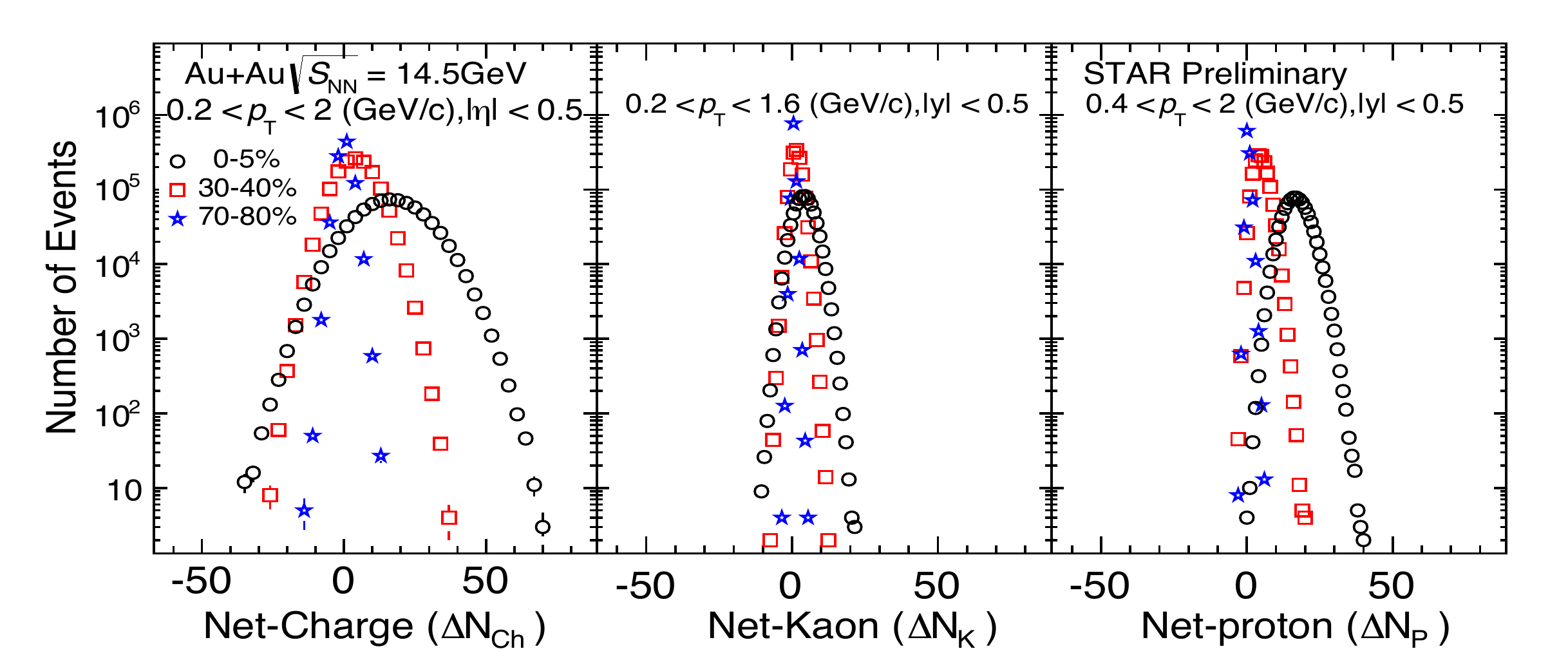}
\caption{Uncorrected raw event-by-event net-charge (left), net-kaon (middle) and net-proton (right) multiplicity distributions for Au+Au collisions at $\sqrt{s_{NN}}$ = 14.5 GeV for 0-5\% top central (black circles), 30-40\% central (red squares), and 70-80\% peripheral collisions (blue stars)~\cite{WWND2016_Xuj,2015_Jochen_QM}.}
\label{fig:distributions}
\end{figure*}
\section{Experimental Results}  \label{sec:results}
One of the main goals of the beam energy scan program at RHIC is to explore the phase structure of the hot dense nuclear matter created in the relativistic heavy-ion collisions, especially searching for the QCD critical point and mapping out the first order phase boundary. From the year of 2010 to 2014, RHIC has finished the first phase of BES program, in which two gold nuclei collide at \sNN\ = 7.7, 11.5, 14.5 (taken at 2014), 19.6, 27, 39, 62.4 and 200 GeV. The STAR experiment has published the energy dependence of cumulants (up to fourth order) of net-proton~\cite{2010_NetP_PRL,STAR_BES_PRL} and net-charge~\cite{netcharge_PRL} multiplicity distributions in Au+Au collisions at {\sNN}= 7.7, 11.5, 19.6, 27, 39, 62.4 and 200 GeV.  For net-proton analysis, the protons and anti-protons are identified with ionization energy loss in the Time Projection Chamber (TPC) of the STAR detector within the transverse momentum range $0.4<p_{T}<0.8$ GeV/c and at mid-rapidity $|y|<0.5$.  For the net-charge, the charged particles are measured within transverse momentum range $0.2<p_{T}<2$ GeV/c and pseudo-rapidity range $|\eta|<0.5$. 

Figure~\ref{fig:cumulants_energy} shows the energy dependence of cumulant ratios of net-proton and net-charge distributions of Au+Au collisions for two centralities (0-5\% and 70\%-80\%) at {\sNN}= 7.7, 11.5, 19.6, 27, 39, 62.4 and 200 GeV.  The Skellam (Poisson) expectations shown in the figure reflect a system of totally uncorrelated, statistically random particle production. It predicts the {\KV} and {\SD}/Skellam to be unity for Skellam expectations as well as in the hadron resonance gas model.  For the net-proton results, the most significant deviation of {\SD} and {\KV} from Skellam distribution is observed at 19.6 and 27 GeV for 0-5\% Au+Au collisions.  At energies above 39 GeV, the results are close to Skellam expectation.  As the statistical errors are large at low energies (7.7 and 11.5 GeV), more statistics is necessary to quantitatively understand the energy dependence of {\SD} and {\KV}. To understand the effects of baryon number conservation etc., UrQMD model calculations (a transport
model which does not include a CP) for 0-5\% are presented and the results show a monotonic decrease with decreasing beam energy. For more details on baseline comparison, one can see~\cite{QM2014_baseline}. For the net-charge results,  we did not observe non-monotonic behavior for {\SD} and {\KV} within current statistics. The expectations from negative binomial distribution can better describe the net-charge data than the Poisson (Skellam) distribution. More statistics is needed for the measurements of net-charge moments.

In the CPOD2014~\cite{2014_Luo_CPOD} and QM2015 conferences~\cite{QM_XFLUO, 2015_Jochen_QM},  the STAR experiment reported the preliminary results of net-proton fluctuations with wider transverse momentum coverage ($0.4<p_{T}<2$ GeV/c). In the new results, the $p_{T}$ range of  (anti-)protons are extended from $0.4<p_{T}<0.8$ to $0.4<p_{T}<2$ GeV/c. This is realized by using the Time of Flight (ToF) detector to identify the high $p_{T}$ ($0.8<p_{T}<2$ GeV/c) (anti-)protons. At low $p_{T}$ region ($0.4<p_{T}<0.8$ GeV/c), only Time Projection Chamber (TPC) is used to identify the (anti-)protons whereas the (anti-)protons at high $p_{T}$ ($0.8<p_{T}<2$ GeV/c) are jointly identified by TPC and ToF. Fig.~\ref{fig:PID_PS} left show the particle identification (PID) plot for TPC and ToF detector. The white dashed boxes in the ToF PID plot denote the protons (upper) and kaons (lower) PID cuts region, respectively.  Fig.~\ref{fig:PID_PS} right show the proton phase space in Au+Au collisions at \sNN\ = 14.5 GeV measured by the STAR experiment. The protons and anti-protons in the regions covered by the blue dashed boxes are used in the net-proton fluctuation analysis. Figure~\ref{fig:distributions} shows the uncorrected event-by-event net-charge, net-kaon and net-proton multiplicity distributions in three centralities (0-5\%, 30-40\% and 70-80\%) for Au+Au collisions at $\sqrt{s_{NN}}$ = 14.5 GeV. Those raw distributions from a wide centrality bin can not be used to calculate the various order cumulant directly due to the effects of finite efficiency and volume variation. However, there are some theoretically works about using those distributions to extract criticality~\cite{kenji,kenji_netp,PBM_netq,PBM_netpdis}. The shape of net-particle multiplicity distributions for different centralities are different. The standard deviation $\sigma$ of the net-particle distributions get bigger for central collisions than peripheral and mid-central. We also observed that the net-charge multiplicity distributions have the largest standard deviation, $\sigma$, comparing with the net-proton and net-kaon distributions at fixed centrality. As shown in Eq. (\ref{eq:err_cum}), the statistical errors of the $r^{th}$ order cumulants are proportional to the $r^{th}$ power of the standard deviation ($\sigma^{r}$). This indicates that with the same number of events, the net-charge fluctuations measurements will have much lager statistical errors than the results of net-proton and net-kaon fluctuations. Detailed discussions about the efficiency correction and error estimation can be found in~\cite{technique,Unified_Errors}.

\begin{figure*}[htbp]
\begin{center}
%\hspace{-0.1cm}
%\vspace{-1.1cm}
\includegraphics[scale=0.7]{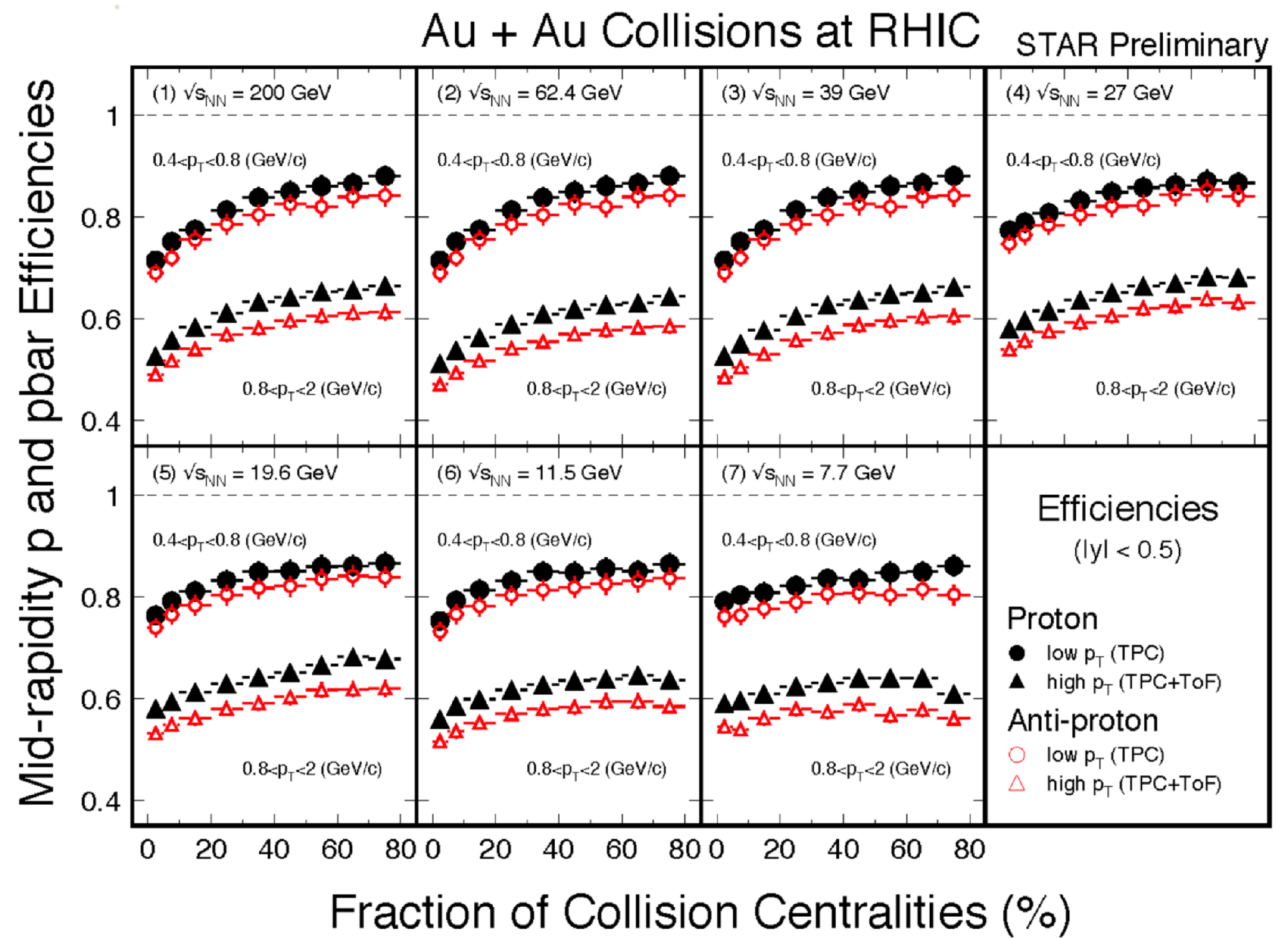}
\vspace{-0.2cm}
\caption[]{(Color online) Centrality dependence of mid-rapidity detecting efficiency for protons and anti-protons in two $p_{T}$ ranges, $0.4<p_{T}<0.8$ GeV/c (circles) and  $0.8<p_{T}<2$ GeV/c (triangles), in Au+Au collisions at {\sNN}=7.7 , 11.5, 19.6, 27, 39, 62.4 and 200 GeV. Black solid points represent efficiency of protons and red empty points are the efficiency of anti-protons~\cite{2014_Luo_CPOD}.  } \label{fig:peff_BES}
\end{center}
\end{figure*}
\begin{figure*}[htbp]
\begin{center}
\hspace{-0.3cm}
\includegraphics[scale=0.9]{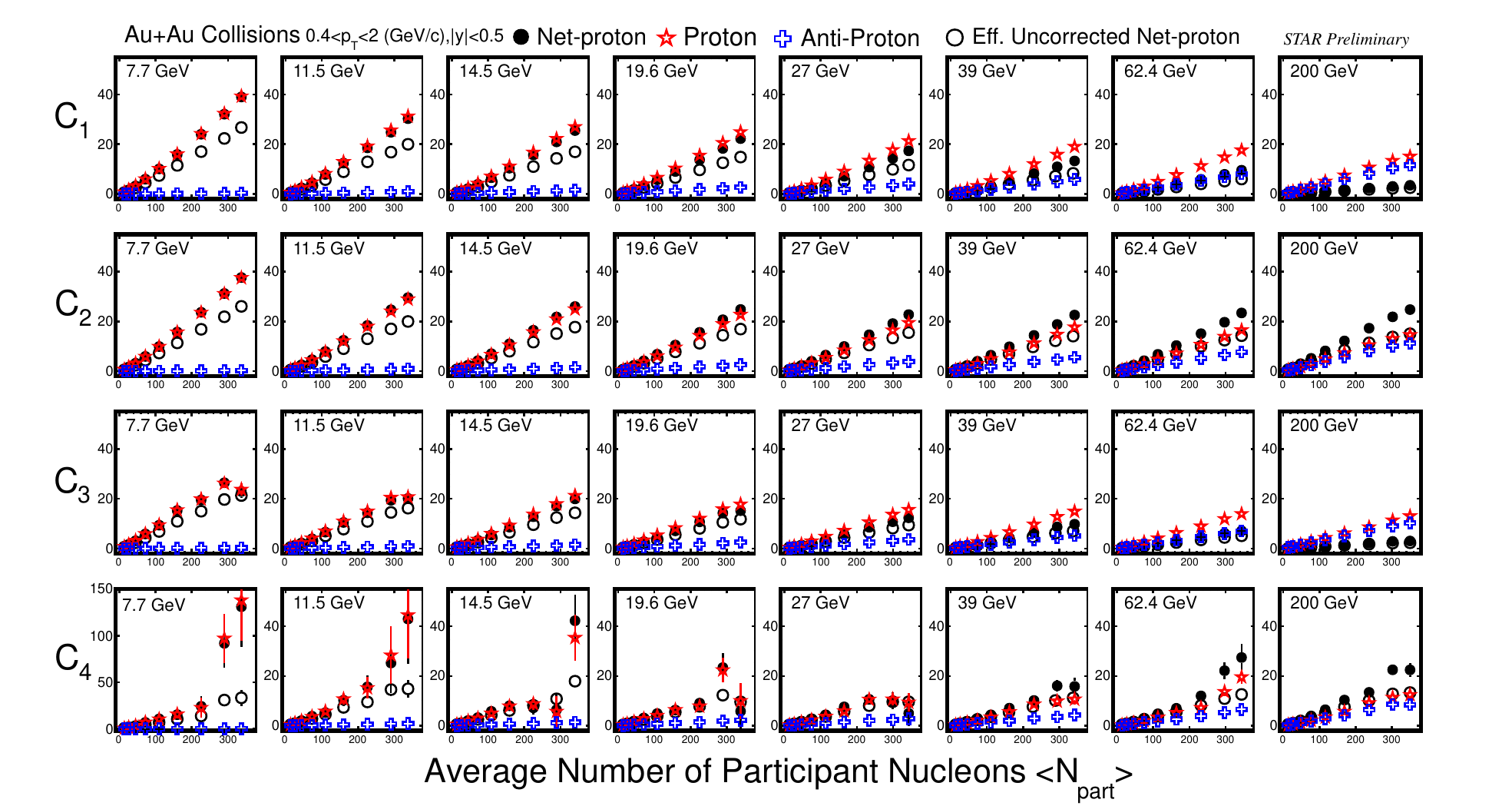}
\caption[]{(Color online) Centrality dependence of various order efficiency corrected cumulants  ($C_{1}\sim C_{4}$) for net-proton, proton and anti-proton distributions in 
Au+Au collisions at {\sNN}=7.7 , 11.5, 19.6, 27, 39, 62.4 and 200 GeV. Error bars in the figure are statistical errors only.   Blue empty circles represent the efficiency uncorrected cumulants of net-proton distributions~\cite{2014_Luo_CPOD,Xuji_AGS_2016}. } \label{fig:Cumulants}
\end{center}
\end{figure*}

Figure \ref{fig:peff_BES} shows the centrality dependence of detection efficiency for (anti-)protons in two $p_{T}$ ranges ( $0.4<p_{T}<0.8$ and $0.8<p_{T}<2$ GeV/c) in Au+Au collisions at {\sNN}=7.7 , 11.5, 19.6, 27, 39, 62.4 and 200 GeV. The efficiency of protons and anti-protons at high $p_{T}$, $0.8<p_{T}<2$ GeV/c is smaller than that of low $p_{T}$, $0.4<p_{T}<0.8$ GeV/c. This is because, besides the time projection chamber (TPC),  the time of flight (ToF) detector is used to identify the high $p_{T}$ (anti-)protons and the ToF matching efficiency is introduced in addition to the TPC tracking/acceptance efficiencies. While at low $p_{T}$, only TPC is used to identify protons and anti-protons. Thus, the average efficiency for protons or anti-protons at low $p_{T}$ and high $p_{T}$ can be calculated as:
\begin{equation}< \varepsilon  >  = \frac{{\int\limits_{{p_{{T_1}}}}^{{p_{{T_2}}}} {\varepsilon ({p_T})f({p_T})d{p_T}} }}{{\int\limits_{{p_{{T_1}}}}^{{p_{{T_2}}}} {f({p_T})d{p_T}} }}
\end{equation}
where the $\varepsilon ({p_T}) = {\varepsilon _{tpc}}({p_T})$ for $0.4<p_{T}<0.8$ GeV/c and $\varepsilon ({p_T}) = {\varepsilon _{tpc}}({p_T}){\varepsilon _{tof}}({p_T})$ for $0.8<p_{T}<2$ GeV/c.  The efficiency corrected $p_T$ distribution function $f({p_T})$ is defined as $f({p_T}) = dN/d{p_T}$.
The TPC efficiency ($\varepsilon _{tpc}({p_T})$) of protons or anti-protons are obtained from the so-called embedding simulation techniques and the ToF matching efficiency ($\varepsilon _{tof}({p_T})$) can be calculated from the real data. The average efficiencies of protons and anti-protons have centrality (multiplicity) dependence and increase from central to peripheral collisions for all energies.  Due to material absorption of anti-protons in the detector, the efficiencies of anti-protons are always slightly lower than protons. 

\begin{figure*}[htbp]
\centering
\hspace{-0.5cm}
\includegraphics[scale=0.8]{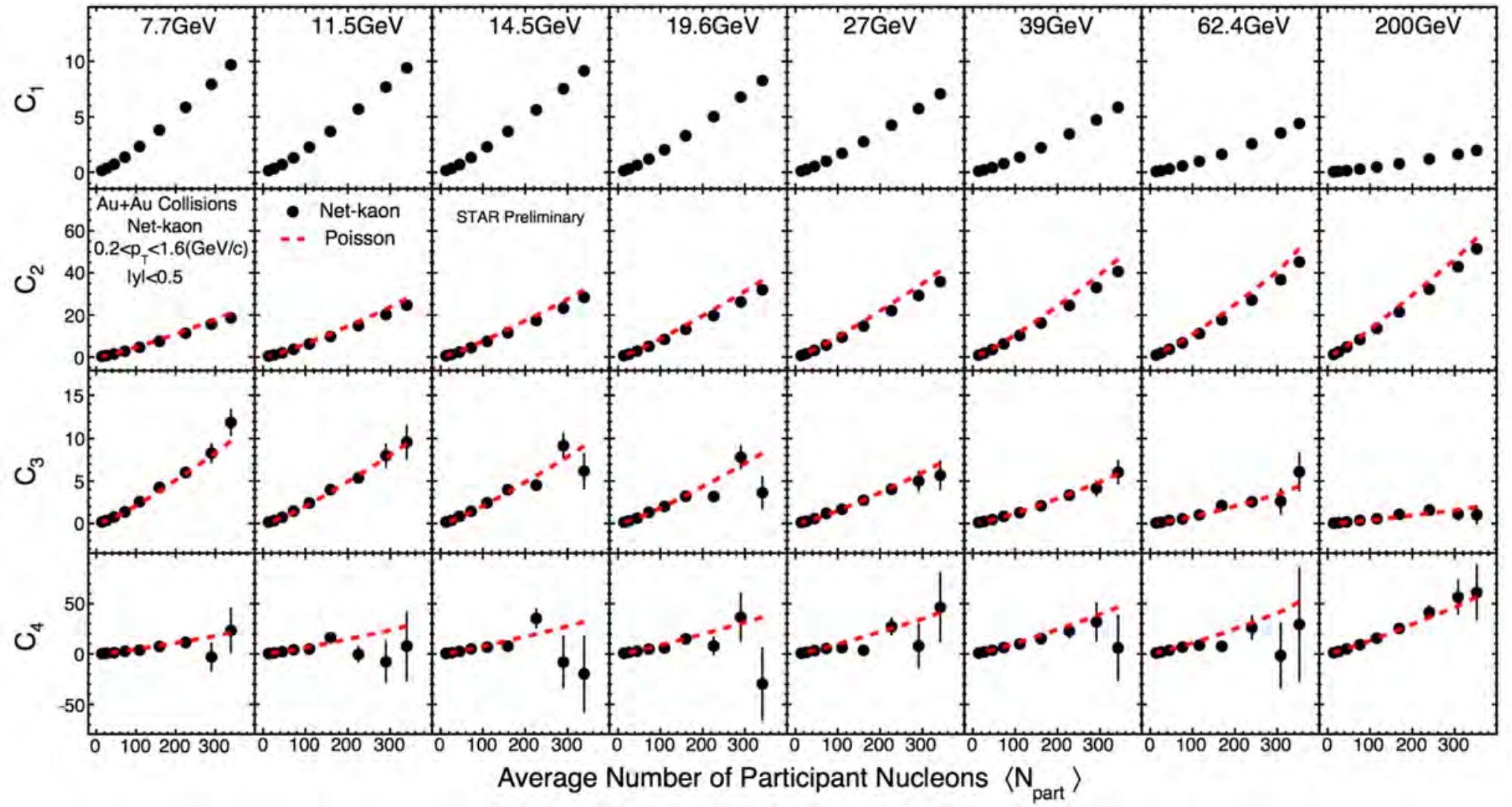}
\caption{Centrality dependence of cumulants $(C_{1}, C_{2}, C_{3}$, and $C_{4})$ of net-kaon multiplicity distributions for Au+Au collisions at $\sqrt{s_{\rm NN}}$ = 7.7, 11.5, 14.5, 19.6, 27, 39, 62.4, and 200GeV~\cite{SQM2016_Xuj}. The Poisson expectations are denoted as dotted lines. The error bars are statistical errors.}
\label{cumu_egy}
\end{figure*}

\begin{figure*}
\includegraphics[width=3.4in]{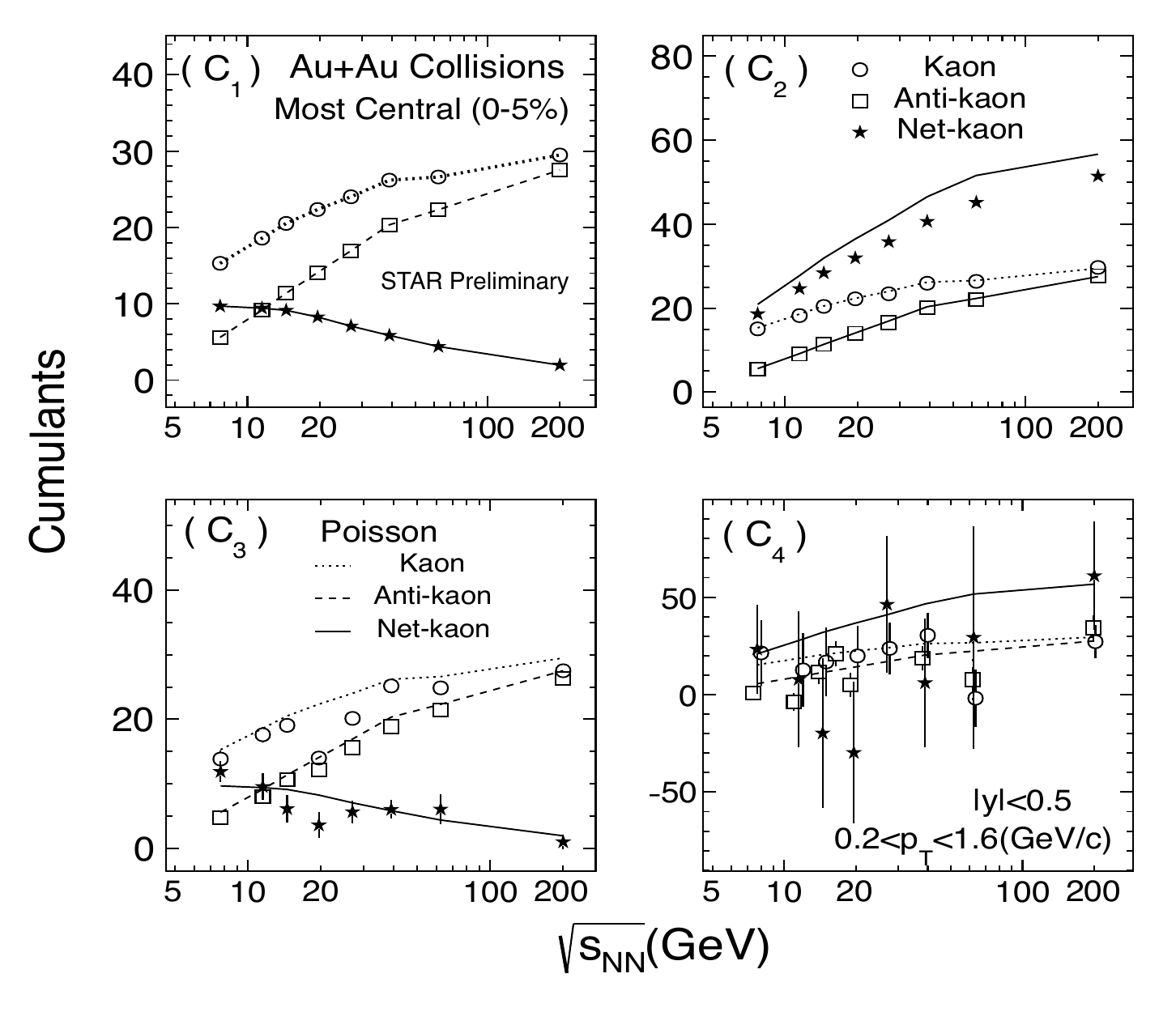}
\includegraphics[width=3.3in]{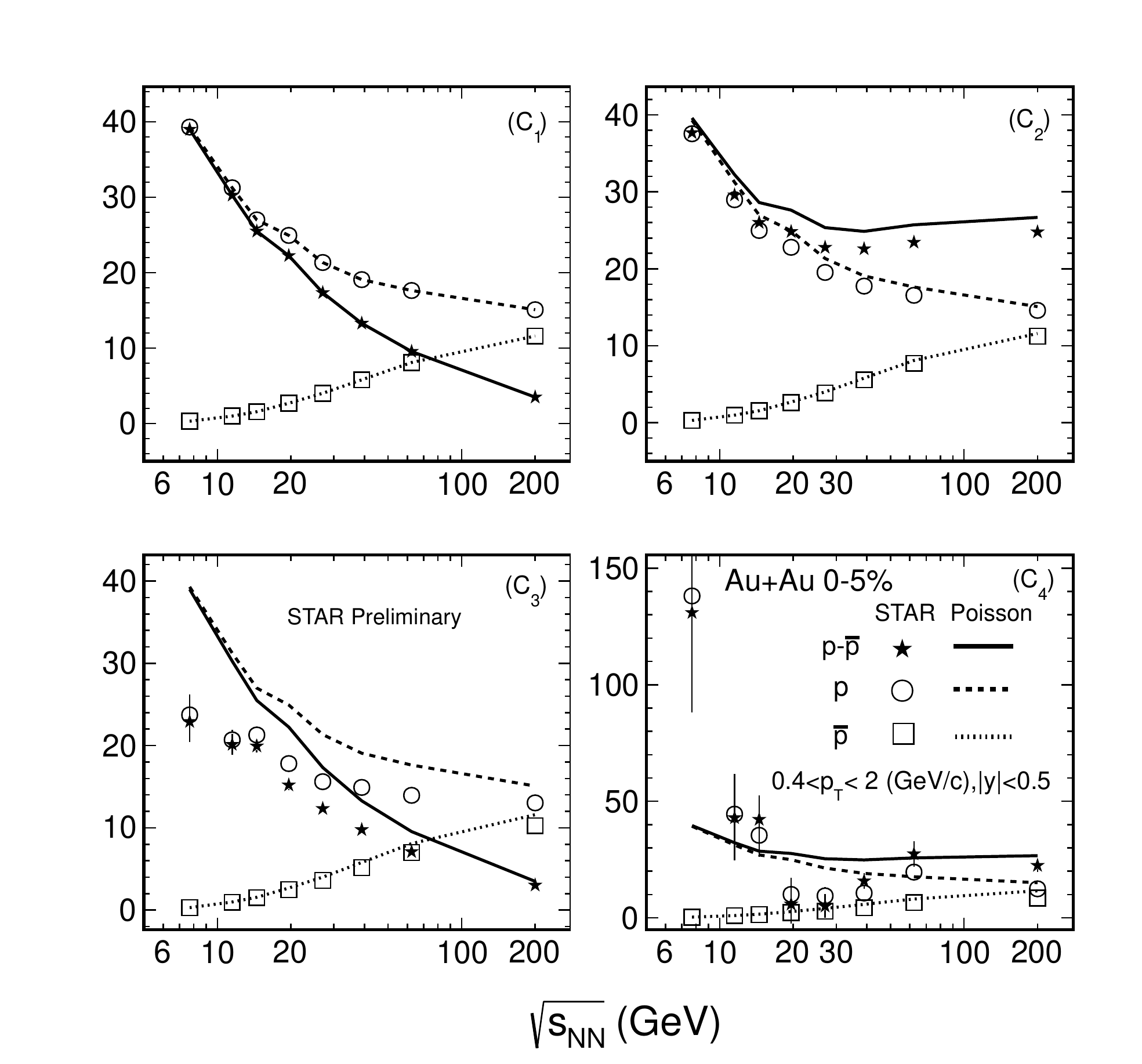}
\caption{ (Left) Energy dependence of cumulants ($C_{1}\sim C_{4}$) for net-kaon, $K^{+}$ and $K^{-}$ multiplicity distributions in 0-5\% most central Au+Au collisions. (Right): Energy dependence of cumulants ($C_{1}\sim C_{4}$) for net-proton, proton, and anti-proton multiplicity distributions in 0-5\% most central Au+Au collisions~\cite{2014_Luo_CPOD,WWND2016_Xuj}.}
\label{cumu_egy2}
\end{figure*}

Figure \ref{fig:Cumulants} shows the centrality dependence of efficiency corrected cumulants ($C_{1}\sim C_{4}$) of net-proton, proton and anti-proton distributions in Au+Au collisions at {\sNN}=7.7, 11.5, 14.5, 19.6, 27, 39, 62.4 and 200 GeV. The protons and anti-protons are measured within transverse momentum $0.4<p_{T}<2$ GeV/c and at mid-rapidity ($|y|<0.5$). At high energies, the cumulants (up to fourth order) of net-proton, proton and anti-proton distributions show a linear dependence on the average number of participant nucleons ($\la N_{part} \ra$).  This is consistent with the additive properties of the cumulants that the system consists of many multi-independent emission sources of protons and anti-protons and those emission sources are linear dependent on the system volume (centralities).  The proton cumulants are always larger than the anti-proton cumulants and the difference between proton and anti-proton cumulants are larger in low energies than high energies. The cumulants of net-proton distributions closely follow the proton cumulants when the colliding energy decreases. These observations can be explained as the interplay between the baryon stopping and pair production of protons and anti-protons. At high energies,  protons and anti-protons mainly come from the pair production and the number of protons and anti-protons are very similar. At low energies,  the production of protons is dominated by baryon stopping and the number of protons is much larger than the number of anti-protons. The efficiency corrected fourth order net-proton and proton cumulants ($C_{4}$)  of 7.7 and 11.5 GeV significantly increase in the $0\sim5\%$ and $5\sim10\%$ centrality bins with respect to the efficiency uncorrected results. It means the efficiency corrections are big effects, especially for the high order cumulants.  Furthermore, the efficiency correction not only affects the values but also lead to increasing of the statistical errors for the various order cumulants, as $error(C_{n})\sim \sigma^{n}/\varepsilon^{\alpha}$, where the $\sigma$ in numerator is the standard deviation of the particle distributions and the denominator $\varepsilon$ is the efficiency number,  $\alpha$ is a positive real number~\cite{Unified_Errors}.

\begin{figure*}
\includegraphics[width=2.2in]{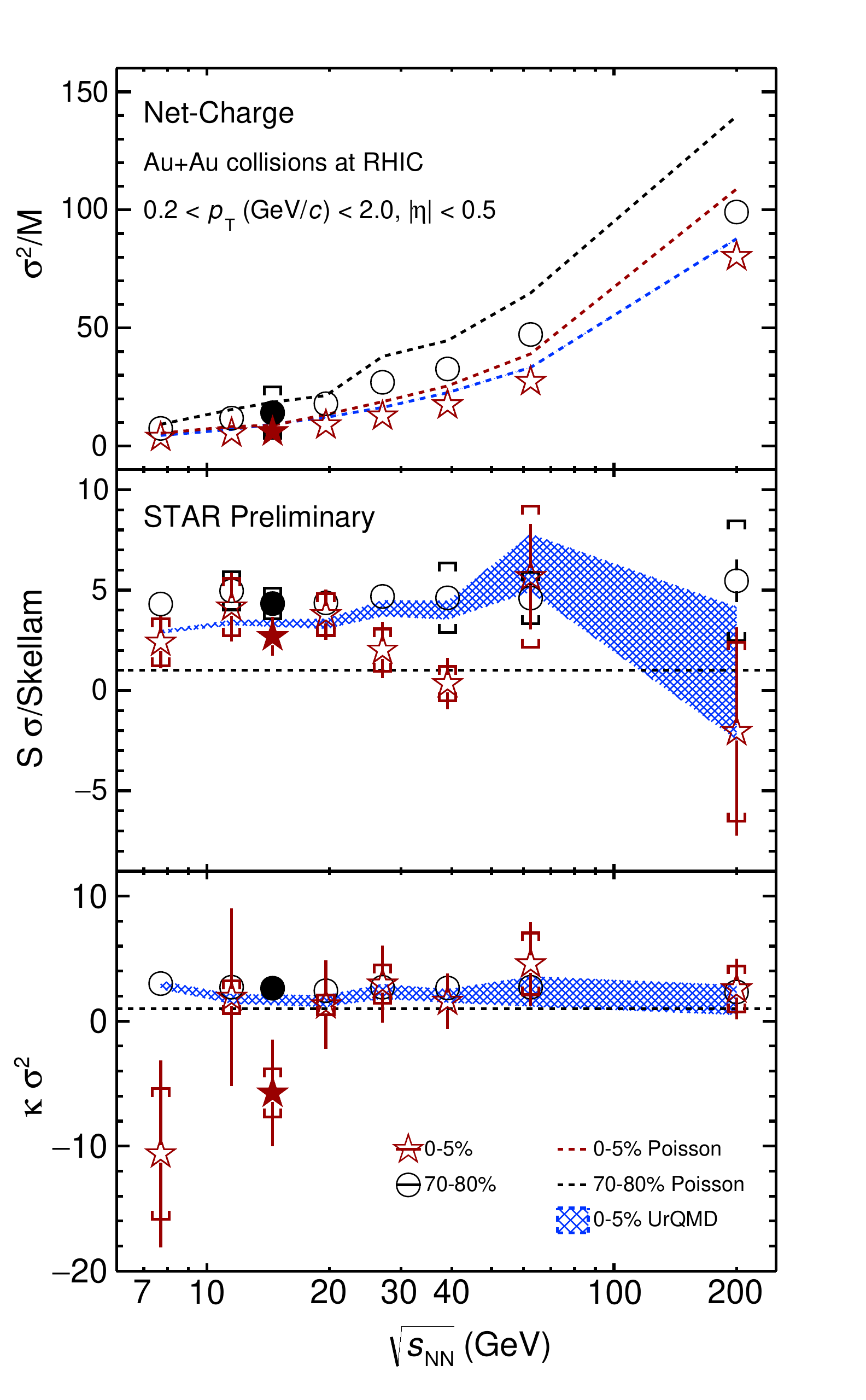}
\includegraphics[width=2.2in]{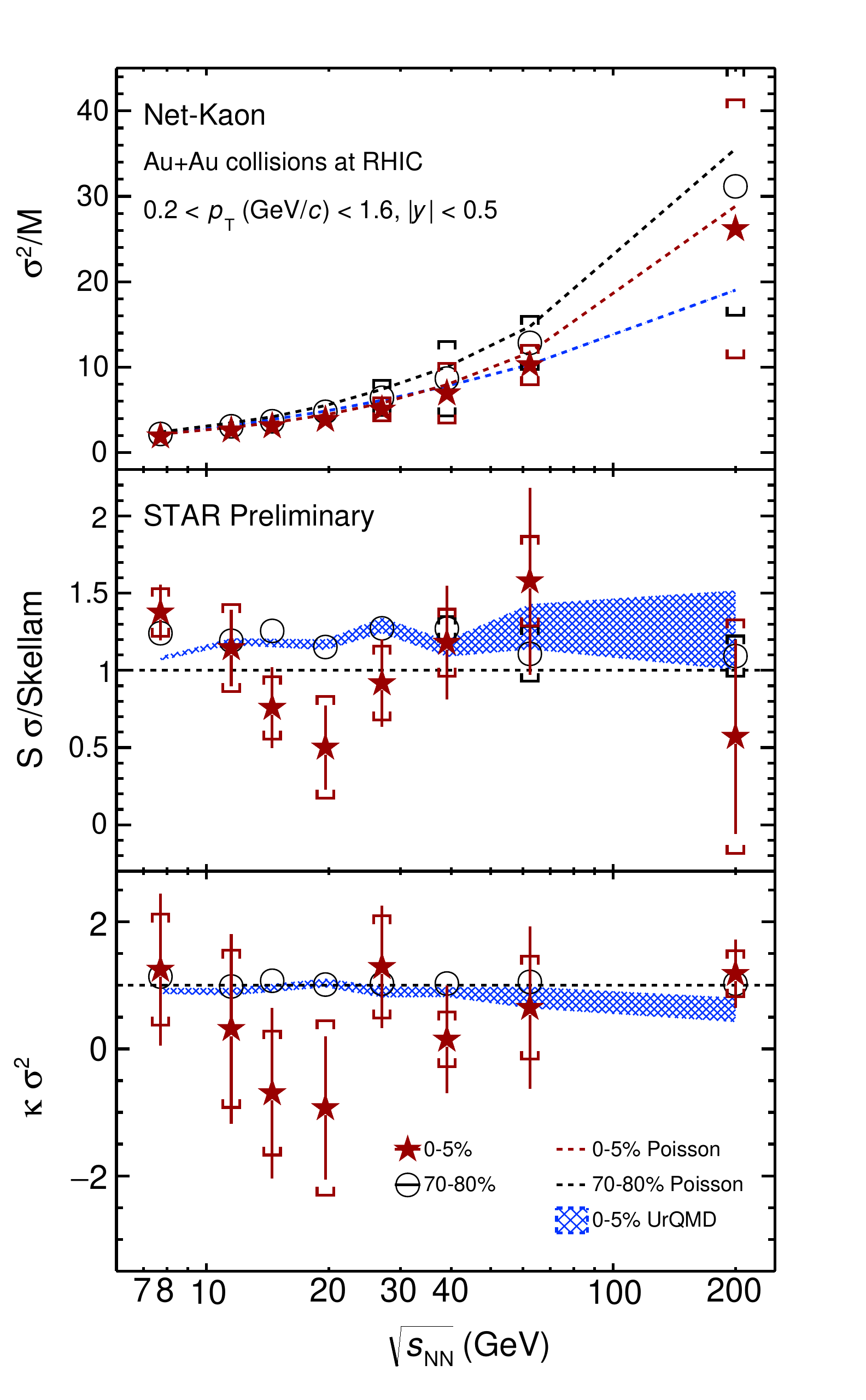}
\includegraphics[width=2.21in]{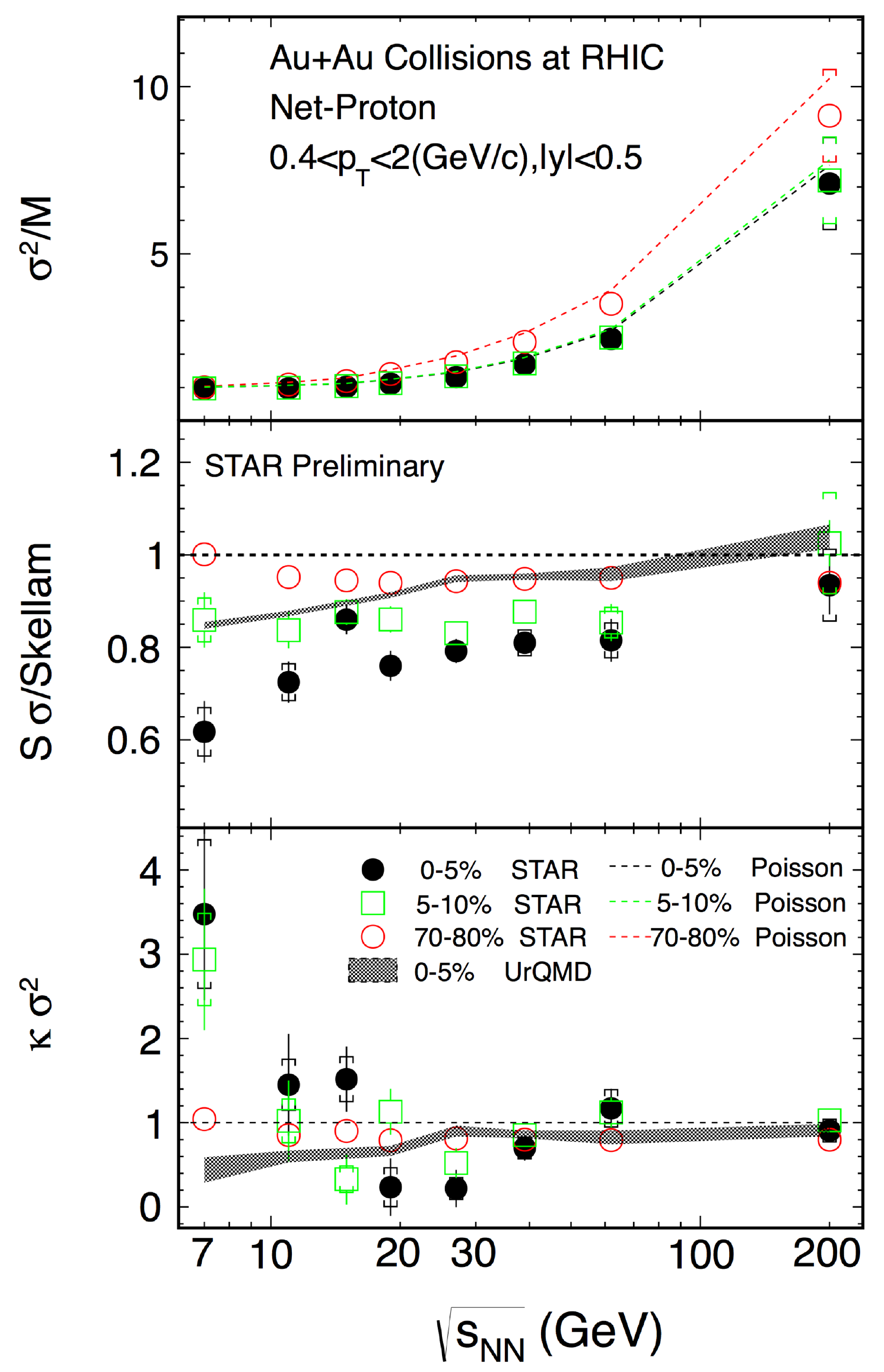}
\caption{ Energy dependence of cumulant ratios ($\sigma^{2}/M$, $S\sigma/$Skellam, $\kappa\sigma^2$) of net-charge, net-kaon and net-proton multiplicity distributions for top 0-5\% , 5-10\% central (green squares), and 70-80\% peripheral collisions. The Poisson expectations are denoted as dotted lines and UrQMD calculations are shown as bands. The statistical and systematical error are shown in bars and brackets, respectively~\cite{WWND2016_Xuj,2015_Jochen_QM, 2014_Luo_CPOD, QM_XFLUO}.} 
\label{egyDependence}
\end{figure*}
The STAR Collaboration reported preliminary results of cumulants of net-kaon distributions in Au+Au collisions at {\sNN}= 7.7, 11.5, 14.5, 19.6, 27, 39, 62.4 and 200 GeV in the QM2015 conference~\cite{2015_Jochen_QM}. The net-kaon fluctuations is used to approximate the fluctuations of net-strangeness,  a conserved charge in strong interaction. The susceptibilities of net-strangeness can be computed in Lattice QCD.  The $K^{+}$ and $K^{-}$ are measured with transverse momentum $0.2<p_{T}<1.6$ GeV/c and at mid-rapidity $|y|<0.5$. At low $p_T$ region ($0.2<p_{T}<0.4$ GeV/c), the charged kaons are identified by TPC only whereas at high $p_T$ ($0.4<p_T<2$ GeV/c), ToF is also used in addition with TPC. To avoid auto-correlation, the collision centrality is determined by measured charged particles within $|\eta|<1$ excluding charged kaons. Fig.~\ref{cumu_egy} shows the efficiency corrected centrality dependence of cumulants ($C_{1}\sim C_{4}$) of net-kaon multiplicity distributions in Au+Au collisions at $\sqrt{s_{\rm NN}}$ =7.7$\sim$200 GeV. The red dashed lines represent the Poisson expectations, where the probability distributions of the $K^{+}$ and $K^{-}$ are assumed to be the independent Poisson distributions. In general, various order cumulants show a linear variation with the averaged number of participant nucleons ($\la N_{part} \ra$). The variance are systematically below the Poisson expectations, especially at high energies. It means that the $K^{+}$ and $K^{-}$ are correlated with each other due to the pair productions. However, the $C_{3}$ and $C_{4}$ are consistent with Poisson expectation within uncertainties.  The large uncertainties observed in the $C_{3}$ and $C_{4}$ are due to the low detection efficiency of kaons ($\sim 40\%$). 

\begin{figure*}[htb]
\hspace{-1cm}
    \includegraphics[scale=0.6]{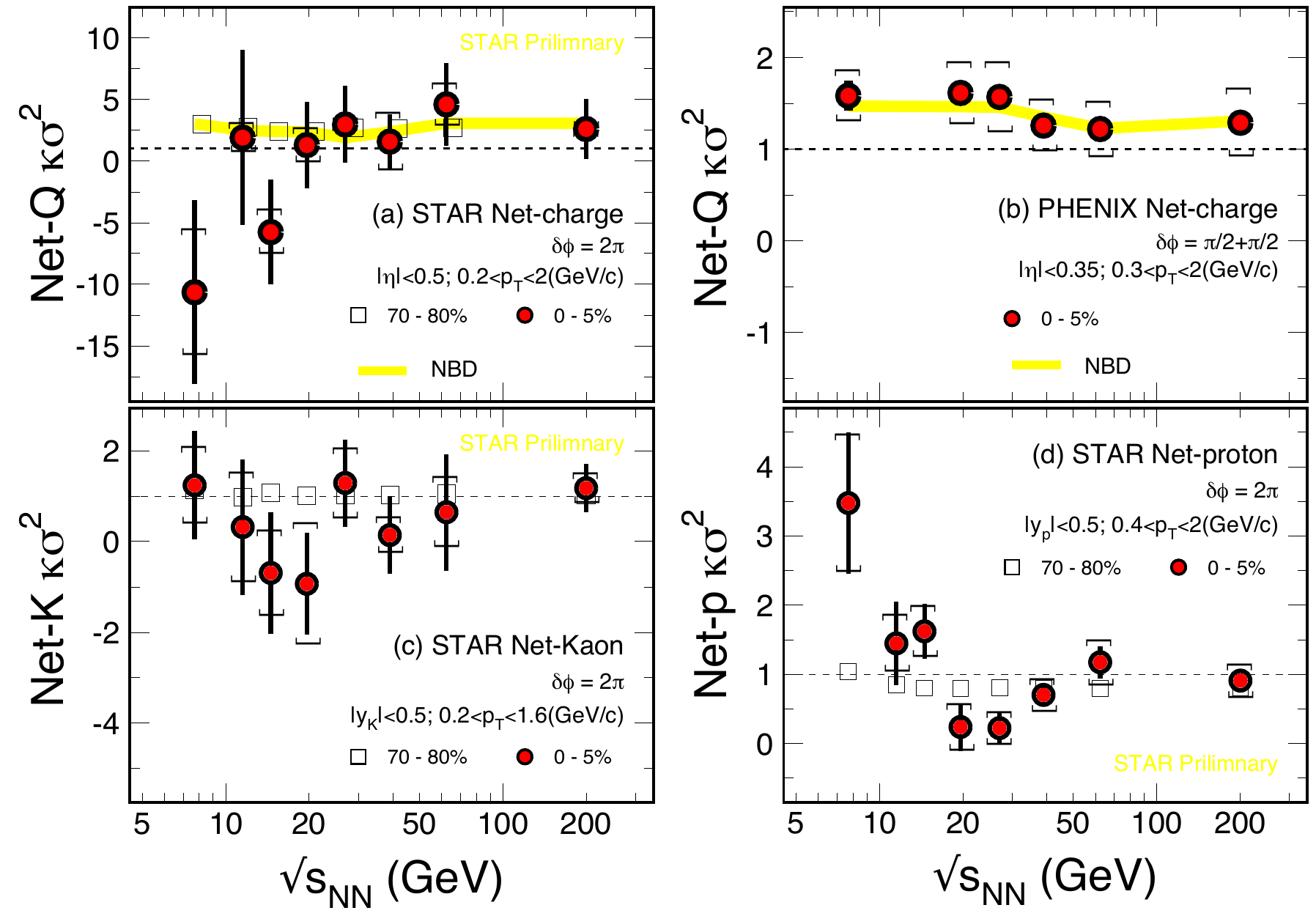}
       \caption{(Color online) The STAR measured energy dependence of {\KV} of net-proton, net-charge (top left) and net-kaon distributions in Au+Au collisions at {\sNN}= 7.7, 11.5, 14.5, 19.6, 27, 39, 62.4 and 200 GeV. The net-charge fluctuations measured by the PHENIX experiment in Au+Au collisions at {\sNN}= 7.7, 19.6, 27, 39, 62.4 and 200 GeV are shown in top right panel.The statistical and systematical error are shown in bars and brackets, respectively~\cite{WWND2016_Xuj,2015_Jochen_QM, 2014_Luo_CPOD, QM_XFLUO, NetQ_PHENIX}. } \label{fig:KV_energy}
\end{figure*}

Figure~\ref{cumu_egy2} left shows the energy dependence of cumulants ($C_1 \sim C_4$) for net-kaon, $K^{+}$, and $K^{-}$ multiplicity distributions in Au+Au collisions measured by the STAR experiment. The mean values of the $K^{+}$ and $K^{-}$ show monotonic decreasing trends when the energy decrease. Furthermore, the mean values of $K^{+}$ is always above $K^{-}$, and the difference between these two values are bigger at lower energies. These two observations are due to interplay of the pair and associate production for $K^{+}$, and $K^{-}$ as a function of collisions energies. In addition to the pair production of $K^{+}$ and $K^{-}$, the $K^{+}$ is also produced by the associate production with $\Lambda$ hyperon and the fraction of $K^{+}$ from associate production is lager at low energies than at high energies. It also leads to the increasing of the net-kaon mean values when decreasing the energies. The corresponding Poisson expectations are also plotted as different lines for comparison. In general, the cumulants of $K^{+}$, and $K^{-}$ distributions are consistent with the Poisson baseline within uncertainties. Due to the correlation between $K^{+}$ and $K^{-}$ , the variance of the net-kaon distributions are smaller than its Poisson expectations, in which one assumes the independent of the $K^{+}$ and $K^{-}$. The higher order net-kaon cumulants are consistent with Poisson expectations within uncertainties. Fig.~\ref{cumu_egy2} right shows the energy dependence of cumulants ($C_1 \sim C_4$) of net-proton, proton and anti-proton multiplicity distributions in Au+Au collisions measured by the STAR experiment. The mean values of protons and net-protons show monotonic increasing trends when decreasing the colliding energy whereas the mean values of anti-protons show opposite trend. Those can be understood in terms of the interplay between the baryon stopping and pair production for proton and anti-proton as a function of collision energy. At low energies, the baryon stopping becomes more dominate while at high energies, the pair production is the main production mechanism of the proton and anti-protons. In the figure, it also shows the comparison between the cumulants of net-proton, proton and anti-proton distributions and the corresponding Poisson expectations. We found that the higher the order of the cumulant, the larger the deviations from the Poisson expectation for the net-proton and proton.  Largest deviations are found for $C_{4}$ at 7.7 GeV.  The cumulants of anti-proton distributions can be described by the Poisson expectations very well. More baselines discussions from Hadronic Resonance Gas model, transport model UrQMD, binomial and negative binomial have been also discussed. 

Figure~\ref{egyDependence} shows the energy dependence of cumulant ratios ($\sigma^{2}/M$, $S\sigma/$Skellam, $\kappa\sigma^2$) of net-charge~\cite{WWND2016_Xuj,2015_Jochen_QM}, net-kaon~\cite{WWND2016_Xuj,SQM2016_Xuj}, and net-proton~\cite{2014_Luo_CPOD} multiplicity distributions in Au+Au collisions measured by the STAR experiment. The black solid circles on the left figure represent the results from Au+Au collisions at $\sqrt{s_{NN}}$ = 14.5 GeV, which is taken in the year 2014 and added into the trend of the published net-charge results~\cite{netcharge_PRL} (open stars). The bands are the results from UrQMD calculations without including the critical physics. The Poisson expectations are displayed as dashed lines. The {\SD} values have been normalized by the Poisson expectations, the Skellam distributions. Thus, the Poisson expectations for both the $S\sigma/$Skellam and {\KV} are unity. It can be found that the {\VM} of net-charge, net-kaon and net-proton monotonically increase when increasing the collision energy. On the other hand, both the $S\sigma/$Skellam and $\kappa\sigma^2$  show weak energy dependence for net-charge and net-kaon measurements. No significant deviations from the Poisson expectations and UrQMD calculation are observed for net-charge and net-kaon cumulant ratios $S\sigma/$Skellam and {\KV} within uncertainties. We observe a clear non-monotonic energy dependence of net-proton $\kappa\sigma^2$ in top 0-5\% central Au+Au collisions. The 0-5\% net-proton {\KV} values are close to unity for energies above 39 GeV and show large deviations below unity around 19.6 and 27 GeV, and then increasing above unity below 19.6 GeV. The UrQMD calculations of net-proton {\KV} displaying a strong suppression below unity at lower energies is due to the effects of baryon number conservation. However, this suppression are not observed at low energies in the STAR data. 

\begin{figure*}[htb]
\centering 
    \includegraphics[scale=0.37]{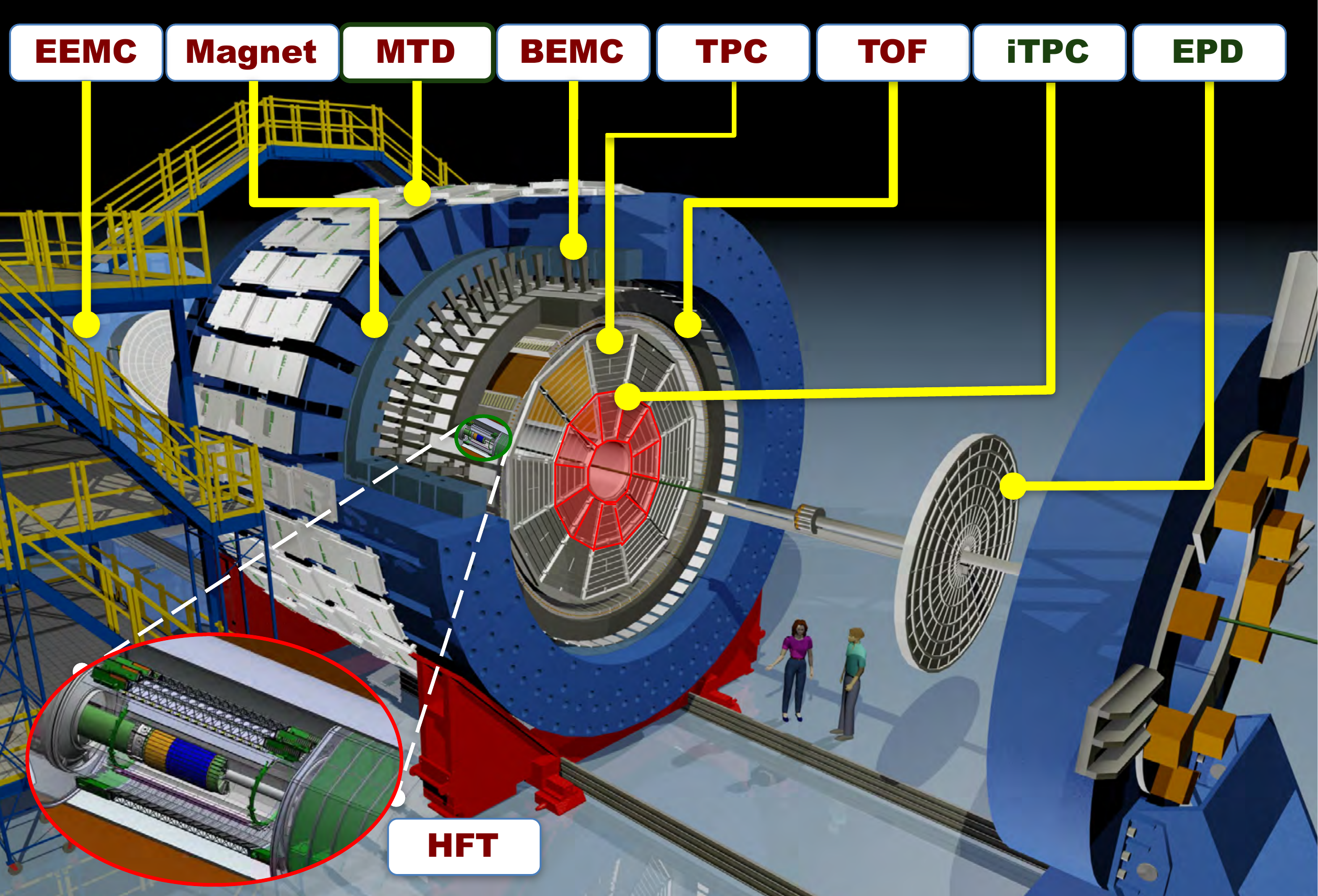}
    \hspace{0.5cm}
      \includegraphics[scale=0.7]{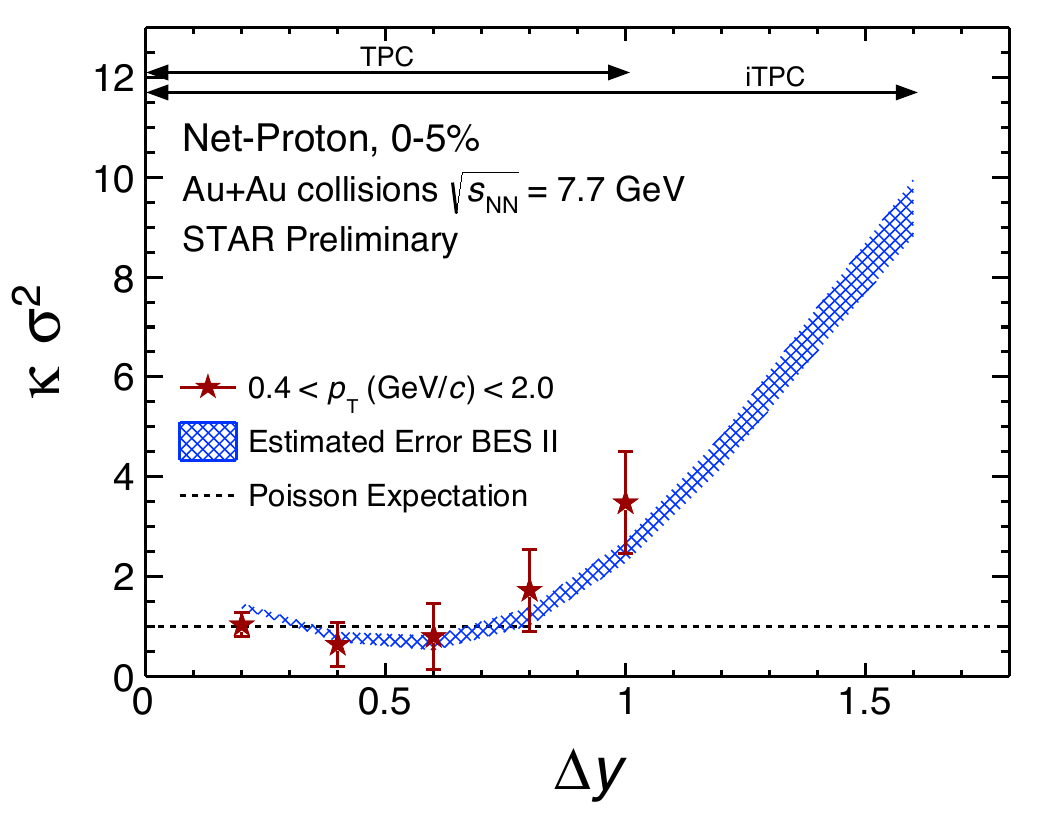}
      \caption{ (Color online) (Left) iTPC and EPD upgrades of the STAR detector for the second phase of beam energy scan at RHIC. Right: Rapidity coverage dependence of the {\KV} of net-proton distribution in 0-5\% central Au+Au collisions at \sNN=7.7 GeV. The blue band shows the expecting trend and statistical error for net-proton {\KV} at BES-II. For this analysis, the rapidity coverage can be extend to $|y|<0.8$ with iTPC upgrades~\cite{BESII_WhitePaper}. } \label{fig:BESII}
\end{figure*}

Figure~\ref{fig:KV_energy} panels (a), (c), (d) show the energy dependence of {\KV} of net-charge, net-kaon and net-proton multiplicity distributions in Au+Au collisions measured by the STAR experiment for two centralities (0-5\% and 70\%-80\%) at {\sNN}= 7.7, 11.5, 14.5, 19.6, 27, 39, 62.4 and 200 GeV.  The {\KV} of net-charge distributions in Au+Au collisions {\sNN}= 7.7, 19.6, 27, 39, 62.4 and 200 GeV measured by the PHENIX experiment~\cite{NetQ_PHENIX} are shown in the panel (b) of Fig.~\ref{fig:KV_energy}. It is observed that the {\KV} of the net-charge and net-kaon distributions measured by the STAR experiment are with larger statistical errors than the errors of net-proton {\KV}. This is because the statistical errors of {\KV} depend on the width ($\sigma$) of the multiplicity distributions ($error({\KV}) \propto \sigma^{n-2}/(\sqrt{N} \epsilon^{n})$) and the net-charge distributions are much wider than the net-proton and net-kaon. On the other hand, due to decay of kaons, the efficiency of kaon ($\sim40\%$) is much lower than proton ($\sim80\%$), this also leads to larger statistical errors for net-kaon fluctuations. For the STAR net-charge and net-kaon results,  we did not observe non-monotonic behavior for {\KV} within current statistics. The Poisson expectations shown as dashed lines in the figure with unity value reflect a system of totally uncorrelated, statistically random particle production. It predicts the {\KV} to be unity for Poisson expectations as well as in the hadron resonance gas model. However, the expectations from negative binomial distribution can better describe the net-charge and net-kaon data than the Poisson expectations. The PHENIX net-charge {\KV} are with smaller errors than the results measured by the STAR experiment. This is because the PHENIX detector has much smaller acceptance than the STAR detector and thus the width of the net-charge distributions measured by the PHENIX experiment is much narrower.  We observe a clear non-monotonic energy dependence for net-proton {\KV} in the most central ( 0-5\%) Au+Au collisions with a minimum around 19.6 and 27 GeV.  This non-monotonic behavior cannot be described by various model calculations without including CP physics~\cite{Xuji_PRC,ShuHe_PLB}.  Another model calculation with volume fluctuations also failed to describe this increasing at low energies~\cite{bzdak2016correlated}. At energies above 39 GeV, the 0-5\% net-proton {\KV} are close to Poisson expectations while at energies below 19.6 GeV, it shows large increasing above unity.  This large increase  in 0-5\% net-proton {\KV} at low energies 

We want to make several remarks : (1) One needs to remember that the resonance decay effects are not excluded in the current experimental measurements of fluctuations of net-proton, net-kaon and net-charge. Based on the hadron resonance gas model calculation~\cite{HRG_baseline}, the decay effects for net-proton {\KV} is small and at 2\% level. While for the net-charge, the decay effects are large. (2) The statistical error of cumulants ($\Delta(C_n$)) are related to the width of the distribution as $\Delta(C_n)$ $\sim$ O($\sigma^{n}$)~\cite{Delta_theory,Unified_Errors}. Thus, the wider is the distribution, the larger are statistical errors for the same number of events.  (4)  It is predicted by theoretical calculations that the net-baryon fluctuations are more sensitive to the criticality than the net-charge and net-strangeness~\cite{Wenkai_NJL,ratioCumulant}. (5) The measurements of fluctuations of conserved quantities can be used to determine the freeze-out conditions in heavy-ion collisions by comparing with the Lattice calculations and/or HRG calculations~\cite{bazavov2012freeze,Borsanyi_freeze,KaonFlu_Lattice,Alba_freeze}. 

\section{Beam Energy Scan Phase-II and STAR Detector upgrades} \label{sec:BESII}
A second phase of the beam energy scan (BES-II) program at RHIC has been planned in the years 2019-2020 and focusing on energy rang \sNN=$7.7 \sim 20$ GeV~\cite{BESII_WhitePaper}. The long beam bunches and stochastic electron cooling technique will be used to accelerate gold beams, which will increase the luminosity about by a factor of  5-15 for corresponding collision energies compared to the BES-I.  
Since the luminosity will decrease as decreasing the colliding energy, the increasing of the luminosity is much more necessary and important at low energies, such as 7.7 GeV. 
This enable us to collect more number of events ($\sim 10-20$ times) to confirm the non-monotonic trends observed in the BES-I data. Furthermore, it will allow us to draw a solid conclusion and have more complete physical pictures from various experimental measurements. To study the QCD phase structure at high baryon density, operating the STAR detector at fixed-target mode has been also proposed. In the BES-II, fixed-target mode Au+Au collisions allow us to have energy coverage from \sNN=3 GeV ($\mu_{B}$=720 MeV) up to 7.7 GeV.
In Fig.\ref{fig:BESII} left, the inner TPC (iTPC) of STAR is to be upgraded to improve the energy loss resolution and can extend the pseudo-rapidity coverage from $|\eta|<1$ to $|\eta|<1.5$~\cite{iTPC_proposal}. It is also planed to install an end cap Time-of-Flight (eTOF) detector at the west end of the STAR TPC to extend the PID capability in the forward region~\cite{eTOF_STAR}. The iTPC upgrade is very important to search for the criticality and study the dynamical evolution of the fluctuations by looking at the rapidity acceptance dependence for the fluctuations of the conserved quantities~\cite{2015_Kitazawa_Rapidity,Asakawa_review}.  In the forward and backward region of STAR detector, a new Event Plane Detector (EPD) will be also built and used to replace the Beam-Beam Counters (BBC) detector for centrality and event plane determination, which can be used to suppress the volume fluctuation and auto-correlation in the fluctuation analysis.  In Fig.\ref{fig:BESII} right, the blue band is the extrapolating from current measurements by assuming a power law behavior induced by critical fluctuations ($\KV{\propto}N^{3}$~\cite{2015_Acceptance_Misha}). The width of the blue band is the estimated statistical errors with the BES-II statistics and iTPC upgrades. 

\begin{figure}[hptb]
\centering
\hspace{-0.5cm}
\includegraphics[width=0.4\textwidth]{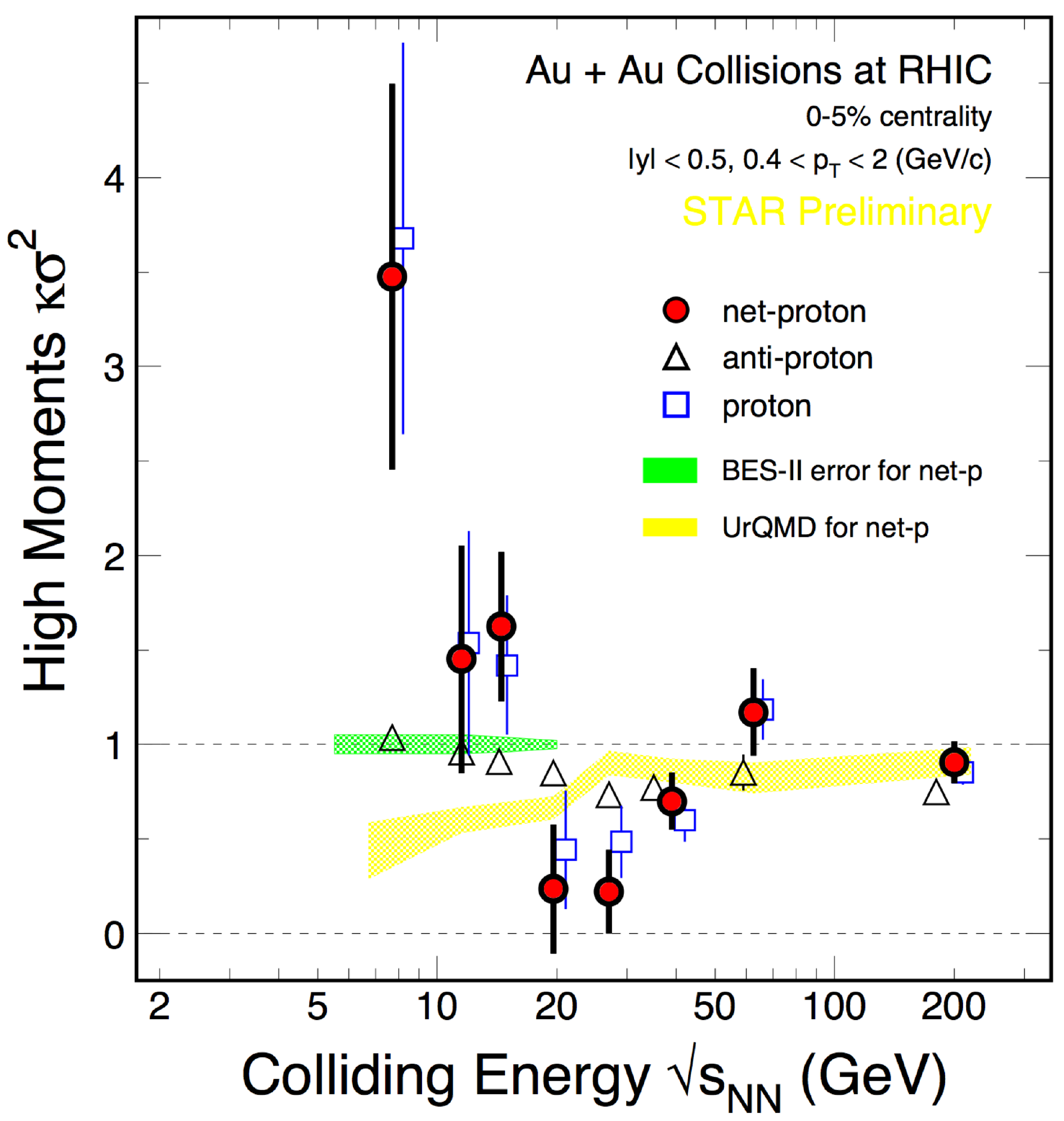}
\caption{(Color online) Energy dependence of the fourth-order fluctuations ({\KV}) of net-protons (filled-circles), anti-proton (open-triangles) an proton (open-squares) from the most top 5\% central Au+Au collisions at RHIC. Those data were taken from the first phase of the RHIC beam energy scan (BES-I) and from the kinetic region of mid-rapidity $|y| < 0.5$ and transverse momentum $0.4 < p_T < 2$ GeV/c. } \label{fig:future_KV}
\end{figure}

Figure \ref{fig:future_KV} shows the STAR preliminary results of energy dependence of the fourth-order fluctuations ({\KV}) of net-proton, proton and anti-proton from the most top 5\% central Au+Au collisions. Those data were taken from the first phase of the RHIC beam energy scan (BES-I) and from the kinematic region of mid-rapidity $|y| < 0.5$ and transverse momentum $0.4 < p_T < 2$ GeV/c. Non-monotonic energy dependence is clearly shown in the {\KV} of net-proton and proton distributions.  Although the statistical errors are large, the data shows a strong enhancement at the highest  $ \mu_B \sim 420$ MeV, corresponding to the Au+Au central collisions at \sNN\ = 7.7 GeV. This indicates an attractive correlation in nature at the large baryon density region. On the other hand, the results from the transport model UrQMD (yellow-line) show a monotonic decrease from low to high baryon density region, reflecting the fact that the baryon number conservation in such high-energy nuclear collisions. All known model calculations have shown just that. It appears that the baryon number conservation is dominant in those model simulations. Note that in the Poisson limit, the absence of criticality or other dynamical correlations, the {\KV} is expected to be unity. 
The green region in the figure is the projected error of the fourth order fluctuations {\KV} of net-protons in the second phase of the RHIC Beam Energy Scan (BES-II) program~\cite{BESII_WhitePaper}. The BES-II program, which is scheduled to take place during the years of 2019 and 2020 for the Au+Au collisions at 7.7-19.6 GeV, will take about 10 to 20 times (depending on energy) higher statistics data to confirm the non-monotonic behavior observed in the fourth order fluctuations ({\KV}) of net-protons in Au+Au collisions in the BES-I measured by STAR. Since no one expects protons freeze-out at the critical point so experimentally one should search for the critical region instead of a point~\cite{ratioCumulant,Neg_Kurtosis}. Assuming the data in the figure is related to the critical region, one must study the net-proton fluctuations at even higher baryon density region, i.e. at lower collision energies. At energy below 7.7 GeV, the collider mode experiments become inefficient so the fixed-target (FXT) mode is the way out. 
 \section{Summary}  \label{sec:summary}
In this review, we summarized the fluctuations (up to fourth order) of net-proton, net-charge and net-kaon in Au+Au collisions at {\sNN}= 7.7, 11.5, 14.5, 19.6, 27, 39, 62.4 and 200 GeV. Those data are taken in the year 2010 to 2014 and in the first phase beam energy scan program at RHIC. The corresponding baryon chemical potential ($\mu_{B}$) coverage is from about 23$\sim$420 MeV. To make precise measurements, a series of data analysis techniques have been built up to suppress the volume fluctuation and auto-correlation backgrounds. We also provide a unified description of the finite detection efficiency correction and error estimation for the various order cumulants of net-particle distributions. The statistical errors of the cumulants are related to the measured standard deviation of distributions ($\sigma$) and the particle detection efficiency as $error({C_n}) \propto \sigma^{n}/(\sqrt{N} \epsilon^{n})$. 

%\newpage

In summary, we have:

{\bf Experimental Observations:}
\begin{enumerate}[(1)]
\item{Due to larger width of the net-charge distribution and lower efficiency of charged kaons, we have bigger statistical errors of cumulants of net-charge and net-kaon than the net-proton cumulants. Within current statistical uncertainties, the energy dependence of the net-charge and net-kaon {\SD} and {\KV} are flat and consistent with Poisson expectations and UrQMD model calculations. }
\item{ In general,  various order cumulants show linear variation with the average number of participant nucleons ($\la N_{part} \ra$). The interplay of the production mechanisms for particle and anti-particle as a function of collision energy have significant impacts on the energy dependence of the cumulants.}
\item { We observed a clear non-monotonic energy dependence for the {\KV} of the net-proton, proton multiplicity distributions in 0-5\% most central Au+Au collisions measured by the STAR experiment. }
\end{enumerate} 

{\bf Theoretical and Model Calculations:}
\begin{enumerate}[(1)]
\item {The non-monotonic behavior observed in the energy dependence of the 0-5\% net-proton, proton {\KV} in Au+Au collisions are consistent with a presence of QCD critical point from model calculations, such as $\sigma$ field, NJL , PNJL and PQM models. Those model calculations suggested an non-monotonic oscillation pattern due to the sign change of the critical contributions in different QCD critical regions. However, it is still not conclusive yet and more works about the dynamical modelling of heavy-ion collisions are needed.}

\item {The large increasing in the net-proton and proton {\KV} at low energies cannot be reproduced by various transport model calculations.  All known transport model calculations show a strong suppression with respect to unity at low energies, which is dominated by the effects of baryon number conservations.}
 
\end{enumerate} 

{\bf Future directions:} 
\begin{enumerate}[(1)]
\item {Experimentally, in order to confirm the observed energy dependence structures in the high moments of net-protons in BES-I, the second phase of the beam energy scan (BES-II) at RHIC has been planned in 2019-2020 with increased luminosity~\cite{BESII_WhitePaper}. This allows us to have 10 to 20 times more statistics at energies \sNN\ = 7.7$\sim$19.6 GeV to explore the phase structure this low energy range with high precision. The upgrades of iTPC and EPD are ongoing in the STAR and will provide large rapidity coverage and forward centrality determination in the BES-II, respectively. The large rapidity coverage is very important for us to perform the rapidity dependence for the fluctuation analysis, which is crucial to test the long range correlation as well as power law behavior induced by QCD critical point. For energies below 7.7 GeV, the fixed-target mode becomes more efficient than collider mode. It also has been tested and proposed to operate the STAR detector under a fixed target mode in the BES-II.  A fixed-target experiment called Compressed Baryonic Experiment (CBM) at FAIR~\cite{CBM_physics} will start in 2024 and, in its first phase SIS100,  will cover the Au+Au collision energy range of \sNN\ =  $2.5 \sim 4.7$ GeV. This will be an ideal experiment to search for the QCD critical point at the high-baryon density region with high precision.}

\item {We have mentioned that the first-order phase boundary, the critical point and the smooth crossover are closely related thermodynamically.  At high net-baryon density region, we are searching for the signatures of the QCD critical point and/or the first-order phase boundary. However, in the near future, at the high-energy frontier, one should also search for the experimental evidence of the smooth crossover. This can be done with higher order fluctuations of conserved quantities. At the vanishing baryon chemical potential, $\mu_{B} \sim 0$, although the transition is a smooth crossover, there should have the remnant criticality of the chiral transition. Higher order fluctuations, cumulants $C_6$ (sixth order) or $C_8$ (eighth order) could show strong oscillation and should be able to pick up the possible signal in heavy-ion collisions at both RHIC and LHC. These results will not only confirm experimentally the smooth crossover nature of the transition, may also provide the information on the width of the crossover, which is one of the key information of the QCD phase diagram at small net-baryon density. On the other hand, the measurements of the various order correlation functions as a function of centrality, rapidity and energy are also very useful to further understand the critical and non-critical physics contributions.}

\item {We also want to point out that due to density fluctuations near the QCD critical point, light nuclei production and/or nucleon-clusters, such as deuteron, $^3$He and $^4$He as well as the energy dependence of the low mass di-lepton yield~\cite{CBM_physics,rapp2013dilepton,Yifei_dilepton,rapp2002chiral} could also be used to aid and complement to the critical point searches at the high baryon density region. Of course these different observables are with different systematics. Details analysis are needed in order to understand these systematic effects.}

\item {Theoretically, careful modellings for the critical fluctuations and dynamical evolution of the thermodynamic medium created in the heavy-ion collision at different energies are needed to understand the phase structure of QCD, in particular the de-confinement transition and possible critical point. Many attempts and progress have been made by physicist worldwide~\cite{2015_Swagato_evolution,Swagato_2016_PRL,Huichao_flu_QM2015, 2016_fluctuations_Huichao}. Those theoretical inputs are particularly important to establish definitive connections between experimental observables and phase structures in the QCD phase diagram.}
\end{enumerate}
\section*{Acknowledgement}
The work was supported in part by the MoST of China 973-Project No.2015CB856901, NSFC under grant No. 11575069.

\bibliography{NST_XFLUO}
\bibliographystyle{unsrt}

\end{document}